\documentclass[12pt]{article}
\usepackage{chngcntr}
\usepackage{xspace}
\usepackage{color}
\usepackage{graphicx}
\usepackage{amsmath}	
\usepackage{amssymb}	
\usepackage{lscape}
\usepackage{longtable}
\usepackage{threeparttable}
\usepackage{multirow}
\usepackage{booktabs}
\usepackage[ ]{hyperref}
\graphicspath{{Images/}}

\newcommand{\numax}{\mbox{$\nu_{\rm max}$}\xspace}
\newcommand{\dnu}{\mbox{$\Delta \nu$}\xspace}

\newcommand{\teff}{\mbox{$T_{\rm eff}$}\xspace}
\newcommand{\logg}{\mbox{$\log(g)$}\xspace}
\newcommand{\feh}{\mbox{$\rm{[Fe/H]}$}\xspace}

\newcommand{\kepler}{\emph{Kepler}\xspace}
\newcommand{\tess}{\emph{TESS}\xspace}
\newcommand{\ktwo}{K2\xspace}

\newcommand{\rtwo}[1]{{#1}}

\renewcommand*{\thefootnote}{\fnsymbol{footnote}}

\begin{document}
\noindent\textbf{\LARGE{Weakened magnetic braking supported by asteroseismic rotation rates of \textit{Kepler} dwarfs}}\\

\noindent Oliver J. Hall$^{1,2,3,*}$,
	Guy R. Davies$^{2,3}$, 
	Jennifer van Saders$^{4}$,
	Martin B. Nielsen$^{2,3}$,
	Mikkel N. Lund$^{3}$, 
	William J. Chaplin$^{2,3}$, 
	Rafael A. Garc\'ia$^{5, 6}$, 
	Louis Amard$^{7}$,
	Angela A. Breimann$^{7}$, 
	Saniya Khan$^{2,3}$, 
	Victor See$^{7}$, 
	Jamie Tayar$^{4, 8}$
	\\
	
	\noindent $^{1}$ European Space Agency (ESA), European Space Research and Technology Centre (ESTEC), Keplerlaan 1, 2201 AZ Noordwijk, The Netherlands

	\noindent 	$^{2}$ School of Physics and Astronomy, University of Birmingham, Edgbaston, Birmingham, B15 2TT, UK

	\noindent 	$^{3}$ Stellar Astrophysics Centre, Department of Physics and Astronomy, Aarhus University, Ny Munkegade 120, 8000 Aarhus C, Denmark

	\noindent 	$^{4}$ Institute for Astronomy, University of Hawai'i, Honolulu, HI 96822

	\noindent 	$^{5}$ IRFU, CEA, Universit\'e Paris-Saclay, F-91191 Gif-sur-Yvette, France

	\noindent 	$^{6}$ AIM, CEA, CNRS, Universit\'e Paris-Saclay, Universit\'e Paris Diderot, Sorbonne Paris Cit\'e, F-91191 Gif-sur-Yvette, France

	\noindent 	$^{7}$ University of Exeter Department of Physics and Astronomy, Stocker Road, Devon, Exeter, EX4 4QL, UK
	
	\noindent $^{8}$ Hubble Fellow
	
	\noindent $^{*}$ Email: \texttt{oliver.hall@esa.int} | Twitter: \texttt{@asteronomer} | GitHub: \texttt{@ojhall94}

\vspace{10mm}

\renewcommand*{\thefootnote}{\arabic{footnote}}
\setcounter{footnote}{0}

\textbf{Studies using asteroseismic ages and rotation rates from star-spot rotation have indicated that standard age-rotation relations may break down roughly half-way through the main sequence lifetime, a phenomenon referred to as weakened magnetic braking. While rotation rates from spots can be difficult to determine for older, less active stars, rotational splitting of asteroseismic oscillation frequencies can provide rotation rates for both active and quiescent stars, and so can confirm whether this effect really takes place on the main sequence.\\
We obtained asteroseismic rotation rates of 91 main sequence stars showing high signal-to-noise modes of oscillation.
Using these new rotation rates, along with effective temperatures, metallicities and seismic masses and ages, we built a hierarchical Bayesian mixture model to determine whether the ensemble more closely agreed with a standard rotational evolution scenario, or one where weakened magnetic braking takes place. The weakened magnetic braking scenario was found to be $98.4\%$ more likely for our stellar ensemble, adding to the growing body of evidence for this stage of stellar rotational evolution. This work represents a large catalogue of seismic rotation on the main sequence, opening up possibilities for more detailed ensemble analysis of rotational evolution with \textit{Kepler}.}

Gyrochronology is the study of the relationship between a star's rotation period and its age. As a star grows older along the main sequence, magnetic winds will cause it to lose angular momentum, slowing its rotation. The loss rate is thought to depend on the depth of the convection zone, which is a strong function of mass and temperature, and so the rotation period of a young star will rapidly settle on to a plane in age-colour-rotation space \cite{barnes2007}. Knowing the rotation and colour of stars therefore provides an avenue to measure ages, to a precision of $\sim 10\%$ for stars most similar to the Sun \cite{meibom+2015}. Age can otherwise be difficult to infer, and so gyrochronology enables more in-depth studies of stellar populations and individual stars \cite{leiner+2019,claytor+2019}.

Gyrochronology was previously calibrated on stellar clusters, which have well constrained ages, but only up to roughly $4\, \rm{Gyr}$ \cite{meibom+2015, barnes+2016}. Recently, ages of main sequence field stars (i.e. not in clusters) observed by \kepler \cite{borucki+2010} became more widely available through asteroseismology, the study of stellar variability \cite{silvaaguirre+2015}. When looking at these stars, disagreements were found between observations and gyrochronology beyond the middle of stars' main sequence lifetimes, which could not be reconciled with existing theories \cite{angus+2015, nielsen+2015, davies+2015}. It was proposed that at some stage in a star's evolution it undergoes \textit{weakened magnetic braking} (WMB, \cite{vansaders+2016}), where the efficiency of angular momentum loss rapidly drops. This would cause stars to maintain relatively fast rotation rates at old ages, which we would not expect from the extrapolation of existing gyrochronology relations.

The mechanism by which weakened magnetic braking occurs is subject of debate, and may be connected to changes in the magnetic field morphology \cite{vansaders+2016,reville+2015,garraffo+2016, metcalfe+2016, metcalfe+2019, see+2019}. It may also be explained from an observational point of view. A large survey of stellar rotation rates measured using star-spots found a lack of slowly rotating stars older then Sun \cite{mcquillan+2014}. As activity reduces with age, older stars with fewer spots are less likely to have their rotation measured \cite{matt+2015, reinhold+2020}. The point at which the detection probability drops appears to lie at a similar level of activity as the expected onset of weakened magnetic braking, placing these two possibilities in opposition to one another \cite{vansaders+2019}.

Determining whether weakened magnetic braking is a true phenomenon or a bias in star-spot observations requires new measurements of rotation in quiescent stars, which can be provided by asteroseismology. \rtwo{Rotation causes certain modes of oscillation -- which lie at the same frequency in a non-rotating case -- to shift to higher or lower frequencies. The size of this `splitting' provides a means of independently measuring the rotation of the star \cite{ledoux1951}}. 

Asteroseismic rotation requires a faster observing cadence and higher signal-to-noise than spot rotation, but does not require spots to be visible, meaning that we can measure rotation for older, quiescent stars that would not have been present in existing rotation catalogues. 
Here, we measured asteroseismic rotation rates of 91 main sequence stars across a broad range of  ages, and determined whether \rtwo{these new data agree} more closely with a classical rotational evolution scenario, or one in which weakened magnetic braking takes place.\\

\textbf{Asteroseismic main sequence targets:} In order to obtain robust asteroseismic rotation rates for solar-like oscillators, we required detections of multiples of both dipole (denoted as $\ell = 1$) and quadrupole ($\ell = 2$) p mode (\textit{Pressure} modes, the type of oscillation visible in main sequence stars) oscillations for each star, of which the latter have significantly lower signal-to noise. Radial oscillations ($\ell = 0$) are also visible, but do not split. In this work, we studied a sample of 95 of the highest signal-to-noise targets observed by \textit{Kepler}, combining the `Kages' \cite{silvaaguirre+2015,davies+2016} and LEGACY \cite{lund+2017, silvaaguirre+2017} catalogues. For these catalogues the mode extraction through frequency fitting \cite{davies+2016, lund+2017} and modelling using mode frequencies to obtain stellar parameters \cite{silvaaguirre+2015, silvaaguirre+2017} are covered in separate papers. Further details on these data can be found in \textbf{Methods}. 

\begin{figure}[h!]
	\centering
	\includegraphics[width=0.6\textwidth]{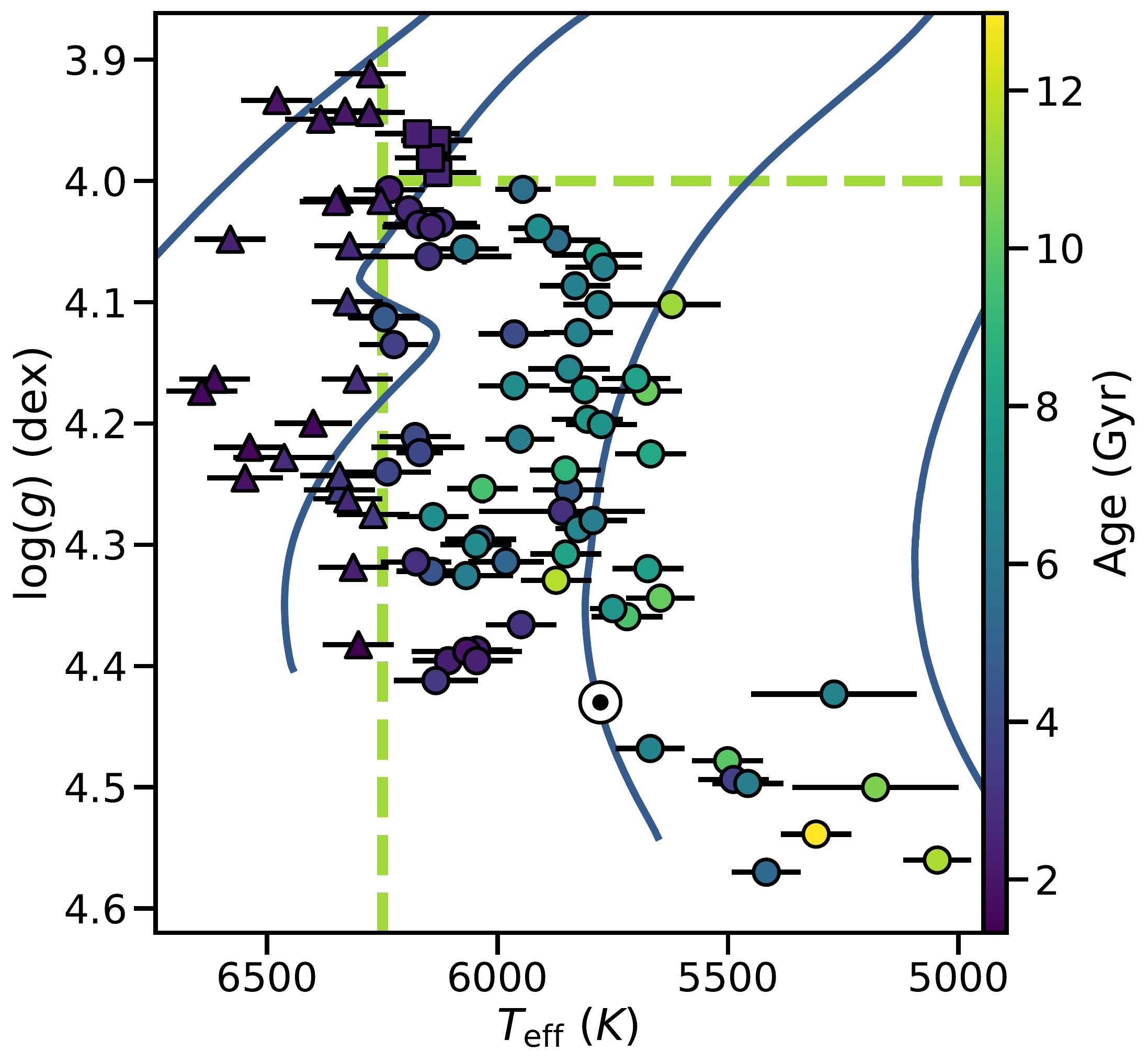}
	\caption{\textbf{Our sample of 95 stars from the `Kages' and LEGACY catalogues.} Data are coloured by asteroseismic age  \cite{silvaaguirre+2015, silvaaguirre+2017}. The dashed lines indicate our classification boundaries: main sequence (circles, $\teff < 6250\, K$, $\logg < 4\, \rm{dex}$), potential sub-giants (squares, $\teff < 6250\, K$, $\logg > 4\, \rm{dex}$), and `hot' stars (triangles, $\teff > 6250\, K$). The Sun is shown with the `$\odot$' symbol, and has an age of $4.6\, \rm{Gyr}$ \cite{bonanno+frohlich2015}. The solid lines are evolutionary tracks generated using MESA \cite{paxton+2017}, for a metallicity of $Z = 0.01493$ and helium content of $Y = 0.26588$. Left to right, they represent masses of $1.5$, $1.25$, $1$ and $0.75\, M_\odot$. Data points represent median values. Horizontal error bars represent the standard deviation on \teff. Vertical error bars represent the 68\% confidence interval on \logg, and are too small to be visible \cite{hunter2007}.}
	\label{fig:sample}
\end{figure}

We separated our sample into three categories \cite{garcia+2014}: 64 main sequence stars (MS, $\rm{log(g)} > 4.0\, \rm dex$, $T_{\rm eff} < 6250\, \rm K$); 4 potential sub-giant stars (SG, $\rm{log(g)} < 4.0\, \rm dex$, $T_{\rm eff} < 6250\, \rm K$), which may have begun to evolve off the main sequence, complicating their rotational history. ; and 24 `hot' stars (H, $T_{\rm eff} > 6250\, \rm K$), which lie above the Kraft break, the point at which convective envelopes are thin, and angular momentum loss through magnetic winds becomes inefficient \cite{kraft1967}. It should be noted that none of these stars have identified mixed dipole modes, which indicate an evolved structure (see \cite{bedding+2010}). Our conclusions are largely insensitive to these categorical assignments, and they are intended solely to help explore results category-by-category. The sample, which can be seen in Figure \ref{fig:sample}, spans surface gravities of $3.8\, \rm{dex} < \rm{log(g)} < 4.6\, \rm{dex}$, temperatures of $5000\, \textrm{K} < T_{\rm eff} <  6700\, \rm{K}$, and ages from $1$ to $\sim 13\, \rm{Gyr}$.\\

\textbf{Results of asteroseismic fitting:} The stars in `Kages' and LEGACY which form our stellar sample have been subject to detailed asteroseismic frequency fitting in the catalogue papers, but did not have reported asteroseismic rotation. Here, we repeated the mode frequency fitting with a Bayesian model that accounted for rotation in more detail, and improved the treatment of existing asteroseismic relations in frequency fitting techniques (see \textbf{Methods} and \textbf{Supplementary Information}). We fit our new model to \kepler power-spectra of 95 stars, which was successful in 94 cases. We report summary statistics used throughout the rest of this paper as the median of the posterior distribution on the model parameters, with the uncertainties being the $15.9^{\rm th}$ and $84.1^{\rm st}$ percentiles, meaning that some reported parameters (such as period) have asymmetric uncertainties.

We compared our results with both published and unpublished asteroseismic rotation rates for stars in our sample \cite{nielsen+2015,davies+2015,davies+2016, lund+2017, benomar+2018}, and found that the three stars with the lowest angles of inclination ($< 10^\circ$) were inconsistent with multiple other studies. For stars viewed near pole-on the split modes of oscillation have extremely low power, making them hard to detect reliably \cite{lund+2014}. We flagged our rotation rates for these stars, and did not use them in the next steps in our analysis, leaving a sample of 91 stars for which we considered our asteroseismic rotation measurements to be robust (for a full justification, see \textbf{Supplementary Information}).\\

\textbf{Seismic versus surface rotation rates:} Different methods of rotation measurement probe different depths of stars. Asteroseismology of main sequence stars probes internal rotation in the near surface layers, where the observed modes of oscillation are most sensitive \cite{lund+2014}. Measurements of star-spot modulation instead probe the rotation rates of star-spots on the surface, at active latitudes. Previous studies have shown that seismic rotation rates show no statistically significant deviation from spot rotation rates \cite{nielsen+2015, benomar+2015}. We confirm that this holds true for our larger sample, in order to ensure that our seismic rotation periods can be reliably used to draw conclusions on the evolution of surface rotation.

\begin{figure}[h!]
	\centering
	\includegraphics[width=0.6\textwidth]{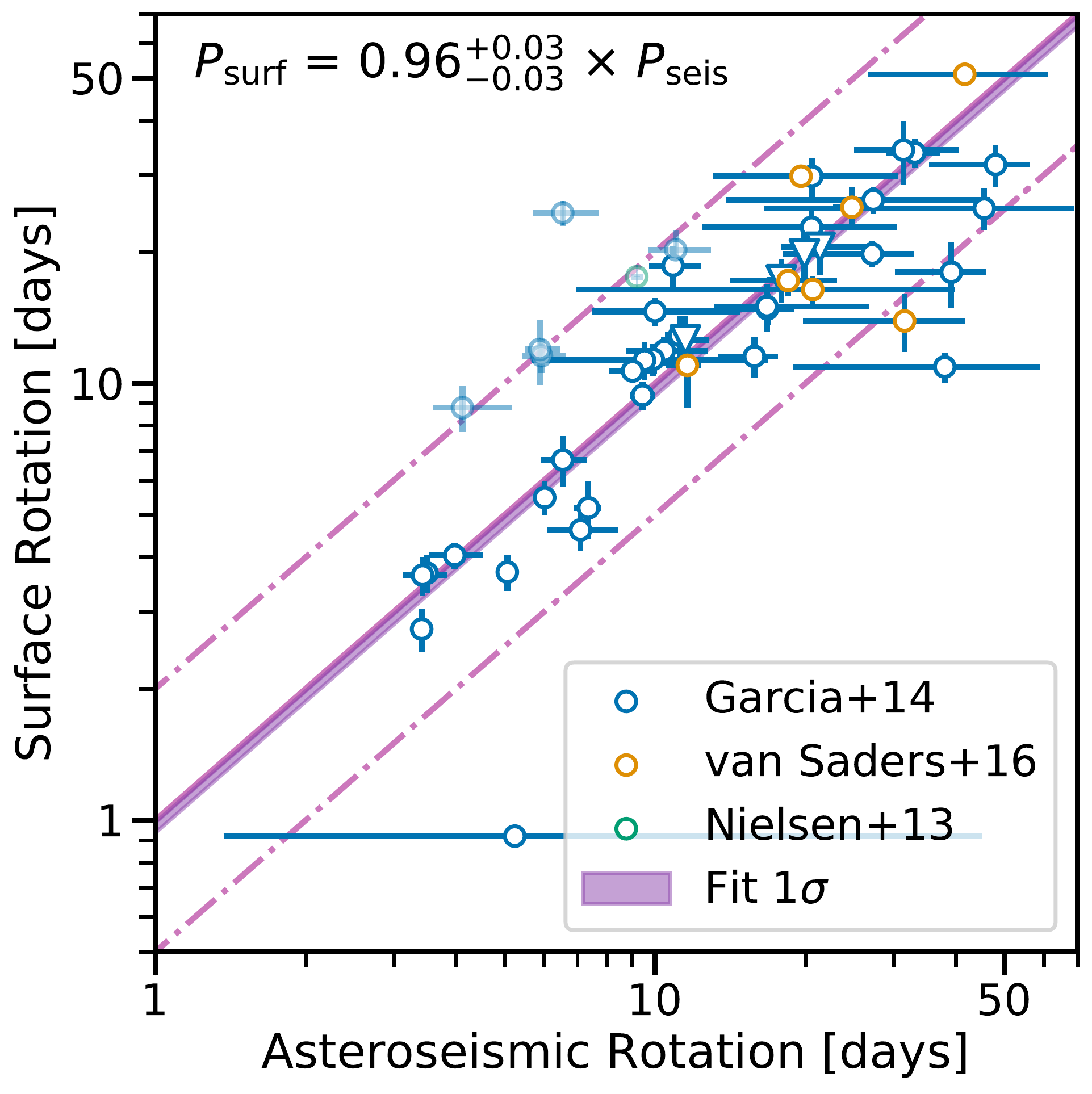}
	\caption{\textbf{Comparisons between asteroseismic and photometric measures of stellar rotation, for 48 stars.} Literature values were taken Garcia et al. (2014) \cite{garcia+2014} and Nielsen et al. (2013) \cite{nielsen+2013} (if a value was reported in both, the Garcia et al. 2014 value was used). Stars used in the original van Saders et al. (2019) \cite{vansaders+2016} study that first proposed weakened magnetic braking are highlighted. The four triangles represent stars included in the Nielsen et al. (2015) study \cite{nielsen+2015} that previously measured this relationship. The solid line indicates 1:1 line, while the dash-dotted lines represent 2:1 and 1:2 lines. The slim shaded region around the 1:1 line is the $1\sigma$ credible interval of the fit to the data, the result of which is shown as the title of the figure. Stars that were not included in the fitting process are transparent. Data points represent median values. The error bars on the asteroseismic rotation represent the $15.9^{\rm th}$ and $84.1^{\rm st}$ percentiles on the measured posterior distributions, and are smaller than the points in some cases.}
	\label{fig:protlit}
\end{figure}

Figure \ref{fig:protlit} shows a comparison between 48 stars from our sample for which spot  rotation rates had previously been measured. Rotation rates from spot modulation are subject to measuring multiples of the true rotation rate, and also sometimes over-correct for this effect. This is not the case for asteroseismic rotation, and so some stars may appear at the 2:1 and 1:2 lines in the figure. We fit a line of the form $P_{\rm surf} = m \times P_{\rm seis}$, using the larger of the asymmetrical uncertainties on asteroseismic rotation, and excluding stars above the 1.8:1 line to avoid likely multiples of the true rotation rate acting as outliers (transparent in the Figure, see \textbf{Methods} for details). Our fit found a value of $m = 0.96 \pm 0.03$, showing a close agreement ($<2\sigma$) between asteroseismic and surface rotation rates on a population level. This overall agreement between spot and seismic rotation rates is in line with previous studies of Sun-like stars \cite{nielsen+2015, benomar+2015, gizon+2013, chaplin+2013}, and so we concluded that our asteroseismic ensemble can be reliably used to draw inferences about gyrochronology.\\

\textbf{Comparing gyrochronology models:} To evaluate the implications of our \rtwo{new seismic rotation rates} for gyrochronology, we compared \rtwo{the stars in our ensemble} to two population models of rotational evolution \cite{vansaders+2019}: a `standard' model, which assumes a traditional angular momentum transport through magnetically driven stellar winds \cite{vansaders+2016,skumanich1972, kawaler1988}, and one where weakened magnetic braking takes place (hereafter the `WMB model') \cite{vansaders+2016}.  The WMB model was identical to the standard model in its input physics, except for the condition that angular momentum loss ceases above a critical Rossby number of $Ro_{\rm crit} = 1.97$. The Rossby number may be defined as $Ro \equiv P/\tau_{\rm cz}$, where $\tau_{\rm cz}$ is the convective turnover timescale, and $P$ is the rotation. The Rossby number roughly scales inversely with stellar activity (and so increases as stars age). \rtwo{Because of this, the distribution of stars with a Rossby number below this threshold will be identical in both models}.

Instead of comparing our rotation rates directly to models of individual stars, we instead compared them to \rtwo{the distribution of} synthetic TRILEGAL populations based on what the \textit{Kepler} field would look like under these two different braking laws \cite{vansaders+2019, girardi+2012}. These synthetic populations were then altered to match the distribution of existing catalogues of stars in the \kepler field in order to replicate the \kepler selection effects \cite{berger+2020}, and evolved following a standard or WMB braking law. For details on how these models were constructed, see \textbf{Methods}.

\begin{figure*}
	\centering
	\includegraphics[width=.9\textwidth]{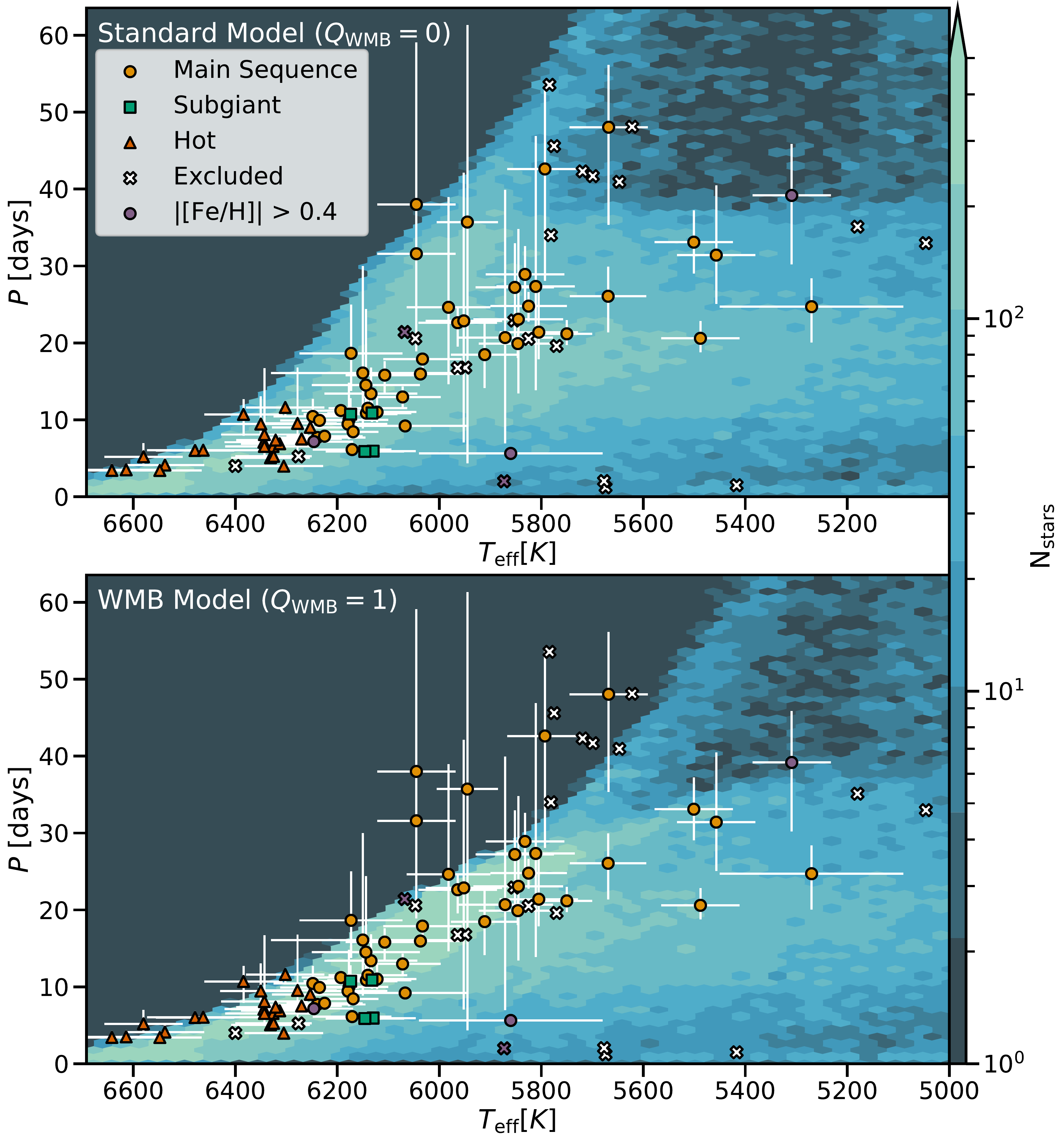}
	\caption{\textbf{Stars for which rotation was measured in this work, plotted over stellar population models evolved under standard (top) and weakened magnetic braking (WMB, bottom) prescriptions of angular momentum evolution.} The model populations were generated using TRILEGAL (see text) \cite{girardi+2012}. \rtwo{The z-axis colour scale indicates the density of stars in $P$-\teff space for the models.} The 21 stars that were excluded from the gyrochronology analysis are marked with crosses and without uncertainties (see text). Circles, squares and triangles denote main sequence, sub-giant, and `hot' stars respectively. Stars with a metallicity of $\bigl\lvert \feh \bigl\rvert > 0.4$ are shown in purple, to indicate that they fall outside the functional range of the stellar models  \cite{vansaders+2019}. Note that this figure shows the projection of a 5-dimensional space onto a 2-dimensional representation, as only period-temperature space is shown here. When evaluating between the two model prescriptions we also considered mass, age and metallicity. Data points represent median values. The error bars on the asteroseismic rotation represent the $15.9^{\rm th}$ and $84.1^{\rm st}$ percentiles on the measured posterior distributions.}
	\label{fig:fullsample}
\end{figure*}

In order to ensure our results were based on the most robust rotation measurements, we focused on the stars with the best convergence metrics in the mode frequency fitting, excluding those with a number of effective samples $n_{\rm eff} < 1000$ and a Gelman-Rubin metric of $\hat{R}<1.1$ \cite{gelman+rubin1992,salvatier+2016} (see  \textbf{Supplementary Table 1}). Two stars with low metallicities ($< -0.8\, \rm {dex}$) that fell outside the boundaries of our stellar models were also excluded. The remaining sample of 73 stars contained 4 potential sub-giants, 22 `hot' stars and 47 main sequence stars. Figure \ref{fig:fullsample} shows these stars plotted over the top of both the standard and WMB population models in rotation-temperature space, alongside the stars not used in the gyrochronology analysis (shown without uncertainties).

We evaluated our ensemble against both population models simultaneously using a Bayesian mixture model, which treated the data as being drawn from a mixture made from Kernel Density Estimates (KDEs) of both populations, modulated by a mixing parameter $Q_{\rm WMB}$. In the limit that $Q_{\rm WMB} \rightarrow 1$, the data are most likely drawn from the WMB model. In the limit $Q_{\rm WMB} \rightarrow 0$, the standard model is more likely. This mixture model was evaluated in a five-dimensional parameter space, accounting for the population distributions in effective temperature ($T_{\rm eff}$), metallicity (\feh), asteroseismic mass ($M$) and age ($t$) and our new rotation periods ($P$). For further details, see \textbf{Methods}.\\

\begin{figure}
	\centering
	\includegraphics[width=\textwidth]{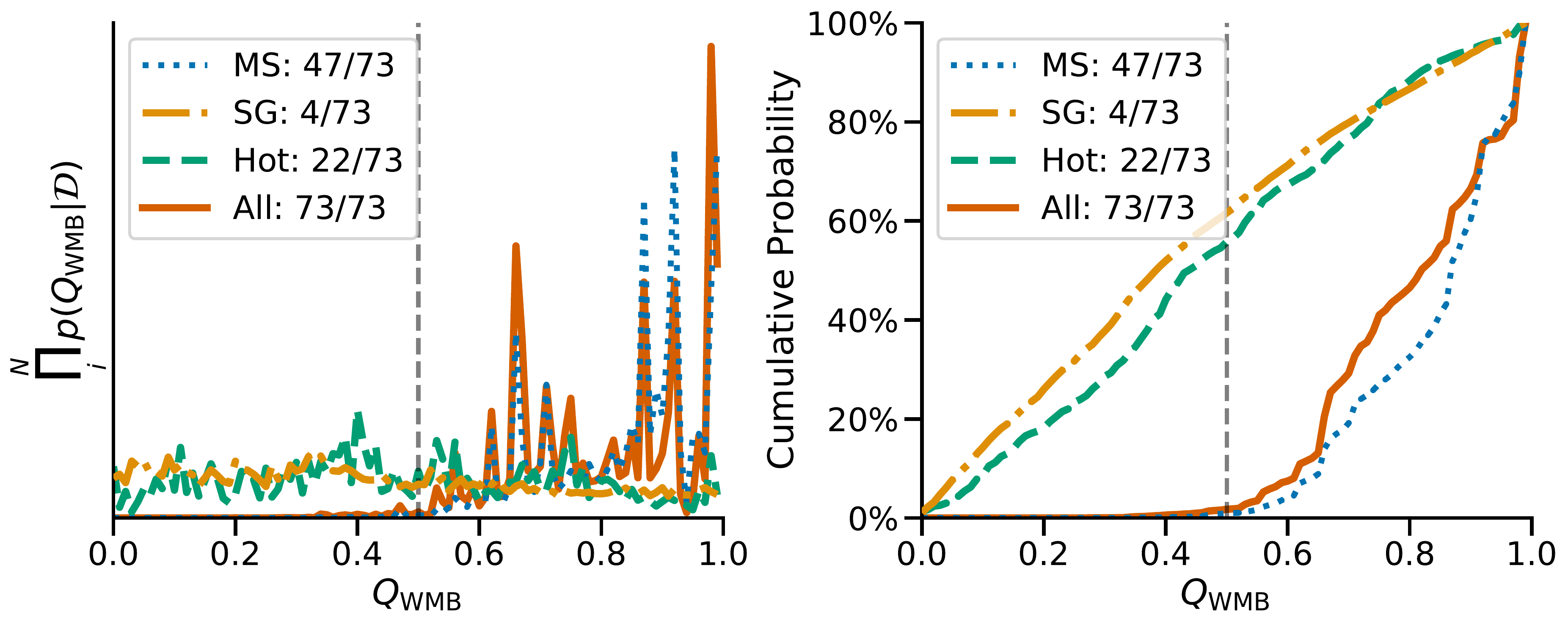}
	\caption{\textbf{Posterior estimates of the mixture model parameter $Q_{\rm WMB}$, broken down by stellar classification.} A value of $Q_{\rm WMB}$ close to 1 indicates that the data are more consistent with a rotational evolution that includes weakened magnetic braking, whereas a value close to 0 indicates that the data are more consistent with a standard rotational evolution scenario. \textit{Left:} \rtwo{the joint posterior probabilities} for 73 stars, or subsets thereof for different stellar classifications. \textit{Right:} the cumulative posterior probability for 73 stars, or subsets thereof for different stellar types. On the y-axis of the left hand plot, $p(Q_{\rm WMB} | \mathcal{D})$ represents the posterior distribution of $Q_{\rm WMB}$ given the data. The product symbol represents the combination of the posteriors for individual stars $i$, for a total of $N$ stars (which differs depending on the stellar type being shown).}
	\label{fig:gyroresults}
\end{figure}

\textbf{Results of comparing models:} We fit our mixture model to each star individually, which yielded a posterior probability density for the value of $Q_{\rm WMB}$ given the data for that star. In order to assess the posterior probability for the full sample, we multiplied the individual posterior probabilities for $Q_{\rm{WMB}}$. \rtwo{This joint posterior then describes the probability distribution of $Q_{\rm WMB}$ given multiple stars.} This was done for the full ensemble of 73 stars together, as well as for the three different stellar types separately. These joint probabilities can be seen in Figure \ref{fig:gyroresults} in the left-hand plot. The cumulative probabilities can be seen in the right-hand plot. In both of these plots, the right-hand side holds 98.4\% of the probability, indicating that the weakened magnetic braking model is preferred. When only considering the 47 main sequence stars, for which braking models are best calibrated, this rises to 99.2\%. 

Both hot and sub-giant stars do not strongly prefer one model over the other (with both cumulative probabilities crossing the $Q_{\rm{WMB}} = 0.5$ line at roughly $60\%$). \rtwo{This indicates that these stars lie below our model's critical Rossby number (and therefore in the parameter space where our two models are identical), or that they are only weakly affected by angular momentum loss (as is the case for hot stars near the Kraft break).} When performing this analysis on all 89 stars for which seismic rotation was measured (excluding the two stars with low metallicity), the total probability above $Q_{\rm{WMB}} = 0.5$ was 96.6\%. \rtwo{The total posterior probability for all stars above $Q_{\rm{WMB}} = 0.6$ and $0.7$ was $93.3\%$ and $78.2\%$  respectively. This does not necessarily mean that a mixture of the two models is preferred, as stars that have not yet undergone weakened braking, or which are near the transition period, will have a flat posterior distribution with no strong preference for either model.}\\

\textbf{Verifying consequences for gyrochronology:} In order to verify that our results presented a meaningful \rtwo{conclusion about} weakened magnetic braking, we ensured that the same results were obtained when using different subsets of the ensemble, and when exploring potential biases. We found that the joint posterior distribution did not change significantly when excluding stars with metallicities that fell only slightly outside the model range, or those with extremely small uncertainties on asteroseismic mass. Stars in our sample with known binary companions that may disrupt their rotational evolution were not found to contribute strongly to the joint posterior probability of $Q_{\rm WMB}$. 

We checked that our results remained the same if the `Kages' and LEGACY catalogues had underestimated the uncertainties on the asteroseismic mass and age, which form part of our mixture model analysis. We reran our analysis after inflating the uncertainties on these properties to align with the expected systematic uncertainty from stellar models \cite{silvaaguirre+2015,silvaaguirre+2017}. The results from this echoed what we found for the unaltered ensemble. We also accounted for the possibility that asteroseismic methods systematically overestimated stellar age, and reran our analysis after shifting the seismic ages younger by their systematic uncertainty ($6-8\%$). This resulted in $95.8\%$ of the posterior probability to lie above $Q_{\rm WMB} = 0.5$, compared to $98.4\%$ for the unaltered ensemble. From this, we can infer that our conclusions are robust against potential issues in seismic analyses. For a full discussion of these tests and other biases, see \textbf{Supplementary Information}.\\

Weakened magnetic braking was first proposed by van Saders et al. (2016) \cite[hereafter the `van Saders' study]{vansaders+2016} in response to stars with spot rotation rates faster than expected from gyrochronology at their asteroseismic ages. This discrepancy was also indicated at around the same time in other studies of asteroseismic ages of main sequence stars \cite{angus+2015, nielsen+2015,  davies+2015}. The theory of weakened magnetic braking has been both reinforced by recent studies \cite{metcalfe+egeland2019}, as well as disputed \cite{lorenzo-oliveira+2019}, at least at the critical Rossby number originally proposed.

A recent study using spectroscopic rotation measurements of 14 solar twins, compared to grids of stellar models, found marginal evidence favouring a standard rotational evolution \cite{lorenzo-oliveira+2019}. If weakened magnetic braking were to take place, this work proposed that it would most likely be outside the range of their sample, at $Ro_{\rm {crit}} \geq 2.29$, compared to the value of $1.97$ used in this work. Our sample does not overlap with theirs, so we can not directly compare to their results.
The statistical agreement between our seismic stars and a model of weakened magnetic braking at a critical Rossby number of $1.97$ over the standard scenario indicates that traces of weakened magnetic braking do occur at Rossby numbers of around $2$. \rtwo{However, we only intend to show that asteroseismic rotation rates indicate a preference for the presence of weakened magnetic braking, instead of claiming a particular critical Rossby number is correct. A measurement of the exact critical Rossby number at which these occurs is the subject of a paper in preparation.}

As a sanity check that the evidence in favour of weakened magnetic braking presented in this work is independently robust, we recreated the joint posterior probability for $Q_{\rm WMB}$, this time looking only at MS stars, and excluding any of the 22 stars that were part of the original van Saders study used to propose weakened magnetic braking. Of the 47 MS stars used to distinguish between stellar models, 16 were included in the van Saders sample. Both the 16 stars in the original van Saders study as well as the remaining MS stars favoured the WMB model, although the van Saders stars did so more strongly, not including any stars that favoured the standard model. Specifically, when considering only the van Saders stars, $96.3\%$ of the total joint probability lay above $Q_{\rm WMB} = 0.5$, compared to $91.7\%$ when considering only those stars \emph{not} included in the van Saders study. The stars initially used to propose the WMB model are still those most strongly in favour of it even when using asteroseismic rotation rates only, a result which is now further reinforced by an expanded sample that includes quiescent stars.\\

\textbf{Conclusions:} In this work, we obtained asteroseismic rotation rates for 91 stars using a hierarchical Bayesian model to fit for oscillation frequencies, convective background, and rotational splitting simultaneously. We validated these results against previously published and unpublished asteroseismic rotation measurements, producing a catalogue with a large self-consistent sample of asteroseismic rotation in main sequence stars. We evaluated the stars in our sample against synthetic population \rtwo{distribution} models of the \textit{Kepler} field evolved under either a `standard' model of rotational evolution, or one where weakened magnetic braking takes place. This evaluation was done using a Bayesian mixture model to determine whether the ensemble preferred one model over another, and was done simultaneously in asteroseismic rotation, mass, age, and spectroscopic temperature and metallicity.\\

We leave the reader with the following conclusions:
\begin{enumerate}
	\item We found that our ensemble was $98.4\%$ more likely to be drawn from a weakened magnetic braking model than a standard model of rotational evolution \cite{vansaders+pinsonneault2013}, at a critical Rossby number of $Ro_{\rm{crit}} = 1.97$. \rtwo{Stars in our ensemble overall showed a weaker rate of braking than one would expect from solar (and extrapolated solar) braking rates.} This work expands upon previous analysis \cite{vansaders+2019} by including quiescent stars older than the Sun. This conclusion was found to be robust against change of ensemble members and potentially underestimated asteroseismic systematic uncertainties. A comparison to other braking laws \cite{matt+2015} and other gyrochronology relations \cite{barnes2010} using our asteroseismic ensemble will take place in future work.
	
	\item We compared our new asteroseismic rotation rates with surface rotation measures from spot modulation. Our findings replicate those in similar studies \cite{nielsen+2015,benomar+2015}, in the sense that we find no statistically significant difference between seismic and spot modulation measures of stellar rotation in our ensemble.
	
	\item The new asteroseismic rotation catalogue presented in this work will act as an entry point for more detailed studies, including comparisons between different braking laws and individually modeled values of stellar rotation at different critical Rossby numbers.
	
\end{enumerate}

In the near future our new asteroseismic rotation catalogue will be further complemented with high-precision surface rotation measurements, activity indicators and atmospheric parameters from spectroscopic surveys such as LAMOST \cite{deng+2012}, 4MOST \cite{dejong+2014}, WEAVE \cite{dalton+2014} and SDSS-V \cite{blanton+2019, kollmeier+2019}, and new asteroseismic measurements of age from the \ktwo and \tess missions. With these surveys, large scale ensemble asteroseismology will continue to increase the possibilities for understanding gyrochronology.\\

\vspace{5mm}

\textbf{Acknowledgments}: The authors would like to thank the anonymous reviewers for their help improving this manuscript. They would also like to thank Sean Matt, Ellis Avallone, Alex Dixon, Warrick Ball and Brett Morris for helpful discussions.
OJH, GRD and WJC acknowledge the support of the UK Science and Technology Facilities Council (STFC). 
JvS acknowledges support from the TESS Guest Investigator Program (80NSSC18K18584).
MBN acknowledges support from the UK Space Agency
This work has received funding from the European Research Council (ERC) under the European Union's Horizon 2020 research and innovation programme (CartographY GA. 804752).
Funding for the Stellar Astrophysics Centre is provided by The Danish National Research Foundation (Grant agreement no.: DNRF106). 
LA, AB and VS acknowledge funding from the European Research Council (ERC) under the European Union's Horizon 2020 research and innovation program (grant agreement No. 682393 AWESoMeStars). AB also acknowledges the support of the College of Engineering, Mathematics and Physical Sciences at the University of Exeter.
RAG acknowledges the support from the PLATO and GOLF CNES grants.
JT acknowledges that support for this work was provided by NASA through the NASA Hubble Fellowship grant $\#51424$ awarded by the Space Telescope Science Institute, which is operated by the Association of Universities for Research in Astronomy, Inc., for NASA, under contract NAS5-26555.
The computations described in this paper were performed using the University of Birmingham's BlueBEAR HPC service.
This paper includes data collected by the Kepler mission and obtained from the MAST data archive at the Space Telescope Science Institute (STScI). Funding for the Kepler mission is provided by the NASA Science Mission Directorate. STScI is operated by the Association of Universities for Research in Astronomy, Inc., under NASA contract NAS 5–26555.\\

\textbf{Author Contributions:} O.J.H. led the project, with help from G.R.D., J.V.S., M.B.N and W.J.C.. J.V.S. also led the development of the stellar population models. M.N.L., R.A.G. and S.K. provided data or stellar models and along with J.T. assessed the validity of our asteroseismic results. L.A., A.B. and V.S. provided the assessment of the theoretical implications of the gyrochronology results. All authors have contributed to the interpretation of the data and the results, and all discussed and provided comments for all drafts of the paper.

\textbf{Competing Interests:} The authors declare that they have no competing financial interests.

\textbf{Correspondence:} Correspondence and requests for materials should be addressed to O.J.H. (email: oliver.hall@esa.int).\\

\clearpage
\section*{\underline{Methods}}
\section{Asteroseismic Methods}
\subsection{Asteroseismic Data}
For our asteroseismic power spectrum data we used the unweighted power spectra from the KASOC pipeline \cite{handberg+lund2014}. We did not apply any additional treatment to these data. For 16 Cyg A \& B (KIC 12069424 and KIC 12069449) we used the KEPSEISMIC lightcurves \cite{garcia+2011}, which have significantly better signal-to-noise for these two stars.

For our asteroseismic ages, we used the ages obtained by \texttt{BASTA} \cite[BAyesian STellar Algorithm]{silvaaguirre+2015} in the `Kages' and LEGACY catalogues. These ages have been obtained by comparisons of measured oscillation properties to stellar models, accounting for an expanded range of metallicities. \texttt{BASTA} is thoroughly compared to four other seismic modelling techniques in \cite{silvaaguirre+2017}. While uncertainties found through \texttt{BASTA} are typically higher than for other techniques, only \texttt{BASTA} and \texttt{ASTFIT} \cite[Aarhus STellar Evolution Code]{christensen-dalsgaard2008} recover the radius, mass and age of the Sun, when applied to solar data. Although the uncertainties on \texttt{ASTFIT} ages are overall lower, they are not published for the `Kages' sample. In order to maintain an internally consistent stellar age sample, we used age results from \texttt{BASTA} for both the `Kages' and LEGACY samples.

For our stellar masses we used asteroseismic model masses obtained by \texttt{BASTA} reported in `Kages' and LEGACY, in order to maintain internal consistency with the age measurements. We note that age and mass posteriors from \texttt{BASTA} are correlated, but chose not to account for the unpublished correlations in this work.

As described in the catalogue papers, for `Kages' stars atmospheric properties (\teff and \feh) were measured through high-resolution spectroscopy \cite{huber+2013a}. For LEGACY stars, atmospheric properties were similarly taken from one study \cite{buchhave+latham2015} for most stars in the catalogue, and complemented by other values from the literature for the remaining stars \cite[see Table 3]{silvaaguirre+2017}.

Other asteroseismic properties used in this study, such as those used as first guesses on the free parameters in our asteroseismic model, were taken from the `Kages' and LEGACY catalogues \cite{astropycollaboration+2013, astropycollaboration+2018, ginsburg+2019}. In cases where both catalogues contained the same target, we used the stellar parameters reported in LEGACY \cite{mckinney2010}. This is also the case for the masses and ages described above.

\subsection{Mode Frequency Fitting}

In order to extract signatures of stellar rotation from the asteroseismic mode frequencies, we built a model that simultaneously treats the convective background, modes of oscillation, and white noise, while accounting for rotational splitting. The foundations of this approach follows best-practices in asteroseismology \cite{davies+2015}, fitting Harvey profiles for the background \cite{harvey1985} and expressing the frequencies of the modes using an asymptotic expression \cite{tassoul1980, vrard+2016}.

The first core improvement in this work is the use of hierarchical latent variable models \cite{hogg+2010, hall+2019}, which account for small-scale deviations in mode frequencies due to effects not explicitly accounted for in our model (such as acoustic glitches \cite{mazumdar+2014}). By improving the inference of mode frequencies in this way, we also improve the ability to resolve the rotational splitting. The second improvement comes in the form of priors on our parameters, and in particular the rotational inclination, which more accurately reflects the true distribution of angles than previous techniques \cite{chaplin+basu2017}. A step-by-step breakdown of the model and all priors can be found in \textbf{Supplementary Information}.\\

We separated our model fitting into two parts. First, we fit a model for the convective background and white noise only, to the region of the power spectra that did not contain modes of oscillation. This was done using \texttt{PyStan} \cite{vanhoey+2013, carpenter+2017}, for 10,000 iterations on each star. These \texttt{PyStan} runs were initiated with a random seed of 11, as were all other random processes in this work. Second, we fit our full model (including the convective background) to the region containing the modes of oscillation. The results of the first fit to the convective background were used as extremely informative priors on the background parameters in this second fit. This was done using \texttt{PyMC3} \cite{vanderwalt+2011, salvatier+2016, thetheanodevelopmentteam+2016} for 2500 iterations each on 4 chains.

From this asteroseismic analysis, we report inclination angle ($i$), rotational splitting ($\nu_{\rm s}$), and rotation period ($P$) in \textbf{Supplementary Table 1}. The summary statistics on these parameters were taken as the median of the posterior distribution, with uncertainties being the $15.9^{\rm th}$ and $84.1^{\rm st}$ percentiles. For inclination angle ($i = \arccos(\cos(i))$) and rotation ($P = 1/\nu_{\rm s}$), the full posterior samples were transformed before taking the summary statistics, as our model sampled in $\cos(i)$ and $\nu_{\rm s}$.

We flagged any sub-optimal conditions of the final fit. We flagged 5 for which the Gelman-Rubin convergence metric, $\hat{R}$, was greater than 1.01 and 2 stars for which it was greater than 1.1 on inclination angle or rotation, where a $\hat{R} = 1$ indicates a converged result \cite{gelman+rubin1992}. We also performed visual checks of the sampled chains on all hyperparameters and of the best-fit model compared to both the raw and smoothed asteroseismic data. We found no issues in the visual investigation of 94 stars. KIC 8478994 is not reported due to both a poor unconverged fit as well as high $\hat{R}$ on rotational parameters. KICs 6603624, 8760414 and 8938364 are reported in  \textbf{Supplementary Table 1}, but were excluded from the gyrochronology analysis due to strong disagreement with multiple studies in the literature (see main body of paper). Finally, we flagged any stars with fewer than 1000 effective samples ($n_{\rm eff}$) of $\nu_s$.

\subsection{Seismic vs Surface rotation}
As described in the main body of the paper, we compared a subsample of our asteroesismic rotation rates to surface rotation rates from spot rotation, where available. Both the surface ($P_{\rm surf}$) and seismic ($P_{\rm seis}$) rotation rates have associated uncertainties. Instead of fitting the slope between these two values directly, we instead fit the distribution

\begin{equation}
	p(\frac{P_{\rm seis}}{P_{\rm surf}}\, |\, m) = \mathcal{N}(m, \sigma_\frac{P_{\rm seis}}{P_{\rm surf}})\, ,
\end{equation}

\noindent where $m$ is the slope, and $\sigma_\frac{P_{\rm seis}}{P_{\rm surf}}$ is the uncertainty corresponding to $\frac{P_{\rm seis}}{P_{\rm surf}}$, following the propagation of the errors on both values. We used the larger of the asymmetrical uncertainties on asteroseismic rotation from our analysis. The model was fit using \texttt{PyMC3} for 2000 iterations each on 4 chains. Five stars for which both surface and seismic rotation were available, but which had a surface rotation above the 1.8:1 line, were excluded from this analysis. This was to exclude likely multiples of the true rotation rate skewing the fit, as measurements of double the true rotation may occur in spot rotation measurements.

Our fit found a value of $m = 0.96 \pm 0.03$ using 48 stars. We reran this analysis using the 21 stars in this sample for which the median of their posterior probability for $Q_{\rm WMB}$ was greater than $0.5$ in our stellar model analysis, preferring the WMB model. A fit to these stars alone found a value of $m = 0.96 \pm 0.04$.

\section{Gyrochronology Methods}\label{s:gyro}
\subsection{Stellar Models}\label{ssec:models}
The braking models used in this work have several parameters that inform the rotational evolution. These are: a normalization factor to reproduce the solar rotation ($f_k$); a disk locking timescale ($T_{\rm disk}$) and period ($P_{\rm disk}$) which together regulate the stellar angular velocity during the pre-main sequence; the critical angular velocity that marks the transition from saturated (rapidly rotating) to unsaturated (slowly rotating) regimes ($\omega_{\rm crit}$), and the critical Rossby number, above which stars conserve angular momentum ($Ro_{\rm crit}$), mentioned above. $T_{\rm disk}$, $P_{\rm disk}$ and $\omega_{\rm crit}$ are calibrated to match the behavior in young open clusters, but have little impact on the rotational evolution beyond $\approx 1\, \rm{Gyr}$ in solar-mass stars. Both $f_k$ and $Ro_{\rm crit}$ affect the late-time evolution. Both models adopt $\omega_{\rm crit} = 3.4  \times 10^{-5}\, \rm{s}^{-1}$, $P_{\rm disk} = 8.1$, $T_{\rm disk} = 0.28$ and $f_k = 6.6$. In the weakened magnetic braking model, $Ro_{\rm crit} = 1.97$. \rtwo{This Rossby number is based on model comparisons to catalogues of spot rotation rates \cite{vansaders+2019}. Its exact value is not critical in our analysis, as any similar value would create a model population significantly different from the standard case. In the model, the Rossby number is defined as the rotation period divided by the convective turnover timescale, which in turn is defined as $\tau_{\rm cz} = H_p / v_{\rm conv}$, where $H_p$ is the pressure scale height at the base of the convection zone, and $v_{\rm conv}$ is the convetive velocity evaluated one pressure scale height above the base in a mixing length theory of convection. Solid body rotation is assumed \cite{nielsen+2015}.} For further details, see \cite{vansaders+pinsonneault2013} and \cite{vansaders+2016, vansaders+2019} for the standard and weakened magnetic braking cases respectively.

To construct a synthetic population of rotating stars, we expanded upon previous forward-models of the \textit{Kepler} field with the purpose of studying weakened magnetic braking \cite{vansaders+2019}, with some important improvements. We started with a TRILEGAL \cite{girardi+2012} Milky-Way simulation of the \kepler field, using the simulation's standard population values intended for this purpose. In order to replicate the \kepler selection effect, the TRILEGAL simulation was matched to the largest current catalogue of temperatures, luminosities and 2MASS $K$-band magnitudes of stars in the \kepler field \cite{berger+2020}. This was done using a nearest-neighbours approach, based on the density of stars on the HR diagram. This ensured that the TRILEGAL population matched the density of stars actually observed by \kepler, replicating its selection effects, and improved upon previous efforts \cite{vansaders+2019} by incorporating contemporary \textit{Gaia} mission data \cite{gaiacollaboration+2018}.

We made further changes to account for possible binarity in the the matching sample \cite{berger+2020}. If the first step is performed blindly, blended binaries in the sample cause an overestimation of the number of old stars. In order to overcome this, we:

\begin{enumerate}
	\item blended the TRILEGAL stars with binary companions drawn from a flat mass-ratio distribution, using a known binary fraction \cite{raghavan+2010},
	\item recalculated the `observed' luminosities and magnitudes assuming that each binary pair was blended, and
	\item shifted these stars' temperatures following a $g$-$K$ magnitude relation \cite{berger+2020}.
\end{enumerate}

This new distribution was used for the nearest-neighbour matching. Once drawn we dropped the binary companion and used the true TRILEGAL properties of those stars. For stars where the mass of the companion was $M < 0.4\, \odot$, binary contributions were ignored. Every binary was assumed to result in a blend, regardless of separation. This results in slightly more young stars than reality, because young, blended binary systems contaminate regions of the HR diagram where one expects to find old stars, and the number of blends is overestimated by assuming every binary system is a blend.

Our asteroseismic sample of stars with short cadence observations are subject to additional selection functions not included in the creation of the model populations above. We did \textit{not} explicitly account for these asteroseismic selection functions in our model, by design. Both the standard and WMB models contain stars with the same fundamental parameters (mass, radius, effective temperature, metallicity) but a different period based on the choice of rotational evolution prescription. Applying an asteroseismic selection function that depends on these fundamental parameters would have an identical effect on both models, therefore providing no net effect on our posterior distribution \cite{chaplin+2011}.  Additionally, we expected any seismic selection function to be relatively flat (and therefore uninformative) on a star-by-star scale, on which we run our model analysis. 

\subsection{Bayesian Mixture Model}
In order to determine whether weakened magnetic braking occurs on the main sequence, we compared our sample of seismic age and rotation, along with temperature, metallicity and mass, to the two stellar population models of the \kepler field \cite{vansaders+2019}, discussed above. Both stellar models were evaluated in a Bayesian framework, with the rationale of determining which of the two models (standard or WMB) is most likely to reproduce our observed data. Each model sample contained temperature (\teff), mass ($M$), age ($t$), metallicity ($\rm{[Fe/H]}$) and rotation ($P$) information.

In order to draw probabilistic inference about the models, we built a five-dimensional Kernel Density Estimate (KDE) of both model populations using the \texttt{statsmodels} package \cite{seabold+perktold2010}. We used a band-width (setting the resolution of the KDE) of $0.02\, \odot$ in mass, $10\, \rm K$ in \teff, $0.01\, \rm{dex}$ in $\ln(t)$, $0.01\, \rm{dex}$ in $\rm{[Fe/H]}$ and $0.01\, \rm{dex}$ in $\ln(P)$. Note that age and rotation were treated in log space, where the posterior estimates from asteroseismology more closely resemble normal distributions. This approach translates the population models to a probability distribution we can use in a Bayesian framework.

We evaluated our data against both models simultaneously by treating the data as being drawn from a mixture of both model KDEs. In this mixture model structure, the two KDEs were modulated by a weighting factor, $Q_{\rm WMB}$. In the limit $Q_{\rm WMB} \rightarrow 1$, the data are most likely drawn from the WMB model. In the limit $Q_{\rm WMB} \rightarrow 0$, the data are most likely drawn from the standard model.

The posterior probability of obtaining $Q_{\rm WMB}$ and additional parameters $\theta$ given our data $\mathcal{D}$ is $p(Q_{\rm{WMB}}, \theta | \mathcal{D})$. Using Bayes equation, we can express this as:

\begin{equation}\label{eq:modelll}
	p(Q_{\rm{WMB}}, \theta | \mathcal{D}) \propto p(\mathcal{D} | \theta)\ p(\theta | Q_{\rm{WMB}}, \kappa_{\rm{s}}, \kappa_{\rm{WMB}})\ p(Q_{\rm{WMB}})\, ,
\end{equation}

\noindent where $\kappa_{\rm{s}}$ and $\kappa_{\rm{WMB}}$ are the KDE functions for the standard and WMB models respectively, and $\theta$ here are parameters, $\theta = \{M, \teff, \ln(t), \feh, \ln(P)\}$. The parameters $\theta$ may be referred to as latent parameters, as they form a step between the parameter we want to infer ($Q_{\rm{WMB}}$) and our data. Using this approach allowed our model to properly take into account the observational uncertainties on the data \rtwo{in five dimensions.}

The second component on the right hand side of Equation \ref{eq:modelll} describes the probability of obtaining our latent parameters $\theta$ given our KDEs and the mixture model weighting parameter $Q_{\rm WMB}$, and is described by the mixture model

\begin{equation}\label{eq:mixturell}
	p(\theta | Q_{\rm{WMB}}, \kappa_{\rm{WMB}}, \kappa_{\rm{s}}) = Q_{\rm WMB} \times \kappa_{\rm WMB}(\theta) + (1 - Q_{\rm WMB}) \times \kappa_{\rm s}(\theta)\, ,
\end{equation}

\noindent where all parameters are as described above. This probability function describes a distribution that is a mixture of both KDEs. While the KDEs are constant, $Q_{\rm WMB}$ is a free parameter, and so the shape of this distribution can vary. The latent parameters $\theta$ are drawn from this distribution, and therefore from some combination of the two stellar models.

The first component in Equation \ref{eq:modelll} describes the likelihood of obtaining the parameters $\theta$ given our data and their observational uncertainty. It takes the form

\begin{equation}
	p(\mathcal{D} | \theta) = \mathcal{N}(\mathcal{D} | \theta, \sigma_{\mathcal{D}})\, ,
\end{equation}

\noindent a normal distribution evaluating the latent parameters $\theta$ against the observations, with observational uncertainty $\sigma_{\mathcal{D}}$. This approach means that in each parameter space (such as age), the age drawn from the stellar model mixture is entered into the likelihood equation with our static observations. The value of this equation (and thus the likelihood) will increase if $\theta$ is closer to the observations, and the mixture model will be modulated in a manner that maximises this probability, inferring whether one stellar model is more likely to produce our data than the other.

The final term, $p(Q_{\rm WMB})$, represents the prior on the mixture model weight, which is uniform between 0 and 1. A visual representation of our model is shown in \textbf{Extended Data} Figure 1.

Typically, this model would evaluate all stars in our sample against the stellar models simultaneously for a single posterior estimate of $Q_{\rm WMB}$. At 95 stars, in 5 parameter spaces, this totals 476 free parameters to marginalise over. This is not an issue for Hamiltonian Monte Carlo \cite[HMC]{betancourt+girolami2013}, however the use of KDE functions, over which a probabilistic gradient can not be measured, reduces HMCs effectiveness. Alternative Markov Chain Monte Carlo techniques \cite[MCMC]{foreman-mackey+2013} can more efficiently sample the KDE functions, but can not treat the large number of hierarchical parameters. To overcome this, we fit our model to each star to obtain an independent individual posterior distribution for $Q_{\rm WMB}$, and multiplied these post hoc to obtain a combined posterior. This comes with the benefit of easily allowing us to calculate the combined posterior for different stellar classifications, at the expense of the ability to marginalise for a single value of $Q_{\rm WMB}$ directly.\\

The parameter space of the stellar model populations were reduced before calculating the KDEs. These cuts were made in $M$, $\teff$, $\ln(t)$ and \feh, removing any stars in the models that fell more than $3 \times \sigma_{\mathcal{D}}$ outside the observations. Our observables $M$, $t$ and $P$ have asymmetric uncertainties from the Bayesian asteroseismic analysis. We used the larger of the reported uncertainties on each parameter as $\sigma_{\mathcal{D}}$ in each parameter space. 

KIC 6278762 was excluded from this stellar model analysis, because its age fell more than $3\sigma$ outside of the highest age in the stellar models (this is a metallicity issue, as the oldest stars have metallicities outside the range of the rotational model grids), and KICs 7106245 and 8760414 were excluded for the same reason due to low metallicities outside the functional range of the stellar models (-0.99 and -0.92 respectively). We also excluded any stars with $n_{\rm eff} < 1000$ for rotational splitting, and those with $\hat{R} > 1.1$. The remaining sample of 73 stars contained 4 sub-giants, 22 `hot' stars, and 47 main sequence stars.

We fit our model Equation \ref{eq:modelll} using \texttt{emcee} \cite{foreman-mackey+2013, foreman-mackey2016}, using 32 walkers for a total of 7500 samples per walker, of which the first 2500 were discarded as a burn-in.

After fitting, we took a normalised histogram of the posterior estimate of $Q_{\rm WMB}$ for each star, using 100 bins. In this histogram, each bin approximated the value of the posterior function for $Q_{\rm WMB}$. The array of 100 bins for all stars were multiplied, resulting in an approximation of the joint posterior probability function for $Q_{\rm WMB}$ given all stars in the ensemble.\\

\textbf{Data availability statement:} The core input data and results are summarised in  \textbf{Supplementary Table 1}, which is published with this article, and is also available on Vizier. Larger data files, such as stellar model populations and individual posterior distribution chains from the asteroseismic and gyrochronology model fitting are fully available on request.

This work made use of publicly available data. \kepler power spectral densities were obtained from the \href{http://kasoc.phys.au.dk/}{KASOC} webpages for the majority of stars, and from the \href{https://archive.
	stsci.edu/prepds/kepseismic/}{MAST} for 16 Cyg A and B. This work used asteroseismic data from Silva Aguirre et al. (2015, 2017), Davies et al. (2016) and Lund et al. (2017) \cite{silvaaguirre+2015, silvaaguirre+2017, davies+2016, lund+2017}. Parameter distributions of the \kepler field used to alter our stellar population models were taken from Berger et al. (2020) \cite{berger+2020}.\\

\textbf{Code availability statement:} The code required to replicate our results has been placed in a curated online repository found here: \url{www.github.com/ojhall94/halletal2021}. 

All code written in the duration of this project, along with a full commit history, can be found in an un-curated online repository here: \url{www.github.com/ojhall94/malatium}. 

The code used to construct the stellar population models used in this work is available upon request.

\renewcommand{\figurename}{Extended Data Figure}

\setcounter{figure}{0}    
\begin{figure}[h!]
	\centering
	\includegraphics[width=0.6 \textwidth]{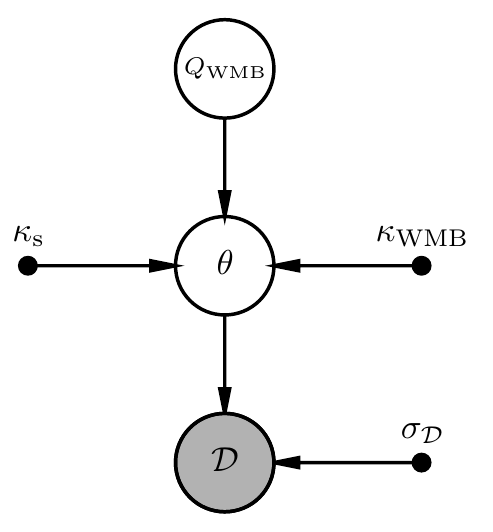}
	\caption{\textbf{A probabilistic graphical model (PGM) represented algebraically in Equation \ref{eq:modelll}.} The shaded circle indicates observed data, and solid black points represent other fixed information, such as the KDEs and observational uncertainties. The remaining circles represent parameters. The underline indicates that the symbol represents a set of parameters or data. Here, $\kappa_{\rm{s}}$ and $\kappa_{\rm WMB}$ represent the KDEs of standard and WMB model populations respectively. $Q_{\rm{WMB}}$ is the mixture model weighting factor. The latent parameters $\theta$, our observations $\mathcal{D}$ and their uncertainties $\sigma_{\mathcal{D}}$ include temperature (\teff), mass ($M$), log-age ($\ln(t)$), metallicity (\feh) and log-rotation ($\ln(P)$). This model is \textit{hierarchical}, as all the latent parameters are drawn from the common probability distribution set by $Q_{\rm{WMB}}$ and described in Equation \ref{eq:mixturell}.}
	\label{fig:pgm}
\end{figure}

\clearpage

\renewcommand{\figurename}{Supplementary Figure}

\section*{\underline{Supplementary Information}}
\tableofcontents

\section{Asteroseismic Model}\label{s:seismo}
In order to extract signatures of stellar rotation from the asteroseismic p mode frequencies, we built a model that simultaneously treats the convective background ($B(\nu)$), the oscillations ($O(\nu)$), and the white noise ($W$), where $\nu$ is frequency \cite{davies+2015}. Our data, observed with \kepler, are also subject to the apodization (attenuation) of signals in the frequency-domain, where we fit our model \cite{chaplin+2011}. The apodization in power is given by

\begin{equation}\label{eq:apodization}
	\eta^2(\nu) = \rm{sinc}^2\left(\frac{\pi}{2}\frac{\nu}{\nu_{\rm nyq}}\right)\, ,
\end{equation}
where $\nu_{\rm nyq}$ is the Nyquist frequency for the \kepler short cadence, which was treated as a free parameter in our model to account for gaps in the data. Apodization only affects signals with characteristic timescales, meaning that it does not affect the white noise level, only the oscillations and convective background components. Given the above, our comprehensive model for the power spectrum is

\begin{equation}\label{eq:model}
	M(\nu) = W + \eta^2(\nu)[O(\nu) + B(\nu)]\, .
\end{equation}

\subsection{Convective Background ($B(\nu)$)}
To model the convective background we used three Harvey components \cite{harvey1985}, which express the background in power as Lorentzian-like functions centered on zero frequency. The Harvey components take the form 

\begin{equation}
	H(\nu, a, b, x) = \frac{4a^2/b}{1 + (2\pi b\nu)^x}\, ,
\end{equation}

\noindent where $a$ and $b$ are the free parameters in our model, and $x$ is fixed. The three Harvey components together form our background function as

\begin{equation}\label{eq:background}
	B(\nu) = H(\nu, a, b, x=4) + H(\nu, c, d, x=4) + H(\nu, j, k, x=2)\, ,
\end{equation}

\noindent where we have labeled parameters for the separate components. The $x = 2$ term here contributes to the background at high frequencies, whereas the $x=4$ terms contribute the background at low frequencies.

\subsection{Modes of Oscillation ($O(\nu)$)}
Modes of oscillation appear in the power spectrum as Lorentzian peaks \cite{chaplin+basu2017}. These peaks can be described by three values: the radial order ($n$, the overtone number of the oscillation), the angular degree ($\ell$) and the azimuthal order ($m$). Due to stellar rotation, each mode with an angular degree of $\ell > 0$ is split into its $(2\ell +1)$ Lorentzian components, labeled by $m$. For all $\ell=(0,1,2)$ modes identified for our targets in the  `Kages'  and LEGACY studies \cite{davies+2016,lund+2017} we add a (set of) Lorentzian(s) to our model, building a composite model representing all visible modes. The construction of our oscillation model takes the form

\begin{equation}
	O(\nu) = \sum_n \sum_\ell \sum_{m=-\ell}^\ell \frac{H_{n,\ell,m}}{1 + \frac{4}{\Gamma^2_{n,\ell}}(\nu - \nu_{n,\ell,m})^2}\, ,
\end{equation}

\noindent where $H_{n,\ell,m}$ is the height of the mode, $\Gamma_{n,\ell}$ is the linewidth of the mode (approximated to be equal for all split azimuthal orders at a single $n$ and $\ell$) and $\nu_{n, \ell, m}$ is the frequency of the mode. The range of $n$ differs per star depending on how many radial orders were reported in LEGACY or Kages, and the range of $\ell$ depends on how many angular degrees were reported for the corresponding radial order.

\subsubsection{Mode Frequencies and Rotational Splitting ($\nu_{n,\ell,m}$)}\label{ssec:frequencies}
The mode frequencies of main sequence stars are described by the asymptotic expression \cite{tassoul1980, vrard+2016}. The asymptotic expression defines the locations of the modes as regularly spaced, with structured deviation around \numax, the frequency of maximum oscillation amplitude. The expression takes the form

\begin{equation}\label{eq:asymptotic}
	\nu_{n,\ell,m} = \dnu\left(n + \epsilon + \delta\nu_{0\ell} + \frac{\alpha}{2}(n - \frac{\numax}{\dnu} + \epsilon)^2\right) + m\nu_{s}\, ,
\end{equation}

\noindent where \dnu is the large frequency separation between two consecutive radial orders $n$, $\epsilon$ is a a phase offset, $\delta\nu_{0\ell}$ is the small frequency separation between two oscillation modes of different $\ell$ at the same radial order, $\alpha$ describes the curvature of the spacing around \numax, and $\nu_s$ is the rotational splitting. Note that here we have expressed the small separation $\delta\nu_{0\ell}$ as a fraction of $\dnu$. In order to improve the computational efficiency of this analysis, we fixed \dnu to the values reported in LEGACY and Kages.

Instead of calculating mode frequencies directly from Equation \ref{eq:asymptotic} for the model, we treated the individual mode frequencies as parameters as well, drawn from Equation \ref{eq:asymptotic}. This is called a `latent parameter' implementation \cite{hogg2012, hall+2019}, as it forms a step between the parameters we want to draw inference on (also called hyperparameters) and our data. 
The parameters $\nu_{n,\ell,m}$ were allowed to vary within an uncertainty $\sigma_{\ell}$, which has a single value for each angular degree and also varied as a free parameter.
This allowed us to account for small shifts in frequency due to sudden changes in the stellar structure \cite{mazumdar+2014}. 

The mode frequency latent parameters were drawn from a normal distribution using Equation \ref{eq:asymptotic} as its mean function, as

\begin{equation}
	\nu_{n, \ell, m} \sim \mathcal{N}(\nu_{n, \ell, m}, \sigma_{\ell})\, ,
\end{equation}

\noindent where the expression $\nu_{n, \ell, m}$ on the right hand side represents the contents of Equation \ref{eq:asymptotic}, and $\mathcal{N}$ represents a normal distribution with a mean equal to $\nu_{n, \ell, m}$ and a standard deviation equal to $\sigma_{\ell}$. The symbol `$\sim$' indicates that the parameters on the left hand side of the equation are drawn from the probability distribution on the right hand side. This notation will be used throughout this work.

\subsubsection{Mode Linewidth ($\Gamma_{n,\ell}$)}
The linewidths of asteroseismic p modes vary roughly as a function of mode frequency, and do so slowly relative to \dnu. This can be expressed as an empirical relation \cite{lund+2017, davies+2014, appourchaux+2016}. However, this relation has six free parameters, none of which are directly relevant to this work. Instead of fitting this relation, we chose to employ a more flexible Gaussian Process \cite[GP]{rasmussen+williams2006} to act as a prior on the linewidths. This can be considered as us modelling the linewidths as correlated measurements, effectively loosely constraining how linewidth varies with frequency.

A GP is defined by a covariance kernel (describing the degree of correlation between linewidths) and a mean function (describing a global trend with frequency). As this approach describes the mode linewidths relative to one another in frequency, the radial orders of each target $n$ were rescaled to be between 0 (for the lowest $n$) and 1 (for the highest $n$). The radial orders of $\ell = 2$ modes were increased by $1$ to ensure this approximation applied to all modes. This approximation was used to describe the change in linewidth as a function of frequency without depending on the exact frequencies of the modes, which themselves were free parameters (see above). Given this, we defined our GP covariance kernel as a Squared Exponential Kernel to capture the slight periodicity of linewidth with frequency, as

\begin{equation}\label{eq:gpkernel}
	K_{i,j} = \rho^2 \exp \left[ -\frac{(n_{\textrm{f}, i} - n_{\textrm{f}, j})^2}{2L^2} \right]\, ,
\end{equation}

\noindent where $n_{\rm f}$ is the fractional radial order of a given mode and  $K_{i,j}$ represents an element of the covariance matrix $\underline{K}$, describing the covariance between two values of linewidth at different fractional radial orders. The GP kernel has two hyperparameters: $\rho$, which determines the spread of the kernel in linewidth, and $L$, which determines the length scale in terms of $n_{\rm f}$. The length scale $L$ was significantly larger than the large frequency separation (\dnu) in all cases, and so we considered the use of fractional radial orders a valid approximation in this model.

A linear function was used for the mean of the GP, as

\begin{equation}\label{eq:gpmean}
	\mu = m \times n_{\textrm{f}} + c\, ,
\end{equation}
where $m$ and $c$ are the slope and intercept of the line. The linewidth latent parameters were then drawn from the multivariate probability distribution

\begin{equation}\label{eq:gammagp}
	\Gamma \sim \mathcal{N}(\mu, \underline{K})\, ,
\end{equation}

\noindent where $\Gamma$ represents the linewidths of all the modes in the model. The parameters $m$, $c$ and $\rho$ were marginalised over, whereas $L$ was fixed to a pre-determined value.

\subsubsection{Mode Heights and Angle of Inclination ($H_{n,\ell,m}$)}
The height in power of each mode, $H_{n, \ell, m}$, varies not only as a function of distance in frequency from \numax, but also due to observation conditions, such as inclination angle and passband. In our model, we treated $H_{n, \ell, m}$ as a deterministic parameter, as

\begin{equation}\label{eq:height}
	H_{n, \ell, m} =  \varepsilon_{\ell, m}(i) \frac{2 (A_{n, \ell})^2}{\pi \Gamma_{n, \ell}}\, ,
\end{equation}

\noindent where $\varepsilon_{\ell, m}(i)$ modulates the height as a function of inclination angle $i$ (see below), and $A_{n, \ell}$ and $\Gamma_{n, \ell}$ are the mode amplitude and linewidth respectively for a given radial order and angular degree. Instead of modeling and modulating height directly, we instead sampled in amplitude and linewidth. This approach mitigates the correlations between height and linewidth in the sampling process \cite{toutain+appourchaux1994}.

As done above for the mode frequencies and linewidths, the mode amplitudes $A_{n,\ell}$ were also treated as latent parameters drawn from a probability distribution governed by hyperparameters. For this, we used a Gaussian function $G(\nu)$, centered on $\numax$, as

\begin{equation}\label{eq:amplitude}
	G(\nu) = A \times \exp\left[-\frac{(\nu - \numax)^2}{2w^2}\right]\, ,
\end{equation}

\noindent where $A$ is the modes' amplitude at \numax, and $w$ is the width of the Gaussian, both free parameters in our model. 
The mode amplitude latent parameters were then drawn from the probability distribution

\begin{equation}\label{eq:amplitwod}
	A_{n, \ell} \sim \mathcal{N}(G(\nu_{n,\ell}) \times V_\ell, \sigma_{A})\, ,
\end{equation}

\noindent where $V_\ell$ is a free parameter for the mode visibility of different angular degrees, which should be consistent for all \kepler observations. These parameters describe the difference in relative height between modes of different radial order. The mode visibility for $V_0$ is fixed at 1, and $V_{1,2}$ are treated as free parameters. The parameter $\sigma_{A}$, the uncertainty on the distribution, is also a free parameter, and takes the same value for all amplitudes regardless of angular degree.\\

The angle of inclination of a star with respect to Earth changes the net perturbation by a given mode when integrated across the stellar disc, changing the amplitudes of modes of different azimuthal orders. This is a geometric problem, and is expressed by $\varepsilon_{\ell, m}(i)$, which takes the form \cite{gizon+solanki2003}

\begin{equation}\label{eq:legendre}
	\varepsilon_{\ell, m}(i) = \frac{(\ell - |m|)!}{(\ell + |m|)!}\left[P_\ell^{|m|}(\cos(i))\right]^2\, ,
\end{equation}

\noindent where $P_\ell^{|m|}$ are associated Legendre functions. For the first three angular degrees, Equation \ref{eq:legendre} takes the form \cite{handberg+campante2011}

\begin{equation}
	\begin{split}
		\varepsilon_{0,0}(i) &= 1\, ,\\
		\varepsilon_{1,0}(i) &= \cos^2(i)\, ,    \\
		\varepsilon_{1,\pm1}(i) &= \frac{1}{2}\sin^2(i)\, ,\\
		\varepsilon_{2,0}(i) &= \frac{1}{4}(3\cos^2(i) - 1)^2\, ,\\
		\varepsilon_{2,\pm1}(i) &= \frac{3}{8}\sin^2(2i)\, ,\\
		\varepsilon_{2,\pm2}(i) &= \frac{3}{8}\sin^4(i)\, ,
	\end{split}
\end{equation}

\noindent where the sum of available components for a single $\ell$ are normalized to one.

\subsection{Likelihood Function for $M(\nu)$}\label{sec:like}
If data have Gaussian noise in the time domain, they will appear in the frequency domain with noise following a $\chi^2$ distribution with two degrees of freedom \cite[$\chi^2_2$ hereafter]{appourchaux+1998}. The noise properties of $\chi^2_2$ distributed data are multiplicative, and require a specific treatment when fitting a model. As our frequency bins can be approximated to be independent, we used the likelihood function \cite{anderson+1990},

\begin{equation}
	\ln p(\mathcal{P} | M(\nu)) = \sum_{j=0}^{N-1} \left[\ln[M_j(\nu)] + \frac{\mathcal{P}_j}{M_j(\nu)}\right]\, , 
\end{equation}

\noindent where $\mathcal{P}$ is the power spectral density (and thus our data), and $M(\nu)$ represents our model. The subscript $j$ denotes an individual datum, for a total of $N$ data. We have omitted the dependence of the model $M(\nu)$ on its parameters, for clarity. This equation is functionally equivalent to the evaluation of a gamma distribution of the form $\gamma(\mathcal{P} | 1, \beta)$, where $\beta = 1/M(\nu)$, which is the implementation we used in the sampling process.

\subsection{Model preparation and hyperparameter priors}
\subsubsection{Fitting the convective background}\label{sec:background}
Fitting the convective background, apodization and white noise component must be done for the full range of the power spectrum in order to be accurately constrained. However, fitting a single model to the full range of frequencies is computationally inefficient when we are interested in the modes of oscillation, as these occupy a relatively small range of frequencies.

In order to speed up this process, the background was first fit independently to a subset of our data for each star. This subset was created by removing all frequencies within a range $0.1 \times \dnu$ below and above the minimum and maximum mode frequencies reported in LEGACY and Kages. For KIC 3427720 we also removed frequencies in the range $90\, \mu\rm{Hz} < \nu < 400\, \mu{\rm{Hz}}$, where there were large peaks not of asteroseismic origin, skewing the background fit.

For each star the model function (see Eq. \ref{eq:model}) was fit, as

\begin{equation}
	M_{B}(\nu) = W + \eta^2(\nu)B(\nu)\, ,
\end{equation}

\noindent where $B(\nu)$ is the background model described in Equation \ref{eq:background}. The parameter components of our background fit are then $\phi_B = \{\log(a), \log(b), \log(c), \log(d), \log(j), \log(k), W, \nu_{\rm nyq}\}$, where the parameters of the Harvey components were sampled in log space. The model was fit to the background data using \texttt{PyStan} \cite{vanhoey+2013}, run for 10,000 iterations on each star. 

\subsubsection{Obtaining First Guesses and Prior Values}
In order to utilise some of the prior measurements of our targets without using them as hard constraints on our parameters, some of our model equations were fit to LEGACY and `Kages' data to obtain first guesses and mean values on hyperparameter priors.

For first guesses for parameters in the asymptotic expression, we fit Equation \ref{eq:asymptotic}, \textit{not} including the rotational component $m\nu_s$, to the $\ell = (0, 1, 2)$ mode frequencies reported in LEGACY and `Kages' for each star, using their reported uncertainties. This yielded estimates of $\hat{\epsilon}$, $\widehat{\delta\nu}_{01}$, $\widehat{\delta\nu}_{02}$ and $\hat{\alpha}$, where the hat symbol `\, $\widehat{}$\, ' indicates a prior value (e.g. $\hat{\nu}_{\rm max}$ is taken from LEGACY or `Kages'). While not precise, as we did not mitigate any perturbations due to acoustic glitches, these rough results act as functional first guesses and prior mean values. The relation was fit to each star using PyMC3 \cite{salvatier+2016} using 5000 iterations on 4 chains.

To obtain first guesses for the parameters used to set the GP prior on linewidth, we fit a GP constructed as in Equation \ref{eq:gammagp} to the linewidths of the $\ell = 0$ modes reported in LEGACY. Linewidths were not reported for the other angular degrees in LEGACY, but the estimates may be generalised to other $\ell$, as linewidth is a strong function of frequency.  The relation was fit to each star using PyMC3 using 2500 iterations on 4 chains.

Fitting the LEGACY linewidths yielded rough estimates of $\hat{m}$, $\hat{c}$, $\hat{\rho}$ and $L$ for each star. As is noted in Equation \ref{eq:gammagp}, $L$ was fixed to this fit value when fitting our full model to our data. For stars in `Kages', for which no linewidths were reported, we instead fixed these prior values to $\hat{m} = 1$, $\hat{c} = 0.5$, $\hat{\rho} = 0.1$, and the length scale to $L = 0.3$. These values were chosen to reflect those found for the LEGACY stars.

Finally, we obtained prior values for the Gaussian function describing the distribution of mode amplitudes around \numax (Eq. \ref{eq:amplitude}). The mode amplitude of the highest peak in the spectrum was used for $\hat{A}$, which was typically at or near \numax. For the width of the Gaussian we used the empirical function \cite{lund+2017}

\begin{equation}
	\hat{w} = 0.25 \times \hat{\nu}_{\rm max},\, .
\end{equation}

\noindent For the mode visibilities, we used $\hat{V}_1 = 1.2$ and $\hat{V}_2 = 0.7$, which reflect the results for these parameters reported in the LEGACY catalogue.

\subsubsection{Priors on our Hyperparameters}
Given our first guesses and measured prior values, we can define the prior probabilities of the hyperparameters on which our model depends. For the mode frequency hyperparameters (Eq. \ref{eq:asymptotic}), these are

\begin{equation}
	\begin{split}
		\numax &\sim \mathcal{N}(\hat{\nu}_{\rm max}, 10)\, ,\\
		\epsilon &\sim \mathcal{N}(\hat{\epsilon}, 1)\, ,\\
		\alpha &\sim \ln\mathcal{N}(\ln(\hat{\alpha}), 0.01)\, ,\\
		\delta\nu_{01} &\sim \ln\mathcal{N}(\ln(\widehat{\delta\nu}_{01}), 0.1)\, ,\\
		\delta\nu_{02} &\sim \ln\mathcal{N}(\ln(\widehat{\delta\nu}_{02}), 0.1)\, ,\\
		\sigma_{0,1,2} &\sim \mathcal{C}_{1/2}(\beta = 2)\, ,
	\end{split}
\end{equation}

\noindent where $\ln\mathcal{N}$ represents a log-Normal distribution and $\mathcal{C}_{1/2}$ represents a half-Cauchy distribution. The half-Cauchy distribution ensures the standard deviations do not inflate to large numbers, and is generally well-behaved close to zero in the case of stars with little deviation from Eq. \ref{eq:asymptotic} \cite{gelman2006}. Other symbols are as described above. All three hyperparameters $\sigma_{0,1,2}$ describing the uncertainty on the latent parameters of different angular degree were subject to the same prior.

For the mode linewidths (Eq. \ref{eq:gpkernel} and \ref{eq:gpmean}), our hyperparameter priors took the form

\begin{equation}
	\begin{split}
		m &\sim \mathcal{N}(\hat{m}, 1)\, ,\\
		c &\sim \mathcal{N}(\hat{c}, 1)\, ,\\
		\rho &\sim \ln\mathcal{N}(\ln(\hat{\rho}), 0.1)\, ,\\
	\end{split}
\end{equation}

\noindent where the conventions are the same as above. For our mode amplitudes (Eq. \ref{eq:amplitude} and \ref{eq:amplitwod}), they took the form

\begin{equation}
	\begin{split}
		w &\sim \ln\mathcal{N}(\ln(\hat{w}), 10)\, ,\\
		A &\sim \ln\mathcal{N}(\ln(\hat{A}), 1)\, ,\\
		V_1 &\sim \ln\mathcal{N}(\ln(\hat{V}_1), 0.1)\, ,\\
		V_2 &\sim \ln\mathcal{N}(\ln(\hat{V}_2), 0.1)\, ,\\
		\sigma_A &\sim \mathcal{C}_{1/2}(\beta = 1)\, .
	\end{split}
\end{equation}

As the convective background had already been fit to our data excluding the region where the modes are present, the results from that fit could be used as extremely informative priors on our fit to the region containing the modes, where there is little information present to constrain the background. To do so, we modeled the background parameters $\phi_B$ in our full model as being drawn from a multivariate normal distribution as

\begin{equation}
	\phi_{B}\sim\mathcal{N}(\hat{\phi}_{B},\underline{\Sigma}_{\hat{\phi}_{B}})\, ,
\end{equation}

\noindent where $\hat{\phi}_B$ are the median values of our posterior distributions from our prior background fit, and $\underline{\Sigma}_{\hat{\phi}_{B}}$ is the full covariance matrix of all the posterior distributions from our prior background fit, taking into account the correlations between the different Harvey components.

Finally, we defined the priors on the rotational parameters: the mode splitting ($\nu_s$), and the inclination angle ($i$). In order to give these an appropriate treatment, we made two reparameterizations. First, we sampled the projected rotational splitting, $\nu_{\rm{s}}\sin(i)$, which is more efficiently sampled due to the strong correlations between $i$ and $\nu_{\rm s}$ \cite{ballot+2006,ballot+2008a}. A prior was applied over this as

\begin{equation}
	\nu_s\sin(i) \sim \ln\mathcal{N}(\ln(0.75), 0.75)\, ,
\end{equation}

\noindent where conventions are as above. This subjective prior was chosen to reflect that most stars will have a solar-like rotation, with a long tail to allow for the fastest rotators. Second, we sampled in $\cos(i)$, and gave this a prior of

\begin{equation}
	\begin{split}
		\cos(i) &\sim \mathcal{U}(0, 1)\, ,
	\end{split}
\end{equation}
which is equivalent to stating that probability to observe an inclination angle $i$ is equal to $\sin(i)$. Here, the $\mathcal{U}(0,1)$ indicates a uniform prior between 0 and 1. Using a uniform prior on $\cos(i)$ allowed us to account for the geometric effect that stars with a large inclination angle with respect to us are more common \cite{chaplin+basu2017}.

\subsection{Fitting procedure}
Using our prior information and model described above, we fit Equation \ref{eq:model} to our data $\mathcal{P}$, using the likelihood function described in Section \ref{sec:like}.

In order to speed up the fitting process, we only applied our model to the region of the power spectrum that contains visible modes of oscillation. We created this subset by removing all frequencies outside a range $0.25 \times \Delta\nu$ below and above the minimum and maximum mode frequencies reported in LEGACY and Kages. This region overlaps minimally with the data used to fit for our prior information of the convective background (see Section \ref{sec:background}), so that both model fits are independent.

To improve computational efficiency, we reduced the number of oscillation modes being fit in five targets. For 16 Cyg A \& B, KIC 7970740 and KIC 8478994, we excluded any modes with a Bayes Factor ($\ln(K)$) of less than 6, as reported in LEGACY \cite{davies+2016,lund+2017,kass+raftery1995}. For KIC 8478994, which is reported without a value for $\ln(K)$ in Kages, we only included modes of an overtone number that contained a detection for all of $\ell = (0, 1, 2)$, retaining 5 sets of higher signal-to-noise overtones. We do not expect this reduced scope to bias our results, although they may reduce the precision on our measured rotation rates.

We fit our model to our power spectrum data with \texttt{PyMC3}, using 2500 iterations each on 4 chains. An example of our model fit to an asteroseismic power spectrum of 16 Cyg A is shown in Supplementary Figure \ref{fig:modelfit}.

 \begin{figure}
	\centering
	\includegraphics[width=.99\textwidth]{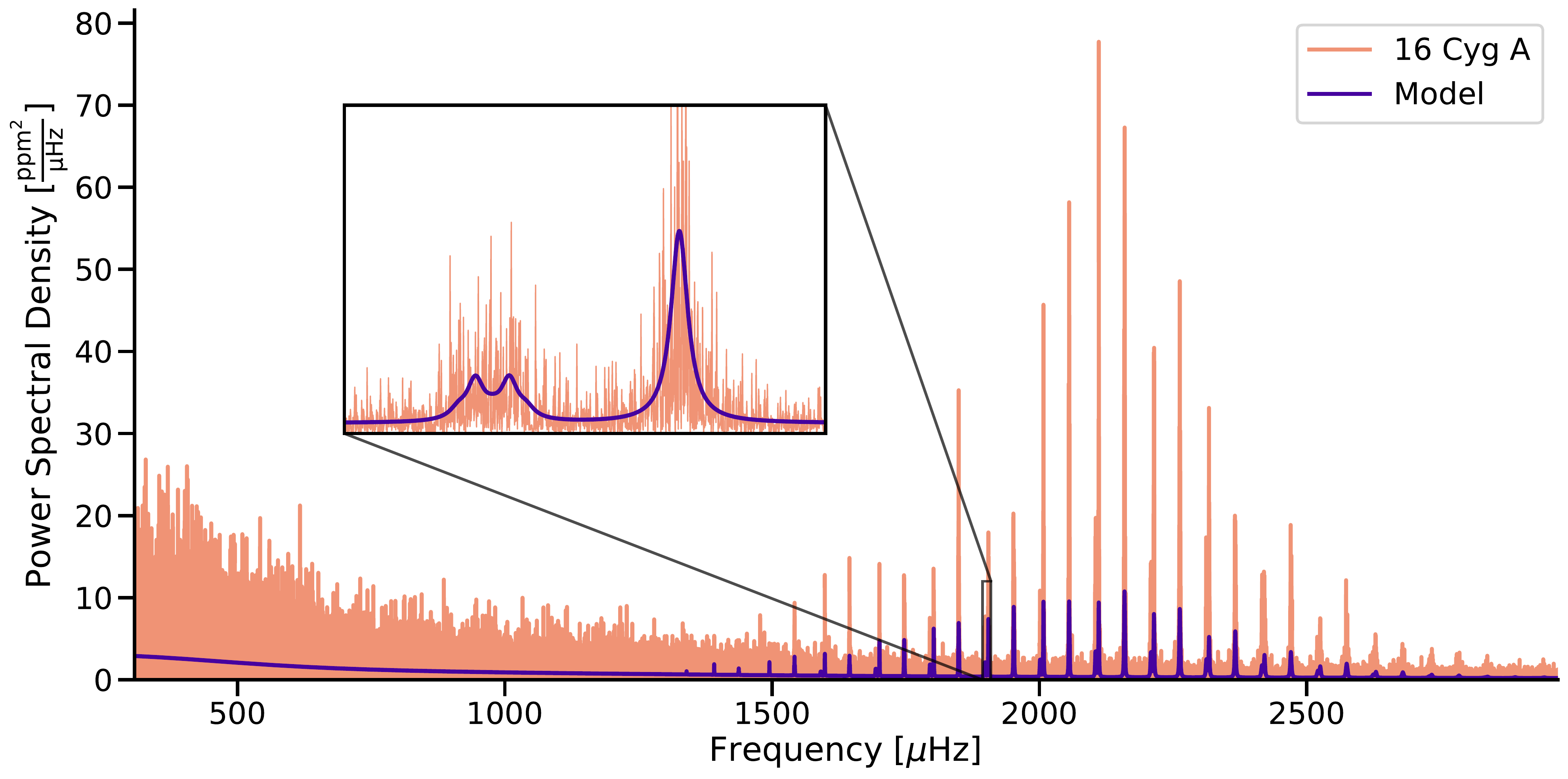}
	\caption{A power spectrum constructed from four years of \kepler observations of 16 Cyg A (KIC 12069424). Plotted over the top is the model resulting from the fit to the data described in this work. The model implements both the mode frequencies, seen on the right hand side of the plot, and the convective background, the effects of which are seen on the left. Low frequencies have been cropped out for clarity. \textit{Inset}: A zoom in on a radial (right) and quadrupole (left) ($\ell = 0, 2$) pair of modes. The quadrupole mode is split into five components by the star's rotation. Due to the star's inclination angle with respect to us, two out of five peaks are more distinct. The height and spacing of the mode components is a function of the star's rotational splitting ($0.56\, \mu\rm{Hz}$, equivalent to $P = 20.5\, \rm{days}$) and angle of inclination ($45^\circ$).}
	\label{fig:modelfit}
\end{figure}

\section{Verifying asteroseismic results}
\subsection{Priors on rotational parameters}
In our Bayesian analysis, we have placed weakly informative priors on our sampled rotational parameters, $\nu_{\rm s}\sin(i)$ and $\cos(i)$. The prior is especially important for the angle of inclination, which is hardest to infer from the data. We are able to validate the robustness of our asteroseismic results by confirming that their posterior distributions are data-dominated, and not prior-dominated. We can do so by comparing the $68\%$ credible regions of the posterior estimates of $\nu_{\rm s}\sin(i)$ and $i$ against the $68\%$ credible regions of their priors.

A comparison between prior and posterior is shown for 94 stars in $\nu_{\rm s}\sin(i)$, $i$ and $P$ in Supplementary Figure \ref{fig:priors}, arranged by age. In the Figure, results with means (symbols) and $68\%$ credible regions (error bars) that are close to those same values for the prior distribution (where the horizontal line is the mean, and the shaded area is the credible region) can be interpreted as prior-dominated (i.e. poorly informed by the data). Cases where the means differ or the credible regions are smaller than the prior distribution are data-dominated. The projected splitting, $\nu_{\rm s}\sin(i)$, is overall well constrained, with only one star being prior dominated. This is expected, as the projected splitting is what we observe on the star before decoupling inclination and rotation. The angle of inclination $i$, sampled as $\cos(i)$, more closely follows the prior distribution in most cases. Combining the two, the rotation period $P$ has no stars directly corresponding to the effective prior on period, and globally follows a trend with increasing age. The three outliers with fast rotation at late ages (KICs 6603624, 8760414 and 8938364) are discussed in more detail in below.

The rotation rates as presented in this work are a product of our Bayesian sampling of both projected splitting and angle of inclination. As seen in the Figure there are instances where $i$ or $\nu_{\rm s}\sin(i)$ closely resemble the prior (and are therefore prior-dominated). There are no cases of this when looking at the resulting period measurements, as they will have been informed by at least one strongly data-driven parameter (commonly $\nu_{\rm s}\sin(i)$). From this, we concluded that our ensemble of asteroseismic rotation is not strongly dominated by the priors imposed on projected splitting and inclination in our Bayesian analysis.

\begin{figure}
	\centering
	\includegraphics[width=\textwidth]{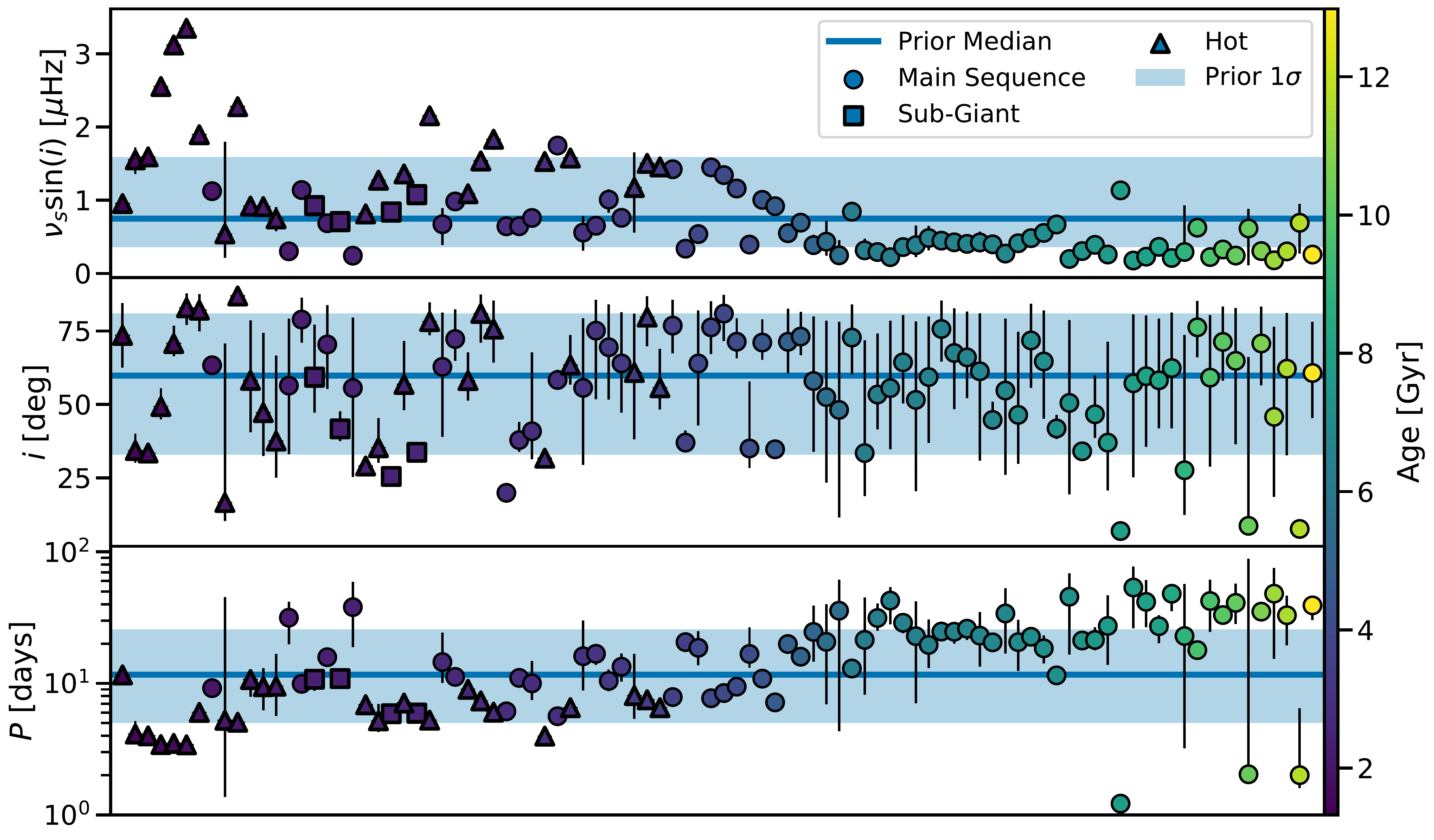}
	\caption{\textbf{Comparisons between posterior estimates of rotational parameters (data points and error bars) and the priors on these parameters (shaded regions).} The data are sorted from young stars (left) to old (right). Shown are projected splitting ($\nu_{\rm s}\sin(i)$), inclination angle ($i$, sampled as $\cos(i)$) and rotation period ($P$). Both the prior and values for $P$ are a transformation of the upper two parameters (see text). In all cases, the extent of the error bars and shaded regions indicate the 68\% credible interval of the posterior and prior distributions respectively. The solid lines indicate the median of the prior distributions. In this figure, results with means and errorbars that closely resemble the prior distribution can be interpreted as prior-dominated (i.e. poorly informed by the data). All stars are sorted and coloured and sorted by age. In the case of inclination angle $i$ and rotation period $P$, the displayed priors are transformed from the priors imposed on the sampled parameters from which their posteriors were derived.}
	\label{fig:priors}
\end{figure}

\subsection{Comparisons to previous studies}\label{ssec:litcomp}
In order to validate our results, we compared our rotational parameters to those obtained in the literature, as well as those resulting from the work presented in LEGACY and `Kages', which were unreported and received through private communication by the authors of the catalogue papers.

Comparisons with LEGACY and `Kages' are shown in Supplementary Figure \ref{fig:legacykages} for projected splitting, inclination angle, and rotation period. In all three cases we show the fractional difference between the values obtained in this work and those from LEGACY and `Kages'. On the right of the Figure, we show the distribution of the fractional differences for the three parameters.
The projected splitting is in good agreement with both studies, however LEGACY finds slightly lower $\nu_{\rm s}\sin(i)$ for the faster rotators, deviating from our work by over $1\sigma$. Neither LEGACY nor `Kages' used a spatially isotropic prior for the inclination angle in their analyses, instead opting for a uniform prior. As posterior estimates of inclination angle are only loosely data-driven, the introduction of an isotropic prior should result in our analysis reporting globally higher inclination angles. This effect is seen in the comparisons of both inclination angles and rotation rates for the LEGACY stars, where we find overall lower rotation rates compared to LEGACY for stars at very similar $\nu_{\rm s}\sin(i)$.

A number of stars are excluded from Supplementary Figure \ref{fig:legacykages} and compared individually as extreme outliers: KICs 5094751, 6196457, 8349582, 8494142, 8554498, 105114430 and 11133306 all have fast rotation rates ($<5\, \rm days$) in `Kages', but are found to have a broader spread of rotation rates in this work. At similar values for $\nu_{\rm s}\sin(i)$, `Kages' found much lower inclination angles with highly asymmetrical uncertainties. Based on a comparison between the summary statistics of these stars, we concluded that the results found in this work have better marginalised over inclination angle, improving our measure of rotation.

Conversely, KICs 6603624, 8760414 and 8938364 have extremely slow rotation periods in LEGACY, but extremely \textit{fast} ($<3\, \rm {days}$) in this work. KICs 8760414 and 8938364 are excluded from the gyrochronology analysis below based on checks for $\hat{R}$ and the number of effective samples. Both stars have ages greater than $10\, \rm Gyr$, making their fast rotation rates highly unlikely under any model of rotational evolution. The posterior estimate of $P$ for KIC 6603624 is well-defined in our analysis, but with an age of $7.8\, \rm Gyr$, its measured rotation of $1.2\, \rm days$ is also highly unlikely under any model of rotational evolution. The LEGACY estimate of rotation is similarly extreme at $378\, \rm days$. These three stars have the lowest inclination angles in our sample ($< 10^\circ$), at which point the power in the split components of the seismic modes is so low that it becomes difficult to probe the measure of splitting. The split components for these stars will have a height roughly 3\% of the central mode. For comparison, for the next lowest inclined star at $17^\circ$, this rises to $10\%$. In cases such as these with lower signal-to-noise, spurious peaks may be interpreted as split components.\\

\begin{figure}[h!]
	\centering
	\includegraphics[width=\textwidth]{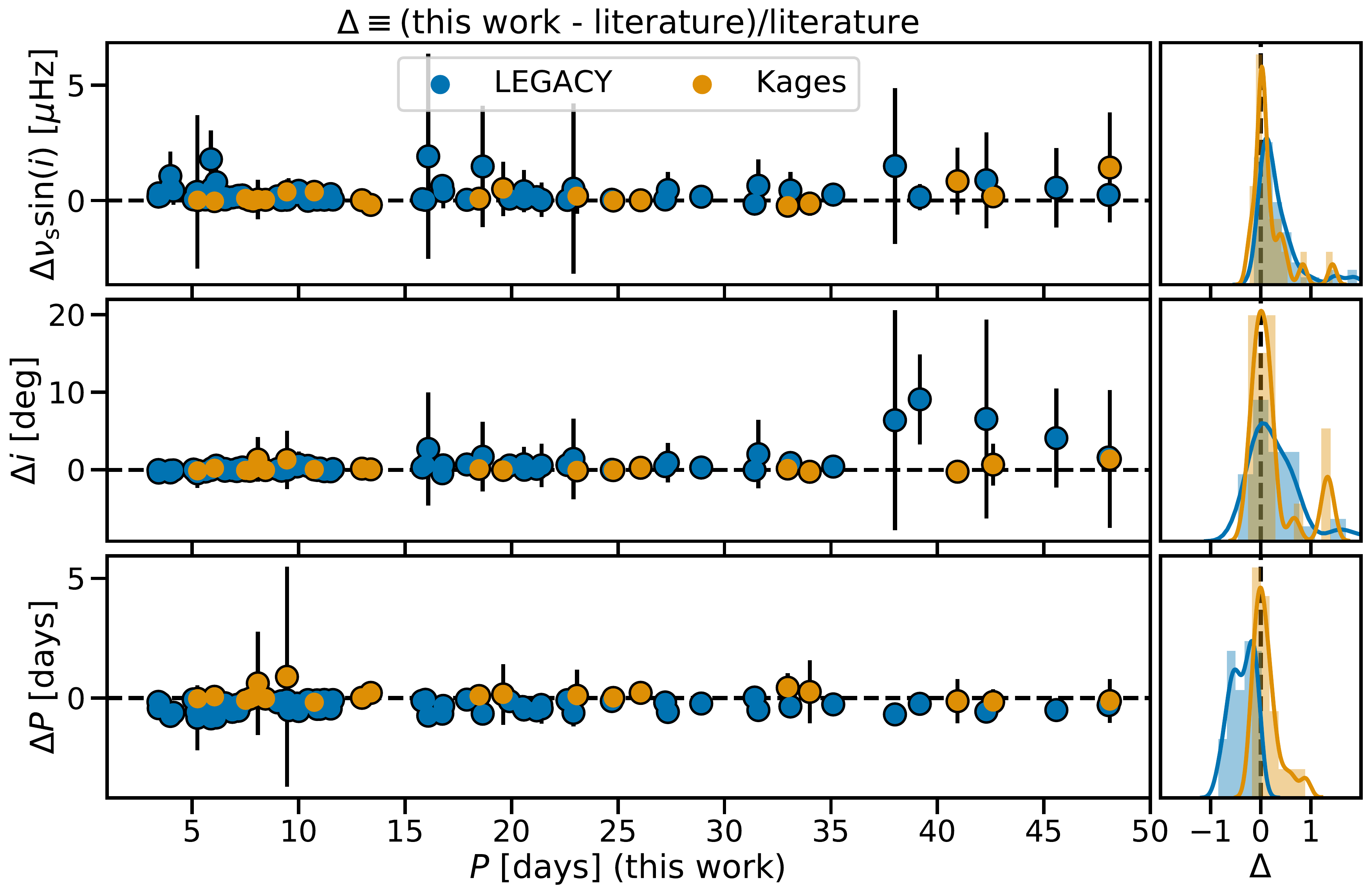}
	\caption{\textbf{Comparisons between posterior estimates of rotational parameters from this work and the literature}. Literature values are taken from LEGACY and `Kages' \cite[private communication]{davies+2016, lund+2017}. Shown are projected splitting ($\nu_{\rm s}\sin(i)$), inclination angle ($i$, sampled as $\cos(i)$) and rotation period ($P$). Fractional differences are plotted against stellar rotation obtained in this work. The $\Delta$ indicates the fractional difference between this work and the literature (i.e. stars above the zero-line have higher values in this work). The right hand panels the distribution of the fractional differences around the zero line. The colour legend is consistent throughout all panels. The x-axis units on the right hand panels are equivalent to the y-axis of the left hand panels. 10 stars have been omitted from this plot and are discussed in more detail in the text: KICs 5094751, 6196457, 8349582, 8494142, 8554498, 105114430 and 11133306 all have extremely low rotation periods in Kages, with high uncertainties. Conversely, KICs 6603624, 8760414 and 8938364 have extremely high rotation periods in LEGACY with low uncertainties. Error bars represent the 68\% confidence intervals. In cases where stars had asymmetric error bars, the larger of the two was used when propagating uncertainty for the purposes of this figure.}
	\label{fig:legacykages}
\end{figure}

We also compared our asteroseismic estimates of stellar rotation with similar studies in the literature, shown in Supplementary Figure \ref{fig:literaturecomp}. These included: a study of the binary solar analogues 16 Cyg A \& B \cite{davies+2015}, ; a study of surface and seismic rotation which our catalogue shares 5 stars with \cite{nielsen+2015}, ; and an asteroseismic study of differential rotation with which our catalogue shares 40 targets \cite{benomar+2018}. For the latter, we used their reported splitting value $a_1$, which is equivalent to $\nu_{\rm s}$. 

Overall,  Supplementary Figure \ref{fig:literaturecomp} shows no strong disagreements between our asteroseismic measurements for stellar rotation and those from the literature. The scatter of the fractional differences lies cleanly around the zero line, with a mean and spread of $0.0_{-15.6}^{+16.4}\, \%$. The increase in uncertainty with period is due to more slowly rotating stars being more difficult to constrain using asteroseismology.

It is of note that 16 Cyg A was found to be rotating slightly faster in our analysis compared to the literature \cite{davies+2015} (deviating within $2\sigma$), despite the fit being performed on the same data. We found an inclination angle that is slightly lower for 16 Cyg A but at a similar projected splitting, which would explain finding a lower value of rotation.

There are three outliers at low period in  Supplementary Figure \ref{fig:literaturecomp}: KICs 6603624, 8760414 and 8938364. These are the same targets found to be outliers in a comparison to the LEGACY and `Kages' measurements (see above), with anomalously fast rotation rates and low inclination angles. As these stars represent the lowest inclination angles in our sample, and were found to disagree with two independent studies, we opted to exclude them from the gyrochronology analysis, and to flag the rotation measurements for these three stars presented in this work.

\begin{figure}[h!]
	\centering
	\includegraphics[width=0.8\textwidth]{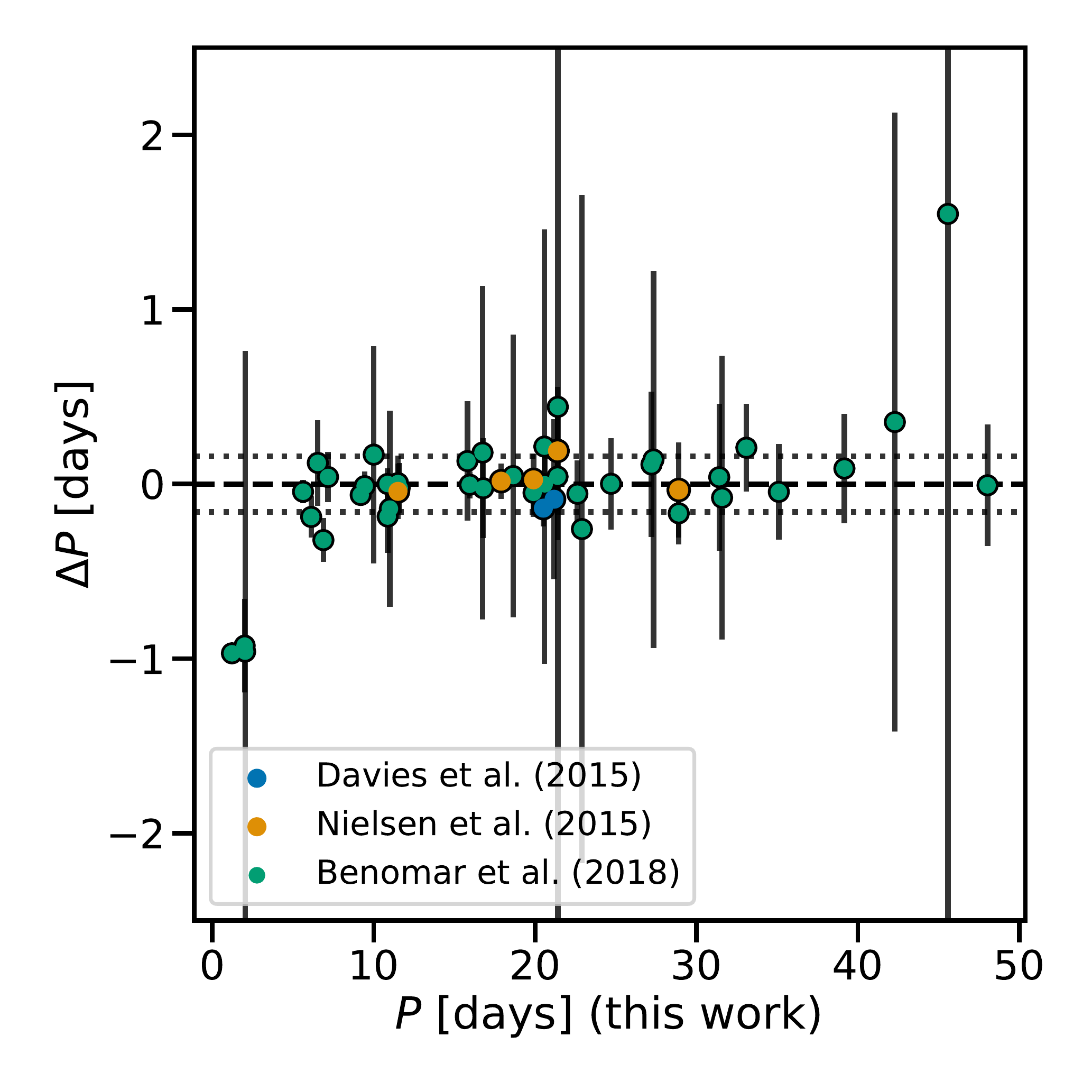}
	\caption{\textbf{Fractional differences between posterior estimates of asteroseismic rotation period from this work.} Literature sources are: Davies et al. (2015) \cite{davies+2015} (16 Cyg A \& B), Nielsen et al. (2015) \cite{nielsen+2015} (5 stars) and Benomar et al. (2018) \cite{benomar+2018} (40 stars). We used the reported parameter $a_{1}$ from the latter, which represents the rotational splitting in the case of uniform latitudinal rotation in their model. The dashed line represents the median of the sample shown, with the dotted lines representing the $15.9^{\rm{th}}$ and $84.1^{\rm{st}}$ percentiles. Error bars represent the 68\% confidence intervals. In cases where stars had asymmetric error bars, the larger of the two was used when propagating uncertainty for the purposes of this figure.}
	\label{fig:literaturecomp}
\end{figure}

\subsection{Seismic vs spectroscopic rotation}
A distinct difference between rotation rates obtained through different techniques may hold information about differential rotation (both latitudinal and radial) of near-surface layers, such as those we observe in the Sun \cite{beck2000}.  A previous comparison of spectroscopic and seismic rotation rates, performed on a sample of 22 stars, found no significant radial differential rotation \cite{benomar+2018}. In this work they not only considered rotation rates from spots, but also spectroscopic measures of the projected surface rotation $\textrm{v}\sin(i)$, which they found to be more reliable.

\begin{figure}[h!]
	\centering
	\includegraphics[width=\textwidth]{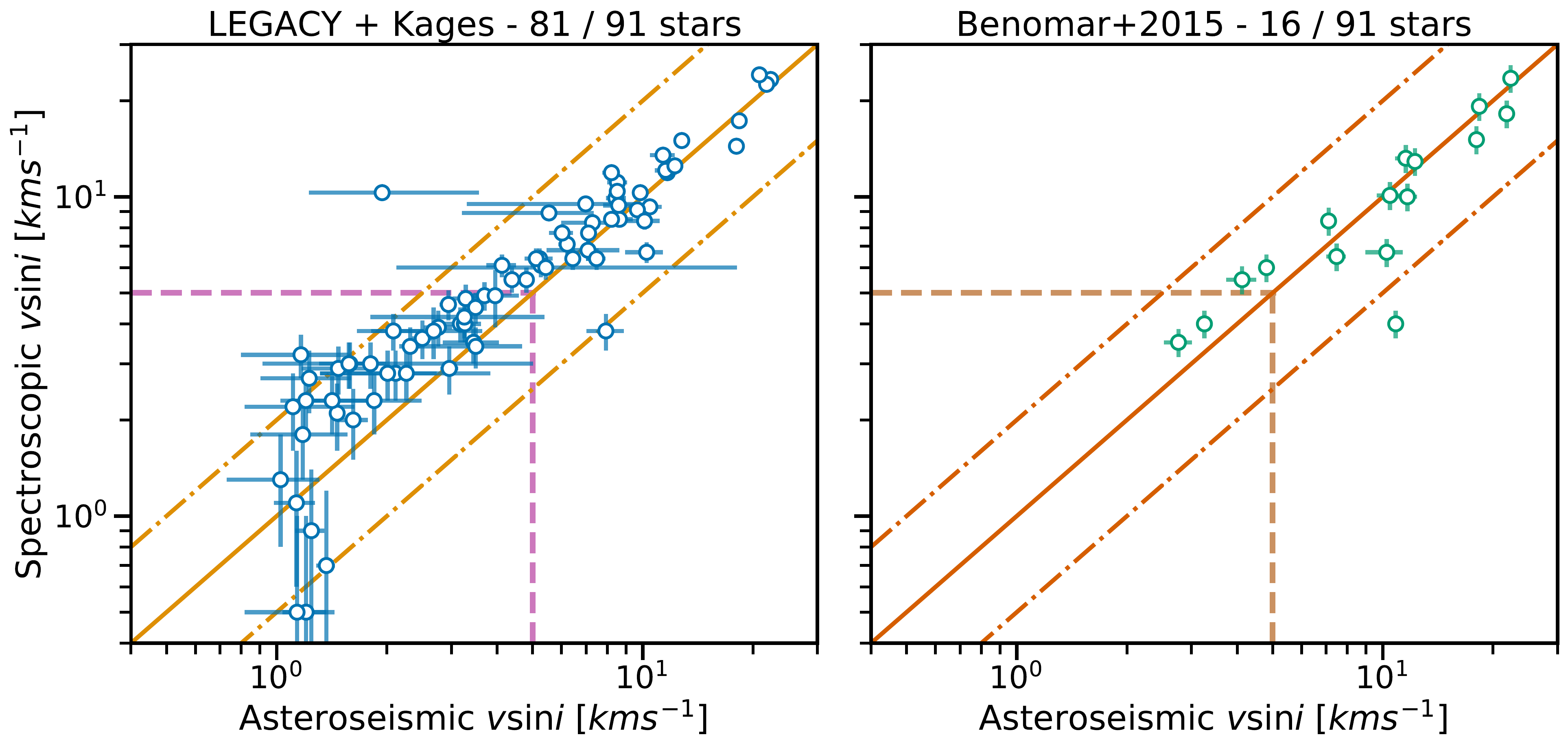}
	\caption{\textbf{Comparisons between asteroseismic and spectroscopic measures of projected surface rotation, $\textrm{v}\sin(i)$.} All asteroseismic (x-axis) values are from this work, all spectroscopic (y-axis) values are from the literature. \textit{Left}: comparisons to 81 stars values reported in LEGACY and Kages. \textit{Right}: comparisons to 16 stars observed by Benomar et al. (2015) \cite{benomar+2015}. Asteroseismic values are transformed from projected splitting ($\nu_s\sin(i)$) using the asteroseismic radius measurements presented in LEGACY and `Kages'. Horizontal error bars represent the 68\% confidence intervals. Vertical error bars represent the formal uncertainty on the published spectroscopic values. The solid lines indicate the 1:1 line, while the dash-dotted lines represent the 2:1 and 1:2 lines. The dashed lines indicate the location of $\textrm{v}\sin(i) = 5\, \rm{kms^{-1}}$, the point at which surface and seismic measures of projected rotation begin to differ strongly \cite{tayar+2015}.}
	\label{fig:vsinilit}
\end{figure}

With our expanded sample of asteroseismic rotation we can perform a similar analysis, to both validate our sample and probe radial differential rotation.  Supplementary Figure \ref{fig:vsinilit} shows a comparison between spectroscopic $\textrm{v}\sin(i)$ measurements as listed in LEGACY and `Kages' (left) and Benomar et al. (2015) \cite{benomar+2015} (right). In these cases the asteroseismic $\textrm{v}\sin(i)$ has been calculated using our measure of asteroseismic $\nu_s\sin(i)$ and the known asteroseismic radii. Three stars (KICs 6603624, 8760414 and 8938364) have been excluded from this figure due to strong disagreements of measured rotation rates with the literature (see above).

For the LEGACY and `Kages' sample, we find no strong deviation from the 1:1 line except at very low velocities, which is likely due to biases inherent to spectroscopic line broadening measurements \cite{doyle+2014, tayar+2015}. For the Benomar et al. (2015) sample stars lie a lot closer to the 1:1 line.

Overall, there appears to be a global offset where spectroscopic measurements of projected rotation appear faster than asteroseismic measures. Based on the LEGACY and `Kages' $v\sin(i)$ values, this offset is roughly $18\%$ (i.e. spectroscopic projected rotation rates are faster than asteroseismic rates). This offset is much smaller ($\sim5\%$) for the \cite{benomar+2015} sample, albeit for far fewer stars. These offsets are within the typical disagreement between spectroscopic methods, based on comparisons of projected rotation measurements for red giant stars \cite[see Figure 2]{tayar+2015}, especially at $< 5\, \rm{km\,s^{-1}}$.

It is worth addressing the impact this comparison would have on our conclusions for gyrochronology, were we to take the spectroscopic rotation rates as if they were the truth, even at velocities of $< 5\, \rm{km\,s^{-1}}$. For the majority of cases, the seismic velocities are slower than the spectroscopic velocities, which would bias our ensemble towards favouring a standard evolution over a weakened magnetic braking scenario (as stars would overall appear to be rotating slower at late ages). Based on the comparison to spot rotation presented in the main body of this paper, and given known issues comparing seismic and spectroscopic rotation at low velocities, we concluded that the divergence seen in  Supplementary Figure \ref{fig:vsinilit} does not undermine our results.

\subsection{Asteroseismic detection biases}

Stars that are rotating very slowly will show small rotational splittings, which may be indistinguishable from a non-rotating case, wheras stars that are spinning very fast may find that their splitting is so wide that they cross over with split modes at different radial orders ($n$). Asteroseismic selection techniques may also introduce additional biases on a population level. We consider the limitations this places on our asteroseismic method.

\begin{itemize}
	\item \textit{Fast rotating stars:} In prinicple, a star could rotate fast enough that the highest-frequency peak in a split dipole ($\ell = 2$) mode would overlap with the radial ($\ell = 0$) mode. This would occur if $2\nu_{\rm s} > \delta\nu_{02}$, the small separation. For the star in our sample with the smallest $\delta\nu_{02}$ (so the highest risk of this happening), this would occur at a rotation of 8.8 days or faster. For the star with the largest $\delta\nu_{02}$, this limit was roughly 2 days.
	
	In practice, the model still properly fits a spectrum where this kind of mode-overlapping occurs, as the $\ell = 0$ modes don't split, meaning that there are no degenerate solutions in this regime, and further constraints can be obtained from the other modes and split components in the spectrum. A more serious issue may be when the $\ell = 1$ mode at lower frequencies overlaps with the $\ell = 2$ mode as well, which would only occur at a rotation rate of 2 days or faster for the star with the smallest $\delta\nu_{01}$.
	
	\item \textit{Slow rotating stars:} Except for the most rapidly rotating cases, asteroseismic mode splitting is detected by a change in the distribution of the power of mode frequencies as the split components begin to separate from the central mode. If there is to be hard limit at which this measurement is not possible, it should be in the case that $\nu_{\rm s} < 2/T$, where $T$ is the observation time, and $1/T$ is the frequency bin width. For the star in our sample with the shortest observing time, this limit would be $\sim 89$ days, which quickly rises for longer observing baselines. While this constitutes a hard limit, slower rotation rates will become harder and harder to constrain as the rate of splitting becomes smaller. A benefit of the Bayesian approach used in this paper is that it should accurately reflect the increased uncertainty associated with the small rotational splitting.
\end{itemize}

While these two limits cause concern when studying the most extreme rotators, the results presented in this work are driven by stars with rotation rates in the regime of 5 to 30 days. Above, we demonstrated that results on period for these stars were data-driven (i.e. do not reflect the priors), indicating that the splitting is well resolved in this regime, and so we do not expect these limits to bias our conclusions.

Finally, we also consider bias in asteroseismic detections. The probability of an asteroseismic detection with \kepler scales with temperature and radius (i.e. scales inversely with \logg, and therefore main sequence age), and detections are unlikely below roughly $5200\, K$ \cite{chaplin+2011, schofield+2019}. Very young ($\lesssim 2\, \rm{Gyr}$), very magnetically active stars may also suppress their oscillation modes. \cite{mathur+2019}. Though these two effects impose a bias on asteroseismic populations, they do so outside the parameter range on which our comparison takes place. As it is older stars, undergoing magnetic braking, that are required to drive the distinction between stellar models, we conclude that our seismic sample can be assumed to be complete for the purposes of this work.

\section{Verifying consequences for gyrochronology}
\subsection{Limits of our stellar models}\label{ssec:limits}
The rotational stellar models used in this work \cite{vansaders+2019} are constructed for metallicities of $-0.4 < \rm{[Fe/H]} < 0.4\, \rm{dex}$, in steps of $0.1\, \rm {dex}$. Our sample of 91 stars contained 4 stars with metallicities below $-0.4\, \rm{dex}$, which are shown as shaded symbols in Figure 3 in the main paper. Of these, 3 were included in our final sample of 73 stars used to evaluate our stellar models: KICs 7970740, 8684723 and KIC 9965715. All three are classed as MS stars, with metallicities of $-0.54 \pm 0.10$, $-0.42 \pm 0.10$ and $-0.44 \pm 0.18\, \rm {dex}$ respectively, placing them within $3\sigma$ of the metallicity limits of our stellar models. KICs 8684723 and 9965715 strongly agree with the WMB model, whereas KIC 7970740 weakly prefers the standard model. Excluding these stars was not found to significantly alter the joint posterior distribution shown in Figure 4 of the main paper.

A recent study \cite{amard+matt2020} compared different rotational evolution models \cite[of which we use the former in this work]{vansaders+pinsonneault2013, matt+2015} while studying the effect metallicity has on rotation. They found that metal-rich stars spin down significantly more effectively than metal poor stars. The population of 73 stars used in our stellar model comparisons is roughly centered on a $\rm{[Fe/H]}$ of 0 $\rm dex$, with a spread of $~0.16\, \rm dex$, with no stars significantly above or below $\pm 0.4\, \rm dex$ (as discussed above). While differences in stellar rotational evolution as a function of metallicity in this region are still somewhat pronounced, they are much less so than for more metal-rich or poor stars \cite[see Figure 2]{amard+matt2020}. However, for stars with $\feh < 0$, the alternative model prescription \cite{matt+2015} sees stars spin down more slowly than the models used in this work (i.e. they will rotate faster at later ages). This work only explores the presence of weakened magnetic braking for a single braking law, and a comparison to alternative braking model will be explored in a future paper.

When constructing KDEs from our stellar model samples, we selected fixed resolutions (or band-widths) for the KDEs. In mass, the band-width of $0.02\, M_\odot$, was larger than the uncertainties of 24 of 73 stars used to construct our joint posterior. For the stars with the smallest uncertainties, this significantly limits the size of the KDE being evaluated (as subsections of the full stellar models are used to evaluate individual stars, for computational efficiency). In order to confirm that these stars do not significantly affect the ensemble's preference towards the WMB model, we recalculated the joint posterior distribution for $Q_{\rm WMB}$, excluding stars with an uncertainty on mass smaller than the KDE band-width. While the 24 stars with small uncertainties do favour the WMB model, they do so very weakly, whereas the remaining 49 stars with larger uncertainties strongly favour the WMB model. Their removal from the total joint posterior probability does not significantly alter it from the distribution shown in Figure 4 of the main paper.

\subsection{Systematic uncertainties from asteroseismology}
In our model analysis, we used asteroseismic mass and age obtained using \texttt{BASTA}, as reported in LEGACY and Kages. Asteroseismic properties obtained through stellar models can be subject to systematic errors arising from differences in input physics and choice of stellar models not included in the reported statistical uncertainties. A quantification of these different systematic effects can be found in the `Kages' catalogue paper \cite{silvaaguirre+2015}. Combining their reported median systematic uncertainties due to input physics results in median uncertainties of $20\%$ (up from $14\%)$ on age and $5\%$ (up from $3\%$) on mass for the `Kages' sample. For LEGACY, the median uncertainties are $18\%$ (up from $10\%)$ and $5.6\%$ (up from $4\%$) for age and mass respectively.

We re-ran our model analysis after inflating uncertainties on mass and age. We increased uncertainties by the fractional difference between the \texttt{BASTA} statistical uncertainties and the median full statistical and systematic uncertainties described above. For example, a LEGACY star with a mass of $2.0 \pm 0.5\, M_\odot$ would have its uncertainty inflated by $1.6\%$ of its mass, to $0.53\, M_\odot$. 

To further test the limits of this analysis, we also reran our mixture model fit, this time only shifting the asteroseismic ages younger by the systematic uncertainty, and retaining the statistical uncertainty (i.e. the ages of LEGACY and `Kages' stars were reduced by $8\%$ and $6\%$ respectively). In this scenario where all asteroseismic ages are overestimated, `true' fast rotators at young ages would have been mistaken for fast rotators at old ages, suggesting the presence of weakened magnetic braking where none existed. 

The results of these tests are discussed further in the main body of the paper.

\subsection{Binaries and Planet Hosts}
For stars to be good probes of existing gyrochronology relations, their rotational evolution must occur in isolation. If a star interacts with a close binary companion (through tides, or a merger) the natural angular momentum loss can be disturbed, causing gyrochronology to mispredict ages \cite{leiner+2019, fleming+2019}. Between LEGACY and Kages, we have 8 known binaries. 

First, KIC 8379927, KIC 7510397, KIC 10454113 and KIC 9025370 are spectroscopic binaries. This does not affect the asteroseismic analysis, but may affect their rotational evolution. Of these, KICs 8379927, 7510397 and 9025370 were included in the gyrochronology analysis. None of them preferred one model strongly over the other, with all three finding flat posteriors for $Q_{\rm WMB}$.

Second, the binary pairs of KIC 9139151 \& 9139163 and 16 Cyg A \& B are individually observed binary components with wide orbital separations, so we do not expect their binarity to have affected their rotational evolution \cite{halbwachs1986, white+2013}. 

While we chose not to account for star-planet tidal interactions in this work, we note that this may disturb the natural stellar rotational evolution when tidal forces are at play \cite{maxted+2015, gallet+delorme2019, benbakoura+2019}, although this has been disputed by observations of asteroseismic planet hosts \cite{ceillier+2016}.\\


\begin{thebibliography}{10}
	\expandafter\ifx\csname url\endcsname\relax
	\def\url#1{\texttt{#1}}\fi
	\expandafter\ifx\csname urlprefix\endcsname\relax\def\urlprefix{URL }\fi
	\providecommand{\bibinfo}[2]{#2}
	\providecommand{\eprint}[2][]{\url{#2}}
	
	\bibitem{barnes2007}
	\bibinfo{author}{{Barnes}, S.~A.}
	\newblock \bibinfo{title}{{Ages for Illustrative Field Stars Using
			Gyrochronology: Viability, Limitations, and Errors}}.
	\newblock \emph{\bibinfo{journal}{Astrophys. J.}}
	\textbf{\bibinfo{volume}{669}}, \bibinfo{pages}{1167--1189}
	(\bibinfo{year}{2007}).
	\newblock \eprint{0704.3068}.
	
	\bibitem{meibom+2015}
	\bibinfo{author}{{Meibom}, S.} \emph{et~al.}
	\newblock \bibinfo{title}{{A spin-down clock for cool stars from observations
			of a 2.5-billion-year-old cluster}}.
	\newblock \emph{\bibinfo{journal}{Nature}} \textbf{\bibinfo{volume}{517}},
	\bibinfo{pages}{589--591} (\bibinfo{year}{2015}).
	\newblock \eprint{1501.05651}.
	
	\bibitem{leiner+2019}
	\bibinfo{author}{{Leiner}, E.}, \bibinfo{author}{{Mathieu}, R.~D.},
	\bibinfo{author}{{Vanderburg}, A.}, \bibinfo{author}{{Gosnell}, N.~M.} \&
	\bibinfo{author}{{Smith}, J.~C.}
	\newblock \bibinfo{title}{{Blue Lurkers: Hidden Blue Stragglers on the M67 Main
			Sequence Identified from Their Kepler/K2 Rotation Periods}}.
	\newblock \emph{\bibinfo{journal}{Astrophys. J.}}
	\textbf{\bibinfo{volume}{881}}, \bibinfo{pages}{47} (\bibinfo{year}{2019}).
	\newblock \eprint{1904.02169}.
	
	\bibitem{claytor+2019}
	\bibinfo{author}{{Claytor}, Z.~R.} \emph{et~al.}
	\newblock \bibinfo{title}{{Chemical Evolution in the Milky Way: Rotation-based
			Ages for APOGEE-Kepler Cool Dwarf Stars}}.
	\newblock \emph{\bibinfo{journal}{Astrophys. J.}}
	\textbf{\bibinfo{volume}{888}}, \bibinfo{pages}{43} (\bibinfo{year}{2020}).
	\newblock \eprint{1911.04518}.
	
	\bibitem{barnes+2016}
	\bibinfo{author}{Barnes, S.~A.}, \bibinfo{author}{Weingrill, J.},
	\bibinfo{author}{Fritzewski, D.}, \bibinfo{author}{Strassmeier, K.~G.} \&
	\bibinfo{author}{Platais, I.}
	\newblock \bibinfo{title}{Rotation {{Periods}} for {{Cool Stars}} in the 4
		{{Gyr}} old {{Open Cluster M67}}, {{The Solar}}-{{Stellar Connection}}, and
		the {{Applicability}} of {{Gyrochronology}} to at least {{Solar Age}}}.
	\newblock \emph{\bibinfo{journal}{Astrophys. J.}}
	\textbf{\bibinfo{volume}{823}}, \bibinfo{pages}{16} (\bibinfo{year}{2016}).
	
	\bibitem{borucki+2010}
	\bibinfo{author}{Borucki, W.~J.}, \bibinfo{author}{Koch, D.} \&
	\bibinfo{author}{Team, K.~S.}
	\newblock \bibinfo{title}{Kepler {{Planet Detection Mission}}: {{Highlights}}
		of the {{First Results}}}.
	\newblock \emph{\bibinfo{journal}{AAS/Division for Planetary Sciences Meeting
			Abstracts \#42}} \textbf{\bibinfo{volume}{42}}, \bibinfo{pages}{47.03}
	(\bibinfo{year}{2010}).
	
	\bibitem{silvaaguirre+2015}
	\bibinfo{author}{{Silva Aguirre}, V.} \emph{et~al.}
	\newblock \bibinfo{title}{{Ages and fundamental properties of Kepler exoplanet
			host stars from asteroseismology}}.
	\newblock \emph{\bibinfo{journal}{Mon. Not. R. Astron. Soc.}}
	\textbf{\bibinfo{volume}{452}}, \bibinfo{pages}{2127--2148}
	(\bibinfo{year}{2015}).
	\newblock \eprint{1504.07992}.
	
	\bibitem{angus+2015}
	\bibinfo{author}{{Angus}, R.}, \bibinfo{author}{{Aigrain}, S.},
	\bibinfo{author}{{Foreman-Mackey}, D.} \& \bibinfo{author}{{McQuillan}, A.}
	\newblock \bibinfo{title}{{Calibrating gyrochronology using Kepler
			asteroseismic targets}}.
	\newblock \emph{\bibinfo{journal}{Mon. Not. R. Astron. Soc.}}
	\textbf{\bibinfo{volume}{450}}, \bibinfo{pages}{1787--1798}
	(\bibinfo{year}{2015}).
	\newblock \eprint{1502.06965}.
	
	\bibitem{nielsen+2015}
	\bibinfo{author}{Nielsen, M.~B.}, \bibinfo{author}{Schunker, H.},
	\bibinfo{author}{Gizon, L.} \& \bibinfo{author}{Ball, W.~H.}
	\newblock \bibinfo{title}{Constraining differential rotation of {{Sun}}-like
		stars from asteroseismic and starspot rotation periods}.
	\newblock \emph{\bibinfo{journal}{Astron. Astrophys.}}
	\textbf{\bibinfo{volume}{582}}, \bibinfo{pages}{A10} (\bibinfo{year}{2015}).
	
	\bibitem{davies+2015}
	\bibinfo{author}{{Davies}, G.~R.} \emph{et~al.}
	\newblock \bibinfo{title}{{Asteroseismic inference on rotation, gyrochronology
			and planetary system dynamics of 16 Cygni}}.
	\newblock \emph{\bibinfo{journal}{Mon. Not. R. Astron. Soc.}}
	\textbf{\bibinfo{volume}{446}}, \bibinfo{pages}{2959--2966}
	(\bibinfo{year}{2015}).
	\newblock \eprint{1411.1359}.
	
	\bibitem{vansaders+2016}
	\bibinfo{author}{{van Saders}, J.~L.} \emph{et~al.}
	\newblock \bibinfo{title}{{Weakened magnetic braking as the origin of
			anomalously rapid rotation in old field stars}}.
	\newblock \emph{\bibinfo{journal}{Nature}} \textbf{\bibinfo{volume}{529}},
	\bibinfo{pages}{181--184} (\bibinfo{year}{2016}).
	\newblock \eprint{1601.02631}.
	
	\bibitem{reville+2015}
	\bibinfo{author}{{R{\'e}ville}, V.}, \bibinfo{author}{{Brun}, A.~S.},
	\bibinfo{author}{{Matt}, S.~P.}, \bibinfo{author}{{Strugarek}, A.} \&
	\bibinfo{author}{{Pinto}, R.~F.}
	\newblock \bibinfo{title}{{The Effect of Magnetic Topology on Thermally Driven
			Wind: Toward a General Formulation of the Braking Law}}.
	\newblock \emph{\bibinfo{journal}{Astrophys. J.}}
	\textbf{\bibinfo{volume}{798}}, \bibinfo{pages}{116} (\bibinfo{year}{2015}).
	\newblock \eprint{1410.8746}.
	
	\bibitem{garraffo+2016}
	\bibinfo{author}{Garraffo, C.}, \bibinfo{author}{Drake, J.~J.} \&
	\bibinfo{author}{Cohen, O.}
	\newblock \bibinfo{title}{The {{Missing Magnetic Morphology Term}} in {{Stellar
				Rotation Evolution}}}.
	\newblock \emph{\bibinfo{journal}{Astronomy \& Astrophysics}}
	\textbf{\bibinfo{volume}{595}}, \bibinfo{pages}{A110} (\bibinfo{year}{2016}).
	\newblock \eprint{1607.06096}.
	
	\bibitem{metcalfe+2016}
	\bibinfo{author}{Metcalfe, T.~S.}, \bibinfo{author}{Egeland, R.} \&
	\bibinfo{author}{{van Saders}, J.}
	\newblock \bibinfo{title}{Stellar {{Evidence That}} the {{Solar Dynamo May Be}}
		in {{Transition}}}.
	\newblock \emph{\bibinfo{journal}{Astrophys. J. Letters}}
	\textbf{\bibinfo{volume}{826}}, \bibinfo{pages}{L2} (\bibinfo{year}{2016}).
	
	\bibitem{metcalfe+2019}
	\bibinfo{author}{{Metcalfe}, T.~S.} \emph{et~al.}
	\newblock \bibinfo{title}{{LBT/PEPSI Spectropolarimetry of a Magnetic
			Morphology Shift in Old Solar-type Stars}}.
	\newblock \emph{\bibinfo{journal}{Astrophys. J.}}
	\textbf{\bibinfo{volume}{887}}, \bibinfo{pages}{L38} (\bibinfo{year}{2019}).
	\newblock \eprint{1912.01186}.
	
	\bibitem{see+2019}
	\bibinfo{author}{See, V.} \emph{et~al.}
	\newblock \bibinfo{title}{Do {{Non}}-dipolar {{Magnetic Fields Contribute}} to
		{{Spin}}-down {{Torques}}?}
	\newblock \emph{\bibinfo{journal}{Astrophys. J.}}
	\textbf{\bibinfo{volume}{886}}, \bibinfo{pages}{120} (\bibinfo{year}{2019}).
	
	\bibitem{mcquillan+2014}
	\bibinfo{author}{McQuillan, A.}, \bibinfo{author}{Mazeh, T.} \&
	\bibinfo{author}{Aigrain, S.}
	\newblock \bibinfo{title}{Rotation {{Periods}} of 34,030 {{Kepler
				Main}}-{{Sequence Stars}}: {{The Full Autocorrelation Sample}}}.
	\newblock \emph{\bibinfo{journal}{Astrophys. J. Supplement Series}}
	\textbf{\bibinfo{volume}{211}}, \bibinfo{pages}{24} (\bibinfo{year}{2014}).
	\newblock \eprint{1402.5694}.
	
	\bibitem{matt+2015}
	\bibinfo{author}{Matt, S.~P.}, \bibinfo{author}{Brun, A.~S.},
	\bibinfo{author}{Baraffe, I.}, \bibinfo{author}{Bouvier, J.} \&
	\bibinfo{author}{Chabrier, G.}
	\newblock \bibinfo{title}{The {{Mass}}-dependence of {{Angular Momentum
				Evolution}} in {{Sun}}-like {{Stars}}}.
	\newblock \emph{\bibinfo{journal}{Astrophys. J.}}
	\textbf{\bibinfo{volume}{799}}, \bibinfo{pages}{L23} (\bibinfo{year}{2015}).
	
	\bibitem{reinhold+2020}
	\bibinfo{author}{{Reinhold}, T.} \emph{et~al.}
	\newblock \bibinfo{title}{{The Sun is less active than other solar-like
			stars}}.
	\newblock \emph{\bibinfo{journal}{Science}} \textbf{\bibinfo{volume}{368}},
	\bibinfo{pages}{518--521} (\bibinfo{year}{2020}).
	\newblock \eprint{2005.01401}.
	
	\bibitem{vansaders+2019}
	\bibinfo{author}{{van Saders}, J.~L.}, \bibinfo{author}{Pinsonneault, M.~H.} \&
	\bibinfo{author}{Barbieri, M.}
	\newblock \bibinfo{title}{Forward {{Modeling}} of the {{Kepler Stellar Rotation
				Period Distribution}}: {{Interpreting Periods}} from {{Mixed}} and {{Biased
				Stellar Populations}}}.
	\newblock \emph{\bibinfo{journal}{Astrophys. J.}}
	\textbf{\bibinfo{volume}{872}}, \bibinfo{pages}{128} (\bibinfo{year}{2019}).
	
	\bibitem{ledoux1951}
	\bibinfo{author}{Ledoux, P.}
	\newblock \bibinfo{title}{The {{Nonradial Oscillations}} of {{Gaseous Stars}}
		and the {{Problem}} of {{Beta Canis Majoris}}.}
	\newblock \emph{\bibinfo{journal}{Astrophys. J.}}
	\textbf{\bibinfo{volume}{114}}, \bibinfo{pages}{373} (\bibinfo{year}{1951}).
	
	\bibitem{davies+2016}
	\bibinfo{author}{{Davies}, G.~R.} \emph{et~al.}
	\newblock \bibinfo{title}{{Oscillation frequencies for 35 Kepler solar-type
			planet-hosting stars using Bayesian techniques and machine learning}}.
	\newblock \emph{\bibinfo{journal}{Mon. Not. R. Astron. Soc.}}
	\textbf{\bibinfo{volume}{456}}, \bibinfo{pages}{2183--2195}
	(\bibinfo{year}{2016}).
	\newblock \eprint{1511.02105}.
	
	\bibitem{lund+2017}
	\bibinfo{author}{Lund, M.~N.} \emph{et~al.}
	\newblock \bibinfo{title}{Standing on the {{Shoulders}} of {{Dwarfs}}: The
		{{Kepler Asteroseismic LEGACY Sample}}. {{I}}. {{Oscillation Mode
				Parameters}}}.
	\newblock \emph{\bibinfo{journal}{Astrophys. J.}}
	\textbf{\bibinfo{volume}{835}}, \bibinfo{pages}{172} (\bibinfo{year}{2017}).
	
	\bibitem{silvaaguirre+2017}
	\bibinfo{author}{Silva~Aguirre, V.} \emph{et~al.}
	\newblock \bibinfo{title}{Standing on the {{Shoulders}} of {{Dwarfs}}: The
		{{Kepler Asteroseismic LEGACY Sample}}. {{II}}.{{Radii}}, {{Masses}}, and
		{{Ages}}}.
	\newblock \emph{\bibinfo{journal}{Astrophys. J.}}
	\textbf{\bibinfo{volume}{835}}, \bibinfo{pages}{173} (\bibinfo{year}{2017}).
	
	\bibitem{garcia+2014}
	\bibinfo{author}{Garc{\'i}a, R.~A.} \emph{et~al.}
	\newblock \bibinfo{title}{Rotation and magnetism of {{Kepler}} pulsating
		solar-like stars. {{Towards}} asteroseismically calibrated age-rotation
		relations}.
	\newblock \emph{\bibinfo{journal}{Astron. Astrophys.}}
	\textbf{\bibinfo{volume}{572}}, \bibinfo{pages}{A34} (\bibinfo{year}{2014}).
	
	\bibitem{kraft1967}
	\bibinfo{author}{Kraft, R.~P.}
	\newblock \bibinfo{title}{Studies of {{Stellar Rotation}}. {{V}}. {{The
				Dependence}} of {{Rotation}} on {{Age}} among {{Solar}}-{{Type Stars}}}.
	\newblock \emph{\bibinfo{journal}{Astrophys. J.}}
	\textbf{\bibinfo{volume}{150}}, \bibinfo{pages}{551} (\bibinfo{year}{1967}).
	
	\bibitem{bedding+2010}
	\bibinfo{author}{Bedding, T.~R.} \emph{et~al.}
	\newblock \bibinfo{title}{Solar-like {{Oscillations}} in {{Low}}-luminosity
		{{Red Giants}}: {{First Results}} from {{Kepler}}}.
	\newblock \emph{\bibinfo{journal}{Astrophys. J. Letters}}
	\textbf{\bibinfo{volume}{713}}, \bibinfo{pages}{L176--L181}
	(\bibinfo{year}{2010}).
	
	\bibitem{bonanno+frohlich2015}
	\bibinfo{author}{Bonanno, A.} \& \bibinfo{author}{Fr{\"o}hlich, H.-E.}
	\newblock \bibinfo{title}{A {{Bayesian}} estimation of the helioseismic solar
		age}.
	\newblock \emph{\bibinfo{journal}{Astron. Astrophys.}}
	\textbf{\bibinfo{volume}{580}}, \bibinfo{pages}{A130} (\bibinfo{year}{2015}).
	
	\bibitem{paxton+2017}
	\bibinfo{author}{{Paxton}, B.} \emph{et~al.}
	\newblock \bibinfo{title}{{Modules for Experiments in Stellar Astrophysics
			(MESA): Convective Boundaries, Element Diffusion, and Massive Star
			Explosions}}.
	\newblock \emph{\bibinfo{journal}{Astrophys. J. Supplement Series}}
	\textbf{\bibinfo{volume}{234}}, \bibinfo{pages}{34} (\bibinfo{year}{2018}).
	\newblock \eprint{1710.08424}.
	
	\bibitem{hunter2007}
	\bibinfo{author}{Hunter, J.~D.}
	\newblock \bibinfo{title}{Matplotlib: {{A 2D Graphics Environment}}}.
	\newblock \emph{\bibinfo{journal}{Computing in Science and Engineering}}
	\textbf{\bibinfo{volume}{9}}, \bibinfo{pages}{90--95} (\bibinfo{year}{2007}).
	
	\bibitem{benomar+2018}
	\bibinfo{author}{{Benomar}, O.} \emph{et~al.}
	\newblock \bibinfo{title}{{Asteroseismic detection of latitudinal differential
			rotation in 13 Sun-like stars}}.
	\newblock \emph{\bibinfo{journal}{Science}} \textbf{\bibinfo{volume}{361}},
	\bibinfo{pages}{1231--1234} (\bibinfo{year}{2018}).
	\newblock \eprint{1809.07938}.
	
	\bibitem{lund+2014}
	\bibinfo{author}{Lund, M.~N.}, \bibinfo{author}{Miesch, M.~S.} \&
	\bibinfo{author}{{Christensen-Dalsgaard}, J.}
	\newblock \bibinfo{title}{Differential rotation in main-sequence solar-like
		stars: {{Qualitative}} inference from asteroseismic data}.
	\newblock \emph{\bibinfo{journal}{Astrophys. J.}}
	\textbf{\bibinfo{volume}{790}}, \bibinfo{pages}{121} (\bibinfo{year}{2014}).
	\newblock \eprint{1406.7873}.
	
	\bibitem{benomar+2015}
	\bibinfo{author}{Benomar, O.}, \bibinfo{author}{Takata, M.},
	\bibinfo{author}{Shibahashi, H.}, \bibinfo{author}{Ceillier, T.} \&
	\bibinfo{author}{Garc{\'i}a, R.~A.}
	\newblock \bibinfo{title}{Nearly uniform internal rotation of solar-like
		main-sequence stars revealed by space-based asteroseismology and
		spectroscopic measurements}.
	\newblock \emph{\bibinfo{journal}{Mon. Not. R. Astron. Soc.}}
	\textbf{\bibinfo{volume}{452}}, \bibinfo{pages}{2654--2674}
	(\bibinfo{year}{2015}).
	
	\bibitem{gizon+2013}
	\bibinfo{author}{Gizon, L.} \emph{et~al.}
	\newblock \bibinfo{title}{Seismic constraints on rotation of {{Sun}}-like star
		and mass of exoplanet}.
	\newblock \emph{\bibinfo{journal}{Proceedings of the National Academy of
			Science}} \textbf{\bibinfo{volume}{110}}, \bibinfo{pages}{13267}
	(\bibinfo{year}{2013}).
	
	\bibitem{chaplin+2013}
	\bibinfo{author}{Chaplin, W.~J.} \emph{et~al.}
	\newblock \bibinfo{title}{Asteroseismic {{Determination}} of {{Obliquities}} of
		the {{Exoplanet Systems Kepler}}-50 and {{Kepler}}-65}.
	\newblock \emph{\bibinfo{journal}{Astrophys. J.}}
	\textbf{\bibinfo{volume}{766}}, \bibinfo{pages}{101} (\bibinfo{year}{2013}).
	
	\bibitem{nielsen+2013}
	\bibinfo{author}{Nielsen, M.~B.}, \bibinfo{author}{Gizon, L.},
	\bibinfo{author}{Schunker, H.} \& \bibinfo{author}{Karoff, C.}
	\newblock \bibinfo{title}{Rotation periods of 12 000 main-sequence {{Kepler}}
		stars: {{Dependence}} on stellar spectral type and comparison with v sin i
		observations}.
	\newblock \emph{\bibinfo{journal}{Astron. Astrophys.}}
	\textbf{\bibinfo{volume}{557}}, \bibinfo{pages}{L10} (\bibinfo{year}{2013}).
	
	\bibitem{skumanich1972}
	\bibinfo{author}{{Skumanich}, A.}
	\newblock \bibinfo{title}{{Time Scales for Ca II Emission Decay, Rotational
			Braking, and Lithium Depletion}}.
	\newblock \emph{\bibinfo{journal}{Astrophys. J.}}
	\textbf{\bibinfo{volume}{171}}, \bibinfo{pages}{565} (\bibinfo{year}{1972}).
	
	\bibitem{kawaler1988}
	\bibinfo{author}{Kawaler, S.~D.}
	\newblock \bibinfo{title}{Angular momentum loss in low-mass stars}.
	\newblock \emph{\bibinfo{journal}{Astrophys. J.}}
	\textbf{\bibinfo{volume}{333}}, \bibinfo{pages}{236} (\bibinfo{year}{1988}).
	
	\bibitem{girardi+2012}
	\bibinfo{author}{Girardi, L.} \emph{et~al.}
	\newblock \bibinfo{title}{{{TRILEGAL}}, a {{TRIdimensional modeL}} of {{thE
				GALaxy}}: {{Status}} and {{Future}}}.
	\newblock \emph{\bibinfo{journal}{Astrophysics and Space Science Proceedings}}
	\textbf{\bibinfo{volume}{26}}, \bibinfo{pages}{165} (\bibinfo{year}{2012}).
	
	\bibitem{berger+2020}
	\bibinfo{author}{{Berger}, T.~A.}, \bibinfo{author}{{Huber}, D.},
	\bibinfo{author}{{Gaidos}, E.}, \bibinfo{author}{{van Saders}, J.~L.} \&
	\bibinfo{author}{{Weiss}, L.~M.}
	\newblock \bibinfo{title}{{The Gaia-Kepler Stellar Properties Catalog. II.
			Planet Radius Demographics as a Function of Stellar Mass and Age}}.
	\newblock \emph{\bibinfo{journal}{Astron. J.}} \textbf{\bibinfo{volume}{160}},
	\bibinfo{pages}{108} (\bibinfo{year}{2020}).
	\newblock \eprint{2005.14671}.
	
	\bibitem{gelman+rubin1992}
	\bibinfo{author}{Gelman, A.} \& \bibinfo{author}{Rubin, D.~B.}
	\newblock \bibinfo{title}{Inference from {{Iterative Simulation Using Multiple
				Sequences}}}.
	\newblock \emph{\bibinfo{journal}{Statistical Science}}
	\textbf{\bibinfo{volume}{7}}, \bibinfo{pages}{457--472}
	(\bibinfo{year}{1992}).
	
	\bibitem{salvatier+2016}
	\bibinfo{author}{Salvatier, J.}, \bibinfo{author}{Wiecki, T.~V.} \&
	\bibinfo{author}{Fonnesbeck, C.}
	\newblock \bibinfo{title}{Probabilistic programming in {{Python}} using
		{{PyMC3}}}.
	\newblock \emph{\bibinfo{journal}{PeerJ Computer Science}}
	\textbf{\bibinfo{volume}{2}}, \bibinfo{pages}{e55} (\bibinfo{year}{2016}).
	
	\bibitem{metcalfe+egeland2019}
	\bibinfo{author}{Metcalfe, T.~S.} \& \bibinfo{author}{Egeland, R.}
	\newblock \bibinfo{title}{Understanding the {{Limitations}} of
		{{Gyrochronology}} for {{Old Field Stars}}}.
	\newblock \emph{\bibinfo{journal}{Astrophys. J.}}
	\textbf{\bibinfo{volume}{871}}, \bibinfo{pages}{39} (\bibinfo{year}{2019}).
	\newblock \eprint{1811.11905}.
	
	\bibitem{lorenzo-oliveira+2019}
	\bibinfo{author}{{Lorenzo-Oliveira}, D.} \emph{et~al.}
	\newblock \bibinfo{title}{{Constraining the evolution of stellar rotation using
			solar twins}}.
	\newblock \emph{\bibinfo{journal}{Mon. Not. R. Astron. Soc.}}
	\textbf{\bibinfo{volume}{485}}, \bibinfo{pages}{L68--L72}
	(\bibinfo{year}{2019}).
	\newblock \eprint{1903.02630}.
	
	\bibitem{vansaders+pinsonneault2013}
	\bibinfo{author}{{van Saders}, J.~L.} \& \bibinfo{author}{Pinsonneault, M.~H.}
	\newblock \bibinfo{title}{Fast {{Star}}, {{Slow Star}}; {{Old Star}}, {{Young
				Star}}: {{Subgiant Rotation}} as a {{Population}} and {{Stellar Physics
				Diagnostic}}}.
	\newblock \emph{\bibinfo{journal}{Astrophys. J.}}
	\textbf{\bibinfo{volume}{776}}, \bibinfo{pages}{67} (\bibinfo{year}{2013}).
	
	\bibitem{barnes2010}
	\bibinfo{author}{{Barnes}, S.~A.}
	\newblock \bibinfo{title}{{A Simple Nonlinear Model for the Rotation of
			Main-sequence Cool Stars. I. Introduction, Implications for Gyrochronology,
			and Color-Period Diagrams}}.
	\newblock \emph{\bibinfo{journal}{Astrophys. J.}}
	\textbf{\bibinfo{volume}{722}}, \bibinfo{pages}{222--234}
	(\bibinfo{year}{2010}).
	
	\bibitem{deng+2012}
	\bibinfo{author}{Deng, L.-C.} \emph{et~al.}
	\newblock \bibinfo{title}{{{LAMOST Experiment}} for {{Galactic Understanding}}
		and {{Exploration}} ({{LEGUE}}) \textemdash{} {{The}} survey's science plan}.
	\newblock \emph{\bibinfo{journal}{Research in Astron. Astrophys.}}
	\textbf{\bibinfo{volume}{12}}, \bibinfo{pages}{735--754}
	(\bibinfo{year}{2012}).
	
	\bibitem{dejong+2014}
	\bibinfo{author}{{de Jong}, R.~S.} \emph{et~al.}
	\newblock \bibinfo{title}{{4MOST: 4-metre Multi-Object Spectroscopic
			Telescope}}.
	\newblock In \bibinfo{editor}{{Ramsay}, S.~K.}, \bibinfo{editor}{{McLean},
		I.~S.} \& \bibinfo{editor}{{Takami}, H.} (eds.)
	\emph{\bibinfo{booktitle}{Ground-based and Airborne Instrumentation for
			Astronomy V}}, vol. \bibinfo{volume}{9147} of \emph{\bibinfo{series}{Society
			of Photo-Optical Instrumentation Engineers (SPIE) Conference Series}},
	\bibinfo{pages}{91470M} (\bibinfo{year}{2014}).
	
	\bibitem{dalton+2014}
	\bibinfo{author}{{Dalton}, G.} \emph{et~al.}
	\newblock \bibinfo{title}{{Project overview and update on WEAVE: the next
			generation wide-field spectroscopy facility for the William Herschel
			Telescope}}.
	\newblock In \bibinfo{editor}{{Ramsay}, S.~K.}, \bibinfo{editor}{{McLean},
		I.~S.} \& \bibinfo{editor}{{Takami}, H.} (eds.)
	\emph{\bibinfo{booktitle}{Ground-based and Airborne Instrumentation for
			Astronomy V}}, vol. \bibinfo{volume}{9147} of \emph{\bibinfo{series}{Society
			of Photo-Optical Instrumentation Engineers (SPIE) Conference Series}},
	\bibinfo{pages}{91470L} (\bibinfo{year}{2014}).
	\newblock \eprint{1412.0843}.
	
	\bibitem{blanton+2019}
	\bibinfo{author}{{Blanton}, M.} \emph{et~al.}
	\newblock \bibinfo{title}{{The Sloan Digital Sky Survey as an Archetypal
			Mid-Scale Program}}.
	\newblock In \emph{\bibinfo{booktitle}{Bull. Am. Astron. Soc.}},
	vol.~\bibinfo{volume}{51}, \bibinfo{pages}{196} (\bibinfo{year}{2019}).
	
	\bibitem{kollmeier+2019}
	\bibinfo{author}{{Kollmeier}, J.} \emph{et~al.}
	\newblock \bibinfo{title}{{SDSS-V Pioneering Panoptic Spectroscopy}}.
	\newblock In \emph{\bibinfo{booktitle}{Bull. Am. Astron. Soc.}},
	vol.~\bibinfo{volume}{51}, \bibinfo{pages}{274} (\bibinfo{year}{2019}).
	
	
\end{thebibliography}

\begin{thebibliography}{99}
	\makeatletter
	\addtocounter{\@listctr}{51}
	\makeatother
	\expandafter\ifx\csname url\endcsname\relax
	\def\url#1{\texttt{#1}}\fi
	\expandafter\ifx\csname urlprefix\endcsname\relax\def\urlprefix{URL }\fi
	\providecommand{\bibinfo}[2]{#2}
	\providecommand{\eprint}[2][]{\url{#2}}
	
	
	\bibitem{handberg+lund2014}
	\bibinfo{author}{Handberg, R.} \& \bibinfo{author}{Lund, M.~N.}
	\newblock \bibinfo{title}{Automated preparation of {{Kepler}} time series of
		planet hosts for asteroseismic analysis}.
	\newblock \emph{\bibinfo{journal}{Mon. Not. R. Astron. Soc.}}
	\textbf{\bibinfo{volume}{445}}, \bibinfo{pages}{2698--2709}
	(\bibinfo{year}{2014}).
	
	\bibitem{garcia+2011}
	\bibinfo{author}{Garc{\'i}a, R.~A.} \emph{et~al.}
	\newblock \bibinfo{title}{Preparation of {{Kepler}} light curves for
		asteroseismic analyses}.
	\newblock \emph{\bibinfo{journal}{Mon. Not. R. Astron. Soc.}}
	\textbf{\bibinfo{volume}{414}}, \bibinfo{pages}{L6--L10}
	(\bibinfo{year}{2011}).
	
	\bibitem{christensen-dalsgaard2008}
	\bibinfo{author}{{Christensen-Dalsgaard}, J.}
	\newblock \bibinfo{title}{{{ASTEC}}\textemdash the {{Aarhus STellar Evolution
				Code}}}.
	\newblock \emph{\bibinfo{journal}{Astrophysics and Space Science}}
	\textbf{\bibinfo{volume}{316}}, \bibinfo{pages}{13} (\bibinfo{year}{2008}).
	
	\bibitem{huber+2013a}
	\bibinfo{author}{{Barnes}, S.~A.}
	\newblock \bibinfo{title}{{A Simple Nonlinear Model for the Rotation of
			Main-sequence Cool Stars. I. Introduction, Implications for Gyrochronology,
			and Color-Period Diagrams}}.
	\newblock \emph{\bibinfo{journal}{Astrophys. J.}}
	\textbf{\bibinfo{volume}{722}}, \bibinfo{pages}{222--234}
	(\bibinfo{year}{2010}).
	
	\bibitem{buchhave+latham2015}
	\bibinfo{author}{Buchhave, L.~A.} \& \bibinfo{author}{Latham, D.~W.}
	\newblock \bibinfo{title}{The {{Metallicities}} of {{Stars}} with and without
		{{Transiting Planets}}}.
	\newblock \emph{\bibinfo{journal}{Astrophys. J.}}
	\textbf{\bibinfo{volume}{808}}, \bibinfo{pages}{187} (\bibinfo{year}{2015}).
	
	\bibitem{astropycollaboration+2013}
	\bibinfo{author}{{Astropy Collaboration}} \emph{et~al.}
	\newblock \bibinfo{title}{Astropy: {{A}} community {{Python}} package for
		astronomy}.
	\newblock \emph{\bibinfo{journal}{Astron. Astrophys.}}
	\textbf{\bibinfo{volume}{558}}, \bibinfo{pages}{A33} (\bibinfo{year}{2013}).
	
	\bibitem{astropycollaboration+2018}
	\bibinfo{author}{{Astropy Collaboration}} \emph{et~al.}
	\newblock \bibinfo{title}{The {{Astropy Project}}: {{Building}} an
		{{Open}}-science {{Project}} and {{Status}} of the v2.0 {{Core Package}}}.
	\newblock \emph{\bibinfo{journal}{Astron. J.}} \textbf{\bibinfo{volume}{156}},
	\bibinfo{pages}{123} (\bibinfo{year}{2018}).
	
	\bibitem{ginsburg+2019}
	\bibinfo{author}{Ginsburg, A.} \emph{et~al.}
	\newblock \bibinfo{title}{Astroquery: {{An Astronomical Web}}-querying
		{{Package}} in {{Python}}}.
	\newblock \emph{\bibinfo{journal}{Astron. J.}} \textbf{\bibinfo{volume}{157}},
	\bibinfo{pages}{98} (\bibinfo{year}{2019}).
	
	\bibitem{mckinney2010}
	\bibinfo{author}{McKinney, W.}
	\newblock \bibinfo{title}{Data {{Structures}} for {{Statistical Computing}} in
		{{Python}}}.
	\newblock In \emph{\bibinfo{booktitle}{Proceedings of the 9th {{Python}} in
			{{Science Conference}}}}, \bibinfo{pages}{51--56} (\bibinfo{year}{2010}).
	
	\bibitem{harvey1985}
	\bibinfo{author}{Harvey, J.}
	\newblock \bibinfo{title}{High-{{Resolution Helioseismology}}}.
	\newblock \emph{\bibinfo{journal}{Future Missions in Solar, Heliospheric \&
			Space Plasma Physics}} \textbf{\bibinfo{volume}{235}}, \bibinfo{pages}{199}
	(\bibinfo{year}{1985}).
	
	\bibitem{tassoul1980}
	\bibinfo{author}{{Tassoul}, M.}
	\newblock \bibinfo{title}{{Asymptotic approximations for stellar nonradial
			pulsations.}}
	\newblock \emph{\bibinfo{journal}{Astrophys. J. Supplement Series}}
	\textbf{\bibinfo{volume}{43}}, \bibinfo{pages}{469--490}
	(\bibinfo{year}{1980}).
	
	\bibitem{vrard+2016}
	\bibinfo{author}{Vrard, M.}, \bibinfo{author}{Mosser, B.} \&
	\bibinfo{author}{Samadi, R.}
	\newblock \bibinfo{title}{Period spacings in red giants. {{II}}. {{Automated}}
		measurement}.
	\newblock \emph{\bibinfo{journal}{Astron. Astrophys.}}
	\textbf{\bibinfo{volume}{588}}, \bibinfo{pages}{A87} (\bibinfo{year}{2016}).
	
	\bibitem{hogg+2010}
	\bibinfo{author}{{Hogg}, D.~W.}, \bibinfo{author}{{Bovy}, J.} \&
	\bibinfo{author}{{Lang}, D.}
	\newblock \bibinfo{title}{{Data analysis recipes: Fitting a model to data}}.
	\newblock \emph{\bibinfo{journal}{arXiv e-prints}}
	\bibinfo{pages}{arXiv:1008.4686} (\bibinfo{year}{2010}).
	\newblock \eprint{1008.4686}.
	
	\bibitem{hall+2019}
	\bibinfo{author}{Hall, O.~J.} \emph{et~al.}
	\newblock \bibinfo{title}{Testing asteroseismology with {{Gaia DR2}}:
		{{Hierarchical}} models of the {{Red Clump}}}.
	\newblock \emph{\bibinfo{journal}{Mon. Not. R. Astron. Soc.}}
	\textbf{\bibinfo{volume}{486}}, \bibinfo{pages}{3569--3585}
	(\bibinfo{year}{2019}).
	\newblock \eprint{1904.07919}.
	
	\bibitem{mazumdar+2014}
	\bibinfo{author}{Mazumdar, A.} \emph{et~al.}
	\newblock \bibinfo{title}{Measurement of acoustic glitches in solar-type stars
		from oscillation frequencies observed by {{Kepler}}}.
	\newblock \emph{\bibinfo{journal}{Astrophys. J.}}
	\textbf{\bibinfo{volume}{782}}, \bibinfo{pages}{18} (\bibinfo{year}{2014}).
	\newblock \eprint{1312.4907}.
	
	\bibitem{chaplin+basu2017}
	\bibinfo{author}{Chaplin, W.~J.} \& \bibinfo{author}{Basu}.
	\newblock \emph{\bibinfo{title}{Asteroseismic {{Data Analysis}}:
			{{Foundations}} and {{Techniques}}}} (\bibinfo{publisher}{{Princeton
			University Press}}, \bibinfo{address}{{Princeton, New Jersey}},
	\bibinfo{year}{2017}), \bibinfo{edition}{1st} edn.
	
	\bibitem{vanhoey+2013}
	\bibinfo{author}{Van~Hoey, S.}, \bibinfo{author}{{van der Kwast}, J.},
	\bibinfo{author}{Nopens, I.} \& \bibinfo{author}{Seuntjens, P.}
	\newblock \bibinfo{title}{Python package for model {{STructure ANalysis}}
		({{pySTAN}})}.
	\newblock \emph{\bibinfo{journal}{EGU General Assembly Conference Abstracts}}
	\textbf{\bibinfo{volume}{15}}, \bibinfo{pages}{EGU2013--10059}
	(\bibinfo{year}{2013}).
	
	\bibitem{carpenter+2017}
	\bibinfo{author}{Carpenter, B.} \emph{et~al.}
	\newblock \bibinfo{title}{Stan: {{A Probabilistic Programming Language}}}.
	\newblock \emph{\bibinfo{journal}{Journal of Statistical Software}}
	\textbf{\bibinfo{volume}{76}}, \bibinfo{pages}{1--32} (\bibinfo{year}{2017}).
	
	\bibitem{vanderwalt+2011}
	\bibinfo{author}{{van der Walt}, S.}, \bibinfo{author}{Colbert, S.~C.} \&
	\bibinfo{author}{Varoquaux, G.}
	\newblock \bibinfo{title}{The {{NumPy Array}}: {{A Structure}} for {{Efficient
				Numerical Computation}}}.
	\newblock \emph{\bibinfo{journal}{Computing in Science and Engineering}}
	\textbf{\bibinfo{volume}{13}}, \bibinfo{pages}{22--30}
	(\bibinfo{year}{2011}).
	
	\bibitem{thetheanodevelopmentteam+2016}
	\bibinfo{author}{{The Theano Development Team}} \emph{et~al.}
	\newblock \bibinfo{title}{{Theano: A Python framework for fast computation of
			mathematical expressions}}.
	\newblock \emph{\bibinfo{journal}{arXiv e-prints}}
	\bibinfo{pages}{arXiv:1605.02688} (\bibinfo{year}{2016}).
	\newblock \eprint{1605.02688}.
	
	\bibitem{gaiacollaboration+2018}
	\bibinfo{author}{{Gaia Collaboration}} \emph{et~al.}
	\newblock \bibinfo{title}{{Gaia Data Release 2. Summary of the contents and
			survey properties}}.
	\newblock \emph{\bibinfo{journal}{Astron. Astrophys.}}
	\textbf{\bibinfo{volume}{616}}, \bibinfo{pages}{A1} (\bibinfo{year}{2018}).
	\newblock \eprint{1804.09365}.
	
	\bibitem{raghavan+2010}
	\bibinfo{author}{Raghavan, D.} \emph{et~al.}
	\newblock \bibinfo{title}{A {{Survey}} of {{Stellar Families}}:
		{{Multiplicity}} of {{Solar}}-type {{Stars}}}.
	\newblock \emph{\bibinfo{journal}{Astrophys. J. Supplement Series}}
	\textbf{\bibinfo{volume}{190}}, \bibinfo{pages}{1--42}
	(\bibinfo{year}{2010}).
	
	\bibitem{chaplin+2011}
	\bibinfo{author}{{Chaplin}, W.~J.} \emph{et~al.}
	\newblock \bibinfo{title}{{Ensemble Asteroseismology of Solar-Type Stars with
			the NASA Kepler Mission}}.
	\newblock \emph{\bibinfo{journal}{Science}} \textbf{\bibinfo{volume}{332}},
	\bibinfo{pages}{213} (\bibinfo{year}{2011}).
	\newblock \eprint{1109.4723}.
	
	\bibitem{seabold+perktold2010}
	\bibinfo{author}{Seabold, S.} \& \bibinfo{author}{Perktold, J.}
	\newblock \bibinfo{title}{{S}tatsmodels: {E}conometric and {S}tatistical
		{M}odeling with {P}ython}.
	\newblock In \bibinfo{editor}{{S}t\'efan van~der {W}alt} \&
	\bibinfo{editor}{{J}arrod {M}illman} (eds.)
	\emph{\bibinfo{booktitle}{{P}roceedings of the 9th {P}ython in {S}cience
			{C}onference}}, \bibinfo{pages}{92 -- 96} (\bibinfo{year}{2010}).
	
	\bibitem{betancourt+girolami2013}
	\bibinfo{author}{{Astrophys. J.}, M.~J.} \& \bibinfo{author}{{Girolami}, M.}
	\newblock \bibinfo{title}{{Hamiltonian Monte Carlo for Hierarchical Models}}.
	\newblock \emph{\bibinfo{journal}{arXiv e-prints}}
	\bibinfo{pages}{arXiv:1312.0906} (\bibinfo{year}{2013}).
	\newblock \eprint{1312.0906}.
	
	\bibitem{foreman-mackey+2013}
	\bibinfo{author}{{Foreman-Mackey}, D.}, \bibinfo{author}{Hogg, D.~W.},
	\bibinfo{author}{Lang, D.} \& \bibinfo{author}{Goodman, J.}
	\newblock \bibinfo{title}{Emcee: {{The MCMC Hammer}}}.
	\newblock \emph{\bibinfo{journal}{Publ. Astron. Soc. Pacific}}
	\textbf{\bibinfo{volume}{125}}, \bibinfo{pages}{306--312}
	(\bibinfo{year}{2013}).
	\newblock \eprint{1202.3665}.
	
	\bibitem{foreman-mackey2016}
	\bibinfo{author}{{Foreman-Mackey}, D.}
	\newblock \bibinfo{title}{Corner.py: {{Scatterplot}} matrices in {{Python}}}.
	\newblock \emph{\bibinfo{journal}{The Journal of Open Source Software}}
	\textbf{\bibinfo{volume}{1}} (\bibinfo{year}{2016}).
\end{thebibliography}

\addcontentsline{toc}{section}{References}
\begin{thebibliography}{99}
	\makeatletter
	\addtocounter{\@listctr}{78}
	\makeatother
	\expandafter\ifx\csname url\endcsname\relax
	\def\url#1{\texttt{#1}}\fi
	\expandafter\ifx\csname urlprefix\endcsname\relax\def\urlprefix{URL }\fi
	\providecommand{\bibinfo}[2]{#2}
	\providecommand{\eprint}[2][]{\url{#2}}
	
\bibitem{hogg2012}
\bibinfo{author}{Hogg, D.~W.}
\newblock \bibinfo{title}{Data analysis recipes: {{Probability}} calculus for
	inference}.
\newblock \emph{\bibinfo{journal}{arXiv e-prints}}
\bibinfo{pages}{arXiv:1205.4446} (\bibinfo{year}{2012}).

\bibitem{davies+2014}
\bibinfo{author}{Davies, G.~R.}, \bibinfo{author}{Chaplin, W.~J.},
\bibinfo{author}{Elsworth, Y.} \& \bibinfo{author}{Hale, S.~J.}
\newblock \bibinfo{title}{{{BiSON}} data preparation: A correction for
	differential extinction and the weighted averaging of contemporaneous data}.
\newblock \emph{\bibinfo{journal}{Mon. Not. R. Astron. Soc.}}
\textbf{\bibinfo{volume}{441}}, \bibinfo{pages}{3009--3017}
(\bibinfo{year}{2014}).

\bibitem{appourchaux+2016}
\bibinfo{author}{Appourchaux, T.} \emph{et~al.}
\newblock \bibinfo{title}{Oscillation mode linewidths and heights of 23
	main-sequence stars observed by {{{\emph{Kepler}}}}
	{\emph{(}}{{{\emph{Corrigendum}}}}{\emph{)}}}.
\newblock \emph{\bibinfo{journal}{Astronomy \& Astrophysics}}
\textbf{\bibinfo{volume}{595}}, \bibinfo{pages}{C2} (\bibinfo{year}{2016}).

\bibitem{rasmussen+williams2006}
\bibinfo{author}{Rasmussen, C.~E.} \& \bibinfo{author}{Williams, C. K.~I.}
\newblock \emph{\bibinfo{title}{Gaussian Processes for Machine Learning}}.
\newblock Adaptive Computation and Machine Learning (\bibinfo{publisher}{{MIT
		Press}}, \bibinfo{address}{{Cambridge, Mass}}, \bibinfo{year}{2006}).

\bibitem{toutain+appourchaux1994}
\bibinfo{author}{Toutain, T.} \& \bibinfo{author}{Appourchaux, T.}
\newblock \bibinfo{title}{Maximum likelihood estimators: {{An}} application to
	the estimation of the precision of helioseismic measurements}.
\newblock \emph{\bibinfo{journal}{Astron. Astrophys.}}
\textbf{\bibinfo{volume}{289}}, \bibinfo{pages}{649--658}
(\bibinfo{year}{1994}).

\bibitem{gizon+solanki2003}
\bibinfo{author}{Gizon, L.} \& \bibinfo{author}{Solanki, S.~K.}
\newblock \bibinfo{title}{Determining the {{Inclination}} of the {{Rotation
			Axis}} of a {{Sun}}-like {{Star}}}.
\newblock \emph{\bibinfo{journal}{Astrophys. J.}}
\textbf{\bibinfo{volume}{589}}, \bibinfo{pages}{1009} (\bibinfo{year}{2003}).

\bibitem{handberg+campante2011}
\bibinfo{author}{Handberg, R.} \& \bibinfo{author}{Campante, T.~L.}
\newblock \bibinfo{title}{Bayesian peak-bagging of solar-like oscillators using
	{{MCMC}}: A comprehensive guide}.
\newblock \emph{\bibinfo{journal}{Astron. Astrophys.}}
\textbf{\bibinfo{volume}{527}}, \bibinfo{pages}{A56} (\bibinfo{year}{2011}).

\bibitem{appourchaux+1998}
\bibinfo{author}{Appourchaux, T.}, \bibinfo{author}{Gizon, L.} \&
\bibinfo{author}{{Rabello-Soares}, M.-C.}
\newblock \bibinfo{title}{The art of fitting p-mode spectra. {{I}}. {{Maximum}}
	likelihood estimation}.
\newblock \emph{\bibinfo{journal}{Astron. Astrophys. Supplement Series}}
\textbf{\bibinfo{volume}{132}}, \bibinfo{pages}{107--119}
(\bibinfo{year}{1998}).

\bibitem{anderson+1990}
\bibinfo{author}{{Anderson}, E.~R.}, \bibinfo{author}{{Duvall}, J., Thomas~L.}
\& \bibinfo{author}{{Jefferies}, S.~M.}
\newblock \bibinfo{title}{{Modeling of Solar Oscillation Power Spectra}}.
\newblock \emph{\bibinfo{journal}{Astrophys. J.}}
\textbf{\bibinfo{volume}{364}}, \bibinfo{pages}{699} (\bibinfo{year}{1990}).

\bibitem{gelman2006}
\bibinfo{author}{Gelman, A.}
\newblock \bibinfo{title}{Prior distributions for variance parameters in
	hierarchical models (comment on article by {{Browne}} and {{Draper}})}.
\newblock \emph{\bibinfo{journal}{Bayesian Analysis}}
\textbf{\bibinfo{volume}{1}}, \bibinfo{pages}{515--534}
(\bibinfo{year}{2006}).

\bibitem{ballot+2006}
\bibinfo{author}{Ballot, J.}, \bibinfo{author}{Garcia, R.~A.} \&
\bibinfo{author}{Lambert, P.}
\newblock \bibinfo{title}{Rotation speed and stellar axis inclination from p
	modes: How {{CoRoT}} would see other suns}.
\newblock \emph{\bibinfo{journal}{Mon. Not. R. Astron. Soc.}}
\textbf{\bibinfo{volume}{369}}, \bibinfo{pages}{1281--1286}
(\bibinfo{year}{2006}).

\bibitem{ballot+2008a}
\bibinfo{author}{Ballot, J.}, \bibinfo{author}{Appourchaux, T.},
\bibinfo{author}{Toutain, T.} \& \bibinfo{author}{Guittet, M.}
\newblock \bibinfo{title}{On deriving p-mode parameters for inclined solar-like
	stars}.
\newblock \emph{\bibinfo{journal}{Astronomy \& Astrophysics}}
\textbf{\bibinfo{volume}{486}}, \bibinfo{pages}{867--875}
(\bibinfo{year}{2008}).
\newblock \eprint{0803.0885}.

\bibitem{kass+raftery1995}
\bibinfo{author}{Kass, R.~E.} \& \bibinfo{author}{Raftery, A.~E.}
\newblock \bibinfo{title}{Bayes {{Factors}}}.
\newblock \emph{\bibinfo{journal}{Journal of the American Statistical
		Association}} \textbf{\bibinfo{volume}{90}}, \bibinfo{pages}{773--795}
(\bibinfo{year}{1995}).

\bibitem{beck2000}
\bibinfo{author}{Beck, J.~G.}
\newblock \bibinfo{title}{A comparison of differential rotation measurements -
	({{Invited Review}})}.
\newblock \emph{\bibinfo{journal}{Solar Physics}}
\textbf{\bibinfo{volume}{191}}, \bibinfo{pages}{47--70}
(\bibinfo{year}{2000}).

\bibitem{doyle+2014}
\bibinfo{author}{Doyle, A.~P.}, \bibinfo{author}{Davies, G.~R.},
\bibinfo{author}{Smalley, B.}, \bibinfo{author}{Chaplin, W.~J.} \&
\bibinfo{author}{Elsworth, Y.}
\newblock \bibinfo{title}{Determining stellar macroturbulence using
	asteroseismic rotational velocities from {{Kepler}}}.
\newblock \emph{\bibinfo{journal}{Mon. Not. R. Astron. Soc.}}
\textbf{\bibinfo{volume}{444}}, \bibinfo{pages}{3592--3602}
(\bibinfo{year}{2014}).

\bibitem{tayar+2015}
\bibinfo{author}{Tayar, J.} \emph{et~al.}
\newblock \bibinfo{title}{Rapid {{Rotation}} of {{Low}}-mass {{Red Giants Using
			APOKASC}}: {{A Measure}} of {{Interaction Rates}} on the
	{{Post}}-main-sequence}.
\newblock \emph{\bibinfo{journal}{Astrophys. J.}}
\textbf{\bibinfo{volume}{807}}, \bibinfo{pages}{82} (\bibinfo{year}{2015}).

\bibitem{schofield+2019}
\bibinfo{author}{Schofield, M.} \emph{et~al.}
\newblock \bibinfo{title}{The {{Asteroseismic Target List}} for {{Solar}}-like
	{{Oscillators Observed}} in 2 minute {{Cadence}} with the {{Transiting
			Exoplanet Survey Satellite}}}.
\newblock \emph{\bibinfo{journal}{Astrophys. J. Supplement Series}}
\textbf{\bibinfo{volume}{241}}, \bibinfo{pages}{12} (\bibinfo{year}{2019}).

\bibitem{mathur+2019}
\bibinfo{author}{{Mathur}, S.} \emph{et~al.}
\newblock \bibinfo{title}{{Revisiting the impact of stellar magnetic activity
		on the detection of solar-like oscillations by Kepler}}.
\newblock \emph{\bibinfo{journal}{Frontiers in Astronomy and Space Sciences}}
\textbf{\bibinfo{volume}{6}}, \bibinfo{pages}{46} (\bibinfo{year}{2019}).
\newblock \eprint{1907.01415}.

\bibitem{amard+matt2020}
\bibinfo{author}{{Amard}, L.} \& \bibinfo{author}{{Matt}, S.~P.}
\newblock \bibinfo{title}{{The Impact of Metallicity on the Evolution of the
		Rotation and Magnetic Activity of Sun-like Stars}}.
\newblock \emph{\bibinfo{journal}{Astrophys. J.}}
\textbf{\bibinfo{volume}{889}}, \bibinfo{pages}{108} (\bibinfo{year}{2020}).
\newblock \eprint{2001.10404}.

\bibitem{fleming+2019}
\bibinfo{author}{Fleming, D.~P.}, \bibinfo{author}{Barnes, R.},
\bibinfo{author}{Davenport, J. R.~A.} \& \bibinfo{author}{Luger, R.}
\newblock \bibinfo{title}{Rotation {{Period Evolution}} in {{Low}}-mass
	{{Binary Stars}}: {{The Impact}} of {{Tidal Torques}} and {{Magnetic
			Braking}}}.
\newblock \emph{\bibinfo{journal}{Astrophys. J.}}
\textbf{\bibinfo{volume}{881}}, \bibinfo{pages}{88} (\bibinfo{year}{2019}).

\bibitem{halbwachs1986}
\bibinfo{author}{{Halbwachs}, J.~L.}
\newblock \bibinfo{title}{{Common proper motion stars in the AGK 3.}}
\newblock \emph{\bibinfo{journal}{Astron. Astrophys.s}}
\textbf{\bibinfo{volume}{66}}, \bibinfo{pages}{131--148}
(\bibinfo{year}{1986}).

\bibitem{white+2013}
\bibinfo{author}{{White}, T.~R.} \emph{et~al.}
\newblock \bibinfo{title}{{Interferometric radii of bright Kepler stars with
		the CHARA Array: {\ensuremath{\theta}} Cygni and 16 Cygni A and B}}.
\newblock \emph{\bibinfo{journal}{Mon. Not. R. Astron. Soc.}}
\textbf{\bibinfo{volume}{433}}, \bibinfo{pages}{1262--1270}
(\bibinfo{year}{2013}).
\newblock \eprint{1305.1934}.

\bibitem{maxted+2015}
\bibinfo{author}{Maxted, P. F.~L.}, \bibinfo{author}{Serenelli, A.~M.} \&
\bibinfo{author}{Southworth, J.}
\newblock \bibinfo{title}{Comparison of gyrochronological and isochronal age
	estimates for transiting exoplanet host stars}.
\newblock \emph{\bibinfo{journal}{Astron. Astrophys.}}
\textbf{\bibinfo{volume}{577}}, \bibinfo{pages}{A90} (\bibinfo{year}{2015}).

\bibitem{gallet+delorme2019}
\bibinfo{author}{{Gallet}, F.} \& \bibinfo{author}{{Delorme}, P.}
\newblock \bibinfo{title}{{Star-planet tidal interaction and the limits of
		gyrochronology}}.
\newblock \emph{\bibinfo{journal}{Astron. Astrophys.}}
\textbf{\bibinfo{volume}{626}}, \bibinfo{pages}{A120} (\bibinfo{year}{2019}).
\newblock \eprint{1905.06070}.

\bibitem{benbakoura+2019}
\bibinfo{author}{Benbakoura, M.}, \bibinfo{author}{R{\'e}ville, V.},
\bibinfo{author}{Brun, A.~S.}, \bibinfo{author}{{Le Poncin-Lafitte}, C.} \&
\bibinfo{author}{Mathis, S.}
\newblock \bibinfo{title}{Evolution of star-planet systems under magnetic
	braking and tidal interaction}.
\newblock \emph{\bibinfo{journal}{Astron. Astrophys.}}
\textbf{\bibinfo{volume}{621}}, \bibinfo{pages}{A124} (\bibinfo{year}{2019}).

\bibitem{ceillier+2016}
\bibinfo{author}{Ceillier, T.} \emph{et~al.}
\newblock \bibinfo{title}{Rotation periods and seismic ages of {{KOIs}} -
	comparison with stars without detected planets from {{Kepler}} observations}.
\newblock \emph{\bibinfo{journal}{Mon. Not. R. Astron. Soc.}}
\textbf{\bibinfo{volume}{456}}, \bibinfo{pages}{119--125}
(\bibinfo{year}{2016}).
	
\end{thebibliography}

\renewcommand{\tablename}{Supplementary Table}

\begin{landscape}
\setlength\LTleft{0pt}
\setlength\LTright{0pt}
\footnotesize
\addcontentsline{toc}{section}{Supplementary Table 1}
\begin{longtable}{c|ccccc|ccc|ccc}
	\caption{Parameters for the 94 stars for which seismic rotation rates were obtained in this work. Temperature (\teff), age, mass, metallicity (\feh) and surface gravity ($\log(g)$) are adopted from the LEGACY \cite[L]{lund+2017,silvaaguirre+2017} and`Kages' \cite[K]{silvaaguirre+2015,davies+2016} catalogues, as listed in the Source column. Projected splitting ($\nu_{\rm s}\sin(i)$), inclination angle ($i$) and asteroseismic rotation ($P$) are from this work. Uncertainties were taken using the $15.9^{\rm th}$ and $84.1^{\rm st}$ percentiles of posterior distributions on the parameters, which are frequently asymmetrical in linear space. Reported values are the median of the posteriors. For parameters with no direct posterior samples (e.g. rotation) the full posterior samples were transformed before taking the summary statistics. The stellar type denotes whether a star is roughly classified as belonging to the main sequence (MS), Sub-Giants (SG) or `hot' stars (H) (see text).
		The flags indicate the following: 0; no issues, used in the gyrochronology analysis. 1; has either a number of effective samples $n_{\rm eff} < 1000$ for the asteroseismic splitting, or Gelman-Rubin convergence metric of $\hat{R} > 1.1$ \cite{gelman+rubin1992}, indicating that rotation measurements for these stars are less robust than those with a flag of 0. 2; was found to strongly disagree with multiple literature values, excluded from the gyrochronology analysis. 3; fell outside the model range of the stellar models, and were therefore not used in the gyrochronology analysis. Table is continued on the next page. A machine-readable version of the table is available (as Supplementary Data 1).}\label{tab:results}\\
	\toprule
	KIC & $T_{\rm{eff}}$ & Age & Mass & \feh & $\log(g)$ & $\nu_{\rm{s}}\sin(i)$ & $i$   & $P$   &  Flag & Type & Source \\
	& [$\mathrm{K}$] &  [$\mathrm{Gyr}$] & [$\mathrm{M_{\odot}}$] & [$\mathrm{dex}$] & [$\mathrm{dex}$] & [$\mathrm{\mu Hz}$]  & [$\mathrm{{}^{\circ}}$] & [$\mathrm{days}$]   &  &  &  \\
	\midrule
	\endfirsthead
	\caption[]{\textit{Continued from previous page.}}\\
	\toprule
	KIC & $T_{\rm{eff}}$ & Age & Mass & \feh & $\log(g)$ & $\nu_{\rm{s}}\sin(i)$ & $i$   & $P$   &  Flag & Type & Source \\
	& [$\mathrm{K}$] &  [$\mathrm{Gyr}$] & [$\mathrm{M_{\odot}}$] & [$\mathrm{dex}$] & [$\mathrm{dex}$] & [$\mathrm{\mu Hz}$]  & [$\mathrm{{}^{\circ}}$] & [$\mathrm{days}$]   &  &  &  \\
	\midrule
	\endhead
	\bottomrule \multicolumn{12}{r}{\textit{Continued on next page}}\\
	\endfoot
	\bottomrule
	\endlastfoot
	
	1435467 & 6326$\pm$77    & 3.02$_{-0.35}^{+0.50}$    & 1.32$_{-0.05}^{+0.03}$ & 0.01$\pm$0.10     & 4.100$_{-0.009}^{+0.009}$ & 1.58$_{-0.09}^{+0.10}$ & 63.4$_{-6.6}^{+10.2}$     & 6.5$_{-0.6}^{+0.8}$      & 0 &        H & L \\
	2837475 & 6614$\pm$77    & 1.63$_{-0.18}^{+0.11}$    & 1.43$_{-0.02}^{+0.02}$ & 0.01$\pm$0.10     & 4.163$_{-0.007}^{+0.007}$ & 3.12$_{-0.08}^{+0.08}$ & 70.7$_{-4.4}^{+6.0}$      & 3.5$_{-0.2}^{+0.2}$      & 0 &        H & L \\
	3425851 & 6343$\pm$85    & 3.32$_{-0.64}^{+0.85}$    & 1.18$_{-0.05}^{+0.05}$ & -0.04$\pm$0.10    & 4.243$_{-0.008}^{+0.008}$ & 1.17$_{-0.62}^{+0.48}$ & 60.9$_{-22.7}^{+20.1}$    & 8.1$_{-2.7}^{+8.6}$      & 0 &        H & K \\
	3427720 & 6045$\pm$77    & 2.23$_{-0.24}^{+0.24}$    & 1.11$_{-0.01}^{+0.02}$ & -0.06$\pm$0.10    & 4.387$_{-0.004}^{+0.005}$ & 0.30$_{-0.06}^{+0.06}$ & 56.4$_{-23.4}^{+22.9}$    & 31.6$_{-11.8}^{+10.2}$   & 0 &        MS & L \\
	3456181 & 6384$\pm$77    & 2.09$_{-0.13}^{+0.13}$    & 1.50$_{-0.02}^{+0.03}$ & -0.15$\pm$0.10    & 3.949$_{-0.009}^{+0.008}$ & 0.92$_{-0.08}^{+0.08}$ & 58.2$_{-17.7}^{+20.4}$    & 10.7$_{-2.8}^{+2.0}$     & 0 &        H & L \\
	3544595 & 5669$\pm$75    & 6.63$_{-0.57}^{+0.62}$    & 0.90$_{-0.01}^{+0.01}$ & -0.18$\pm$0.10    & 4.468$_{-0.003}^{+0.003}$ & 0.40$_{-0.04}^{+0.04}$ & 66.0$_{-13.9}^{+15.6}$    & 26.1$_{-4.7}^{+3.9}$     & 0 &        MS & K \\
	3632418 & 6193$\pm$77    & 2.63$_{-0.18}^{+0.18}$    & 1.41$_{-0.02}^{+0.02}$ & -0.12$\pm$0.10    & 4.024$_{-0.008}^{+0.008}$ & 0.98$_{-0.03}^{+0.03}$ & 72.3$_{-7.4}^{+10.0}$     & 11.2$_{-0.7}^{+0.6}$     & 0 &        MS & L \\
	3656476 & 5668$\pm$77    & 8.37$_{-1.57}^{+1.72}$    & 1.04$_{-0.04}^{+0.05}$ & 0.25$\pm$0.10     & 4.225$_{-0.010}^{+0.008}$ & 0.21$_{-0.02}^{+0.02}$ & 62.4$_{-20.5}^{+18.9}$    & 48.0$_{-12.7}^{+8.1}$    & 0 &        MS & L \\
	3735871 & 6107$\pm$77    & 2.35$_{-0.85}^{+1.04}$    & 1.09$_{-0.04}^{+0.04}$ & -0.04$\pm$0.10    & 4.396$_{-0.007}^{+0.007}$ & 0.69$_{-0.05}^{+0.05}$ & 70.4$_{-15.1}^{+13.4}$    & 15.8$_{-2.5}^{+1.8}$     & 0 &        MS & L \\
	4141376 & 6134$\pm$91    & 3.27$_{-0.64}^{+0.59}$    & 1.02$_{-0.03}^{+0.02}$ & -0.24$\pm$0.10    & 4.412$_{-0.003}^{+0.004}$ & 0.76$_{-0.13}^{+0.13}$ & 64.0$_{-16.7}^{+17.5}$    & 13.4$_{-3.0}^{+3.4}$     & 0 &        MS & K \\
	4143755 & 5622$\pm$106   & 11.27$_{-1.35}^{+1.50}$   & 0.92$_{-0.03}^{+0.02}$ & -0.40$\pm$0.11    & 4.102$_{-0.001}^{+0.002}$ & 0.18$_{-0.05}^{+0.08}$ & 45.8$_{-27.3}^{+30.6}$    & 48.1$_{-32.7}^{+27.4}$   & 1 &        MS & K \\
	4349452 & 6270$\pm$79    & 3.45$_{-0.72}^{+0.81}$    & 1.16$_{-0.05}^{+0.04}$ & -0.04$\pm$0.10    & 4.275$_{-0.007}^{+0.008}$ & 1.50$_{-0.09}^{+0.09}$ & 79.7$_{-10.0}^{+7.1}$      & 7.5$_{-0.6}^{+0.5}$     & 0 &        H & K \\
	4914423 & 5845$\pm$88    & 6.67$_{-0.62}^{+0.69}$    & 1.10$_{-0.03}^{+0.02}$ & 0.07$\pm$0.11     & 4.155$_{-0.004}^{+0.004}$ & 0.42$_{-0.14}^{+0.15}$ & 61.3$_{-30.5}^{+19.8}$    & 23.1$_{-9.6}^{+11.7}$    & 0 &        MS & K \\
	4914923 & 5805$\pm$77    & 7.57$_{-1.79}^{+1.66}$    & 1.06$_{-0.05}^{+0.06}$ & 0.08$\pm$0.10     & 4.197$_{-0.010}^{+0.008}$ & 0.39$_{-0.03}^{+0.03}$ & 46.6$_{-8.1}^{+13.3}$     & 21.4$_{-3.5}^{+5.4}$     & 0 &        MS & L \\
	5094751 & 5952$\pm$75    & 6.35$_{-1.05}^{+1.05}$    & 1.07$_{-0.04}^{+0.04}$ & -0.08$\pm$0.10    & 4.213$_{-0.008}^{+0.007}$ & 0.39$_{-0.16}^{+0.27}$ & 51.5$_{-31.1}^{+26.7}$    & 22.9$_{-15.8}^{+19.3}$   & 0 &        MS & K \\
	5184732 & 5846$\pm$77    & 4.85$_{-0.88}^{+1.57}$    & 1.15$_{-0.06}^{+0.04}$ & 0.36$\pm$0.10     & 4.255$_{-0.008}^{+0.010}$ & 0.55$_{-0.02}^{+0.02}$ & 71.3$_{-10.8}^{+11.2}$    & 19.9$_{-1.9}^{+1.3}$     & 0 &        MS & L \\
	5773345 & 6130$\pm$84    & 2.55$_{-0.24}^{+0.26}$    & 1.47$_{-0.03}^{+0.03}$ & 0.21$\pm$0.09     & 3.993$_{-0.007}^{+0.008}$ & 1.08$_{-0.08}^{+0.08}$ & 33.7$_{-2.5}^{+2.8}$      & 5.9$_{-0.5}^{+0.7}$      & 0 &        SG & L \\
	5866724 & 6169$\pm$50    & 3.89$_{-0.48}^{+0.59}$    & 1.20$_{-0.03}^{+0.03}$ & 0.09$\pm$0.08     & 4.224$_{-0.005}^{+0.007}$ & 1.34$_{-0.08}^{+0.07}$ & 80.9$_{-9.4}^{+6.3}$      & 8.4$_{-0.5}^{+0.5}$      & 0 &        MS & K \\
	5950854 & 5853$\pm$77    & 8.93$_{-1.15}^{+1.12}$    & 0.97$_{-0.03}^{+0.03}$ & -0.23$\pm$0.10    & 4.238$_{-0.007}^{+0.007}$ & 0.29$_{-0.12}^{+0.64}$ & 27.6$_{-15.2}^{+46.0}$    & 22.9$_{-19.7}^{+34.2}$   & 1 &        MS & L \\
	6106415 & 6037$\pm$77    & 5.03$_{-1.12}^{+1.28}$    & 1.07$_{-0.04}^{+0.05}$ & -0.04$\pm$0.10    & 4.295$_{-0.009}^{+0.009}$ & 0.69$_{-0.02}^{+0.02}$ & 73.1$_{-6.3}^{+8.4}$      & 16.0$_{-0.8}^{+0.7}$     & 0 &        MS & L \\
	6116048 & 6033$\pm$77    & 9.58$_{-1.90}^{+2.16}$    & 0.94$_{-0.05}^{+0.05}$ & -0.23$\pm$0.10    & 4.254$_{-0.012}^{+0.009}$ & 0.63$_{-0.02}^{+0.02}$ & 76.3$_{-10.2}^{+9.0}$     & 17.9$_{-1.2}^{+0.8}$     & 0 &        MS & L \\
	6196457 & 5871$\pm$94    & 5.52$_{-0.48}^{+0.51}$    & 1.21$_{-0.03}^{+0.02}$ & 0.17$\pm$0.11     & 4.049$_{-0.004}^{+0.005}$ & 0.43$_{-0.19}^{+0.28}$ & 52.5$_{-29.1}^{+26.0}$    & 20.7$_{-13.7}^{+19.2}$   & 0 &        MS & K \\
	6225718 & 6313$\pm$76    & 2.41$_{-0.43}^{+0.53}$    & 1.16$_{-0.03}^{+0.03}$ & -0.07$\pm$0.10    & 4.319$_{-0.007}^{+0.005}$ & 0.81$_{-0.03}^{+0.03}$ & 29.1$_{-1.8}^{+2.1}$      & 6.9$_{-0.5}^{+0.6}$      & 0 &        H & L \\
	6278762 & 5046$\pm$74    & 11.54$_{-0.94}^{+0.99}$   & 0.74$_{-0.01}^{+0.01}$ & -0.37$\pm$0.09    & 4.560$_{-0.003}^{+0.002}$ & 0.30$_{-0.09}^{+0.09}$ & 62.2$_{-29.5}^{+19.0}$    & 33.0$_{-13.5}^{+13.6}$   & 1, 3 &        MS & K \\
	6508366 & 6331$\pm$77    & 2.06$_{-0.14}^{+0.13}$    & 1.53$_{-0.02}^{+0.03}$ & -0.05$\pm$0.10    & 3.942$_{-0.007}^{+0.005}$ & 2.28$_{-0.04}^{+0.04}$ & 87.0$_{-3.2}^{+2.1}$      & 5.1$_{-0.1}^{+0.1}$      & 0 &        H & L \\
	6521045 & 5825$\pm$75    & 6.50$_{-0.56}^{+0.46}$    & 1.11$_{-0.02}^{+0.02}$ & 0.02$\pm$0.10     & 4.125$_{-0.004}^{+0.004}$ & 0.45$_{-0.02}^{+0.03}$ & 75.7$_{-11.2}^{+9.7}$     & 24.8$_{-2.0}^{+1.9}$     & 0 &        MS & K \\
	6603624 & 5674$\pm$77    & 7.82$_{-0.86}^{+0.94}$    & 1.01$_{-0.02}^{+0.03}$ & 0.28$\pm$0.10     & 4.320$_{-0.005}^{+0.004}$ & 1.13$_{-0.13}^{+0.13}$ & 6.9$_{-0.8}^{+0.8}$       & 1.2$_{-0.0}^{+0.0}$      & 2 &        MS & L \\
	6679371 & 6479$\pm$77    & 1.95$_{-0.16}^{+0.18}$    & 1.53$_{-0.02}^{+0.04}$ & 0.01$\pm$0.10     & 3.934$_{-0.008}^{+0.007}$ & 1.90$_{-0.06}^{+0.05}$ & 82.1$_{-7.2}^{+5.5}$      & 6.0$_{-0.2}^{+0.2}$      & 0 &        H & L \\
	6933899 & 5832$\pm$77    & 6.34$_{-0.62}^{+0.72}$    & 1.13$_{-0.03}^{+0.03}$ & -0.01$\pm$0.10    & 4.087$_{-0.007}^{+0.008}$ & 0.36$_{-0.02}^{+0.02}$ & 64.3$_{-14.0}^{+16.1}$    & 28.9$_{-4.8}^{+3.7}$     & 0 &        MS & L \\
	7103006 & 6344$\pm$77    & 2.47$_{-0.24}^{+0.22}$    & 1.42$_{-0.02}^{+0.04}$ & 0.02$\pm$0.10     & 4.015$_{-0.007}^{+0.007}$ & 1.36$_{-0.09}^{+0.08}$ & 56.8$_{-8.8}^{+14.8}$     & 7.1$_{-1.0}^{+1.3}$      & 0 &        H & L \\
	7106245 & 6068$\pm$102   & 6.27$_{-1.06}^{+1.06}$    & 0.92$_{-0.04}^{+0.02}$ & -0.99$\pm$0.19    & 4.325$_{-0.007}^{+0.007}$ & 0.32$_{-0.10}^{+0.16}$ & 33.4$_{-14.7}^{+38.4}$    & 21.4$_{-13.2}^{+23.8}$   & 1, 3 &        MS & L \\
	7206837 & 6305$\pm$77    & 2.90$_{-0.30}^{+0.42}$    & 1.30$_{-0.03}^{+0.03}$ & 0.10$\pm$0.10     & 4.163$_{-0.007}^{+0.008}$ & 1.53$_{-0.12}^{+0.12}$ & 31.7$_{-2.8}^{+3.2}$      & 4.0$_{-0.4}^{+0.6}$      & 0 &        H & L \\
	7296438 & 5775$\pm$77    & 7.23$_{-1.77}^{+1.49}$    & 1.08$_{-0.05}^{+0.06}$ & 0.19$\pm$0.10     & 4.201$_{-0.010}^{+0.009}$ & 0.20$_{-0.06}^{+0.06}$ & 50.5$_{-31.1}^{+28.2}$    & 45.6$_{-29.0}^{+23.4}$   & 1 &        MS & L \\
	7510397 & 6171$\pm$77    & 2.82$_{-0.16}^{+0.14}$    & 1.37$_{-0.02}^{+0.02}$ & -0.21$\pm$0.10    & 4.036$_{-0.004}^{+0.007}$ & 0.64$_{-0.06}^{+0.06}$ & 19.9$_{-2.0}^{+2.0}$      & 6.1$_{-0.6}^{+0.7}$      & 0 &        MS & L \\
	7670943 & 6463$\pm$110   & 2.78$_{-0.51}^{+0.62}$    & 1.24$_{-0.05}^{+0.04}$ & 0.09$\pm$0.11     & 4.228$_{-0.008}^{+0.008}$ & 1.83$_{-0.14}^{+0.14}$ & 75.7$_{-11.5}^{+9.8}$     & 6.0$_{-0.6}^{+0.6}$      & 0 &        H & K \\
	7680114 & 5811$\pm$77    & 7.68$_{-1.28}^{+1.45}$    & 1.06$_{-0.05}^{+0.04}$ & 0.05$\pm$0.10     & 4.172$_{-0.010}^{+0.008}$ & 0.26$_{-0.04}^{+0.05}$ & 37.0$_{-16.3}^{+34.3}$    & 27.3$_{-13.5}^{+19.6}$   & 0 &        MS & L \\
	7771282 & 6248$\pm$77    & 3.24$_{-0.32}^{+0.35}$    & 1.29$_{-0.03}^{+0.03}$ & -0.02$\pm$0.10    & 4.112$_{-0.007}^{+0.007}$ & 1.01$_{-0.18}^{+0.14}$ & 69.5$_{-17.8}^{+14.6}$    & 10.4$_{-1.7}^{+2.3}$     & 0 &        MS & L \\
	7871531 & 5501$\pm$77    & 9.96$_{-1.77}^{+1.93}$    & 0.83$_{-0.02}^{+0.03}$ & -0.26$\pm$0.10    & 4.478$_{-0.005}^{+0.007}$ & 0.33$_{-0.03}^{+0.03}$ & 71.3$_{-13.2}^{+12.0}$    & 33.1$_{-4.1}^{+4.1}$     & 0 &        MS & L \\
	7940546 & 6235$\pm$77    & 2.33$_{-0.08}^{+0.08}$    & 1.40$_{-0.01}^{+0.03}$ & -0.20$\pm$0.10    & 4.007$_{-0.001}^{+0.003}$ & 1.14$_{-0.03}^{+0.03}$ & 78.9$_{-7.9}^{+7.4}$      & 9.9$_{-0.4}^{+0.3}$      & 0 &        MS & L \\
	7970740 & 5309$\pm$77    & 12.98$_{-2.00}^{+1.36}$   & 0.73$_{-0.01}^{+0.03}$ & -0.54$\pm$0.10    & 4.539$_{-0.005}^{+0.004}$ & 0.26$_{-0.02}^{+0.03}$ & 60.7$_{-15.3}^{+17.4}$    & 39.2$_{-9.0}^{+6.7}$     & 0 &        MS & L \\
	8006161 & 5488$\pm$77    & 3.59$_{-1.45}^{+1.53}$    & 0.98$_{-0.03}^{+0.03}$ & 0.34$\pm$0.10     & 4.494$_{-0.007}^{+0.007}$ & 0.34$_{-0.02}^{+0.02}$ & 37.0$_{-3.4}^{+4.1}$      & 20.6$_{-1.8}^{+2.2}$     & 0 &        MS & L \\
	8077137 & 6072$\pm$75    & 6.23$_{-1.23}^{+0.56}$    & 1.12$_{-0.05}^{+0.04}$ & -0.09$\pm$0.10    & 4.056$_{-0.013}^{+0.010}$ & 0.84$_{-0.07}^{+0.06}$ & 72.8$_{-12.6}^{+11.3}$    & 13.0$_{-1.5}^{+1.3}$     & 0 &        MS & K \\
	8150065 & 6173$\pm$101   & 3.83$_{-0.67}^{+0.99}$    & 1.19$_{-0.05}^{+0.04}$ & -0.13$\pm$0.15    & 4.220$_{-0.008}^{+0.008}$ & 0.54$_{-0.13}^{+0.11}$ & 64.0$_{-21.1}^{+18.0}$    & 18.6$_{-4.9}^{+6.4}$     & 0 &        MS & L \\
	8179536 & 6343$\pm$77    & 3.54$_{-0.81}^{+1.04}$    & 1.16$_{-0.06}^{+0.05}$ & -0.03$\pm$0.10    & 4.255$_{-0.010}^{+0.010}$ & 1.46$_{-0.09}^{+0.10}$ & 55.7$_{-7.4}^{+13.3}$     & 6.5$_{-0.8}^{+1.2}$      & 0 &        H & L \\
	8228742 & 6122$\pm$77    & 2.89$_{-0.18}^{+0.16}$    & 1.38$_{-0.02}^{+0.02}$ & -0.08$\pm$0.10    & 4.035$_{-0.005}^{+0.007}$ & 0.64$_{-0.04}^{+0.04}$ & 37.9$_{-4.0}^{+6.2}$      & 11.0$_{-1.3}^{+2.0}$     & 0 &        MS & L \\
	8292840 & 6239$\pm$94    & 3.85$_{-0.75}^{+0.81}$    & 1.15$_{-0.05}^{+0.05}$ & -0.14$\pm$0.10    & 4.240$_{-0.008}^{+0.008}$ & 1.45$_{-0.07}^{+0.07}$ & 76.2$_{-9.1}^{+8.9}$      & 7.7$_{-0.5}^{+0.5}$      & 0 &        MS & K \\
	8349582 & 5699$\pm$74    & 8.03$_{-0.70}^{+0.80}$    & 1.07$_{-0.02}^{+0.02}$ & 0.30$\pm$0.10     & 4.163$_{-0.003}^{+0.004}$ & 0.23$_{-0.06}^{+0.07}$ & 59.6$_{-24.1}^{+20.7}$    & 41.7$_{-14.9}^{+19.6}$   & 1 &        MS & K \\
	8379927 & 6067$\pm$120   & 1.99$_{-0.75}^{+0.85}$    & 1.12$_{-0.04}^{+0.04}$ & -0.10$\pm$0.15    & 4.388$_{-0.007}^{+0.008}$ & 1.12$_{-0.02}^{+0.02}$ & 63.3$_{-2.3}^{+2.5}$      & 9.2$_{-0.2}^{+0.3}$      & 0 &        MS & L \\
	8394589 & 6143$\pm$77    & 4.45$_{-0.83}^{+0.94}$    & 1.04$_{-0.03}^{+0.04}$ & -0.29$\pm$0.10    & 4.322$_{-0.008}^{+0.008}$ & 1.01$_{-0.03}^{+0.03}$ & 71.1$_{-5.9}^{+7.9}$      & 10.9$_{-0.6}^{+0.6}$     & 0 &        MS & L \\
	8424992 & 5719$\pm$77    & 9.61$_{-1.74}^{+1.92}$    & 0.92$_{-0.04}^{+0.04}$ & -0.12$\pm$0.10    & 4.359$_{-0.007}^{+0.007}$ & 0.22$_{-0.06}^{+0.06}$ & 59.1$_{-30.3}^{+21.3}$    & 42.3$_{-17.7}^{+19.4}$   & 1 &        MS & L \\
	8494142 & 6144$\pm$106   & 2.62$_{-0.24}^{+0.26}$    & 1.42$_{-0.02}^{+0.03}$ & 0.13$\pm$0.10     & 4.038$_{-0.005}^{+0.005}$ & 0.67$_{-0.28}^{+0.22}$ & 62.8$_{-23.8}^{+18.5}$    & 14.5$_{-4.5}^{+9.9}$     & 0 &        MS & K \\
	8554498 & 5945$\pm$60    & 5.60$_{-0.42}^{+0.45}$    & 1.20$_{-0.03}^{+0.02}$ & 0.17$\pm$0.05     & 4.007$_{-0.003}^{+0.003}$ & 0.25$_{-0.09}^{+0.21}$ & 48.1$_{-36.6}^{+30.0}$    & 35.7$_{-31.4}^{+25.6}$   & 0 &        MS & K \\
	8694723 & 6246$\pm$77    & 4.69$_{-0.51}^{+0.48}$    & 1.14$_{-0.02}^{+0.02}$ & -0.42$\pm$0.10    & 4.113$_{-0.009}^{+0.007}$ & 0.92$_{-0.05}^{+0.05}$ & 34.7$_{-2.7}^{+3.4}$      & 7.2$_{-0.6}^{+0.8}$      & 0 &        MS & L \\
	8760414 & 5873$\pm$77    & 11.66$_{-1.61}^{+1.28}$   & 0.81$_{-0.02}^{+0.03}$ & -0.92$\pm$0.10    & 4.329$_{-0.005}^{+0.006}$ & 0.69$_{-0.42}^{+0.26}$ & 7.6$_{-1.7}^{+2.3}$       & 2.0$_{-0.4}^{+4.5}$      & 2, 3 &        MS & L \\
	8866102 & 6325$\pm$75    & 2.60$_{-0.53}^{+0.56}$    & 1.23$_{-0.04}^{+0.04}$ & 0.01$\pm$0.10     & 4.262$_{-0.007}^{+0.008}$ & 2.15$_{-0.04}^{+0.04}$ & 78.2$_{-4.7}^{+6.6}$      & 5.3$_{-0.2}^{+0.2}$      & 0 &        H & K \\
	8938364 & 5677$\pm$77    & 10.25$_{-0.65}^{+0.56}$   & 0.99$_{-0.01}^{+0.01}$ & -0.13$\pm$0.10    & 4.173$_{-0.002}^{+0.007}$ & 0.61$_{-0.50}^{+0.27}$ & 8.7$_{-2.3}^{+57.5}$      & 2.0$_{-0.2}^{+87.0}$     & 2 &        MS & L \\
	9025370 & 5270$\pm$180   & 6.55$_{-1.13}^{+1.26}$    & 0.97$_{-0.03}^{+0.03}$ & -0.12$\pm$0.18    & 4.423$_{-0.004}^{+0.007}$ & 0.43$_{-0.04}^{+0.04}$ & 67.5$_{-19.1}^{+15.2}$    & 24.7$_{-4.7}^{+3.7}$     & 0 &        MS & L \\
	9098294 & 5852$\pm$77    & 8.08$_{-0.73}^{+0.99}$    & 0.97$_{-0.03}^{+0.02}$ & -0.18$\pm$0.10    & 4.308$_{-0.007}^{+0.005}$ & 0.36$_{-0.04}^{+0.04}$ & 58.2$_{-16.3}^{+21.0}$    & 27.2$_{-7.0}^{+5.7}$     & 0 &        MS & L \\
	9139151 & 6302$\pm$77    & 1.32$_{-0.75}^{+0.94}$    & 1.18$_{-0.05}^{+0.04}$ & 0.10$\pm$0.10     & 4.382$_{-0.008}^{+0.008}$ & 0.95$_{-0.04}^{+0.04}$ & 73.5$_{-11.0}^{+11.0}$    & 11.6$_{-1.1}^{+0.8}$     & 0 &        H & L \\
	9139163 & 6400$\pm$84    & 1.60$_{-0.22}^{+0.22}$    & 1.40$_{-0.02}^{+0.03}$ & 0.15$\pm$0.09     & 4.200$_{-0.008}^{+0.009}$ & 1.59$_{-0.08}^{+0.07}$ & 33.5$_{-3.0}^{+3.0}$      & 4.0$_{-0.3}^{+0.3}$      & 1 &        H & L \\
	9206432 & 6538$\pm$77    & 1.53$_{-0.30}^{+0.21}$    & 1.38$_{-0.02}^{+0.04}$ & 0.16$\pm$0.10     & 4.220$_{-0.007}^{+0.005}$ & 1.55$_{-0.20}^{+0.17}$ & 34.3$_{-4.2}^{+5.7}$      & 4.1$_{-0.5}^{+1.0}$      & 0 &        H & L \\
	9353712 & 6278$\pm$77    & 2.15$_{-0.13}^{+0.11}$    & 1.51$_{-0.02}^{+0.03}$ & -0.05$\pm$0.10    & 3.943$_{-0.005}^{+0.007}$ & 0.75$_{-0.17}^{+0.16}$ & 37.6$_{-12.6}^{+29.1}$    & 9.5$_{-3.9}^{+7.3}$      & 0 &        H & L \\
	9410862 & 6047$\pm$77    & 6.93$_{-1.33}^{+1.49}$    & 0.97$_{-0.04}^{+0.05}$ & -0.31$\pm$0.10    & 4.300$_{-0.008}^{+0.009}$ & 0.41$_{-0.08}^{+0.09}$ & 46.4$_{-16.7}^{+28.3}$    & 20.6$_{-8.1}^{+9.8}$     & 1 &        MS & L \\
	9414417 & 6253$\pm$75    & 2.65$_{-0.18}^{+0.16}$    & 1.40$_{-0.03}^{+0.02}$ & -0.13$\pm$0.10    & 4.016$_{-0.005}^{+0.005}$ & 1.09$_{-0.05}^{+0.05}$ & 58.1$_{-6.9}^{+9.7}$      & 9.0$_{-0.9}^{+1.1}$      & 0 &        H & L \\
	9592705 & 6174$\pm$92    & 2.33$_{-0.16}^{+0.18}$    & 1.51$_{-0.02}^{+0.03}$ & 0.22$\pm$0.10     & 3.961$_{-0.004}^{+0.003}$ & 0.93$_{-0.09}^{+0.09}$ & 59.3$_{-12.1}^{+17.8}$    & 10.7$_{-1.9}^{+2.1}$     & 0 &        SG & K \\
	9812850 & 6321$\pm$77    & 2.71$_{-0.35}^{+0.46}$    & 1.37$_{-0.05}^{+0.04}$ & -0.07$\pm$0.10    & 4.053$_{-0.009}^{+0.008}$ & 1.54$_{-0.08}^{+0.09}$ & 81.0$_{-10.0}^{+6.4}$      & 7.4$_{-0.5}^{+0.4}$     & 0 &        H & L \\
	9955598 & 5457$\pm$77    & 6.29$_{-1.84}^{+1.95}$    & 0.90$_{-0.03}^{+0.04}$ & 0.05$\pm$0.10     & 4.497$_{-0.005}^{+0.007}$ & 0.29$_{-0.04}^{+0.04}$ & 53.4$_{-11.9}^{+20.8}$    & 31.4$_{-6.4}^{+9.1}$     & 0 &        MS & L \\
	9965715 & 5860$\pm$180   & 2.92$_{-0.75}^{+0.86}$    & 1.21$_{-0.05}^{+0.04}$ & -0.44$\pm$0.18    & 4.272$_{-0.009}^{+0.008}$ & 1.75$_{-0.05}^{+0.05}$ & 58.3$_{-3.2}^{+3.5}$      & 5.6$_{-0.3}^{+0.3}$      & 0 &        MS & L \\
	10068307 & 6132$\pm$77   & 2.36$_{-0.10}^{+0.08}$    & 1.47$_{-0.02}^{+0.01}$ & -0.23$\pm$0.10    & 3.967$_{-0.004}^{+0.004}$ & 0.71$_{-0.03}^{+0.03}$ & 41.7$_{-4.1}^{+6.0}$      & 10.9$_{-1.1}^{+1.5}$     & 0 &        SG & L \\
	10079226 & 5949$\pm$77   & 3.06$_{-0.65}^{+0.70}$    & 1.12$_{-0.03}^{+0.02}$ & 0.11$\pm$0.10     & 4.366$_{-0.005}^{+0.005}$ & 0.65$_{-0.08}^{+0.07}$ & 75.1$_{-23.4}^{+10.6}$    & 16.8$_{-3.0}^{+2.2}$     & 1 &        MS & L \\
	10162436 & 6146$\pm$77   & 2.46$_{-0.11}^{+0.10}$    & 1.45$_{-0.01}^{+0.02}$ & -0.16$\pm$0.10    & 3.981$_{-0.005}^{+0.005}$ & 0.84$_{-0.05}^{+0.05}$ & 25.5$_{-1.8}^{+2.1}$      & 5.9$_{-0.4}^{+0.6}$      & 0 &        SG & L \\
	10454113 & 6177$\pm$77   & 2.89$_{-0.53}^{+0.56}$    & 1.17$_{-0.03}^{+0.02}$ & -0.07$\pm$0.10    & 4.314$_{-0.005}^{+0.005}$ & 0.76$_{-0.07}^{+0.07}$ & 40.9$_{-9.6}^{+26.9}$     & 10.0$_{-2.5}^{+4.8}$     & 0 &        MS & L \\
	10514430 & 5784$\pm$98   & 7.84$_{-0.91}^{+0.40}$    & 1.06$_{-0.02}^{+0.04}$ & -0.11$\pm$0.11    & 4.061$_{-0.004}^{+0.004}$ & 0.18$_{-0.05}^{+0.05}$ & 57.2$_{-32.0}^{+23.5}$    & 53.6$_{-27.2}^{+23.6}$   & 1 &        MS & K \\
	10516096 & 5964$\pm$77   & 7.01$_{-1.45}^{+1.33}$    & 1.06$_{-0.06}^{+0.05}$ & -0.11$\pm$0.10    & 4.169$_{-0.011}^{+0.010}$ & 0.48$_{-0.03}^{+0.03}$ & 71.8$_{-16.2}^{+12.5}$    & 22.6$_{-3.1}^{+1.9}$     & 0 &        MS & L \\
	10586004 & 5770$\pm$83   & 6.43$_{-0.61}^{+0.64}$    & 1.18$_{-0.03}^{+0.02}$ & 0.29$\pm$0.10     & 4.071$_{-0.005}^{+0.005}$ & 0.48$_{-0.17}^{+0.16}$ & 59.3$_{-22.4}^{+20.6}$    & 19.6$_{-6.5}^{+11.1}$    & 1 &        MS & K \\
	10644253 & 6045$\pm$77   & 2.39$_{-0.96}^{+1.12}$    & 1.10$_{-0.04}^{+0.04}$ & 0.06$\pm$0.10     & 4.396$_{-0.008}^{+0.007}$ & 0.24$_{-0.08}^{+0.08}$ & 55.6$_{-30.3}^{+24.0}$    & 38.0$_{-19.1}^{+21.1}$   & 0 &        MS & L \\
	10666592 & 6350$\pm$80   & 2.11$_{-0.24}^{+0.29}$    & 1.50$_{-0.04}^{+0.04}$ & 0.26$\pm$0.08     & 4.017$_{-0.007}^{+0.009}$ & 0.91$_{-0.11}^{+0.11}$ & 47.2$_{-14.8}^{+27.2}$    & 9.5$_{-3.2}^{+3.6}$      & 0 &        H & K \\
	10730618 & 6150$\pm$180  & 3.05$_{-0.29}^{+0.46}$    & 1.34$_{-0.05}^{+0.04}$ & -0.11$\pm$0.18    & 4.062$_{-0.007}^{+0.008}$ & 0.56$_{-0.25}^{+0.23}$ & 55.6$_{-26.2}^{+23.7}$    & 16.1$_{-7.2}^{+13.9}$    & 0 &        MS & L \\
	10963065 & 6140$\pm$77   & 7.15$_{-1.61}^{+1.92}$    & 0.99$_{-0.06}^{+0.06}$ & -0.19$\pm$0.10    & 4.277$_{-0.011}^{+0.011}$ & 0.67$_{-0.03}^{+0.03}$ & 41.8$_{-3.6}^{+4.7}$      & 11.5$_{-1.0}^{+1.3}$     & 0 &        MS & L \\
	11081729 & 6548$\pm$82   & 1.88$_{-0.42}^{+0.59}$    & 1.30$_{-0.05}^{+0.04}$ & 0.11$\pm$0.10     & 4.245$_{-0.009}^{+0.010}$ & 3.35$_{-0.10}^{+0.10}$ & 82.9$_{-5.9}^{+4.9}$      & 3.4$_{-0.1}^{+0.1}$      & 0 &        H & L \\
	11133306 & 5982$\pm$82   & 5.14$_{-0.88}^{+0.86}$    & 1.06$_{-0.03}^{+0.04}$ & -0.02$\pm$0.10    & 4.314$_{-0.004}^{+0.007}$ & 0.39$_{-0.13}^{+0.14}$ & 58.0$_{-24.1}^{+21.9}$    & 24.6$_{-10.0}^{+14.3}$   & 0 &        MS & K \\
	11253226 & 6642$\pm$77   & 1.60$_{-0.13}^{+0.06}$    & 1.41$_{-0.01}^{+0.02}$ & -0.08$\pm$0.10    & 4.173$_{-0.004}^{+0.005}$ & 2.55$_{-0.12}^{+0.11}$ & 49.3$_{-4.4}^{+6.3}$      & 3.4$_{-0.3}^{+0.4}$      & 0 &        H & L \\
	11295426 & 5793$\pm$74   & 6.31$_{-0.34}^{+0.32}$    & 1.07$_{-0.02}^{+0.01}$ & 0.12$\pm$0.07     & 4.280$_{-0.003}^{+0.003}$ & 0.22$_{-0.03}^{+0.03}$ & 55.6$_{-20.9}^{+22.9}$    & 42.6$_{-14.5}^{+11.5}$   & 0 &        MS & K \\
	11401755 & 5911$\pm$66   & 7.10$_{-0.59}^{+0.61}$    & 1.06$_{-0.02}^{+0.03}$ & -0.20$\pm$0.06    & 4.039$_{-0.004}^{+0.004}$ & 0.55$_{-0.10}^{+0.09}$ & 64.6$_{-19.5}^{+17.4}$    & 18.5$_{-4.3}^{+4.7}$     & 0 &        MS & K \\
	11772920 & 5180$\pm$180  & 10.67$_{-2.97}^{+2.73}$   & 0.83$_{-0.04}^{+0.04}$ & -0.09$\pm$0.18    & 4.500$_{-0.008}^{+0.005}$ & 0.31$_{-0.04}^{+0.03}$ & 70.8$_{-14.3}^{+12.6}$    & 35.1$_{-4.7}^{+5.3}$     & 1 &        MS & L \\
	11807274 & 6225$\pm$75   & 3.59$_{-0.45}^{+0.78}$    & 1.24$_{-0.04}^{+0.04}$ & 0.00$\pm$0.08     & 4.135$_{-0.007}^{+0.009}$ & 1.42$_{-0.07}^{+0.07}$ & 76.8$_{-9.4}^{+8.8}$      & 7.9$_{-0.5}^{+0.5}$      & 0 &        MS & K \\
	11853905 & 5781$\pm$76   & 6.71$_{-0.67}^{+0.77}$    & 1.12$_{-0.03}^{+0.02}$ & 0.09$\pm$0.10     & 4.102$_{-0.005}^{+0.004}$ & 0.27$_{-0.09}^{+0.09}$ & 54.7$_{-28.8}^{+24.5}$    & 34.0$_{-17.2}^{+18.9}$   & 1 &        MS & K \\
	11904151 & 5647$\pm$74   & 10.23$_{-0.67}^{+0.83}$   & 0.92$_{-0.02}^{+0.01}$ & -0.15$\pm$0.10    & 4.344$_{-0.003}^{+0.003}$ & 0.24$_{-0.07}^{+0.06}$ & 64.9$_{-28.4}^{+18.1}$    & 40.9$_{-12.8}^{+16.5}$   & 1 &        MS & K \\
	12009504 & 6179$\pm$77   & 3.97$_{-0.43}^{+0.57}$    & 1.17$_{-0.04}^{+0.02}$ & -0.08$\pm$0.10    & 4.211$_{-0.005}^{+0.007}$ & 1.16$_{-0.04}^{+0.03}$ & 71.3$_{-5.7}^{+8.0}$      & 9.4$_{-0.5}^{+0.5}$      & 0 &        MS & L \\
	12069127 & 6276$\pm$77   & 2.01$_{-0.13}^{+0.11}$    & 1.57$_{-0.02}^{+0.03}$ & 0.08$\pm$0.10     & 3.912$_{-0.004}^{+0.005}$ & 0.54$_{-0.33}^{+1.26}$ & 16.6$_{-6.3}^{+54.2}$     & 5.2$_{-3.9}^{+39.9}$     & 1 &        H & L \\
	12069424 & 5825$\pm$50   & 6.67$_{-0.77}^{+0.81}$    & 1.05$_{-0.02}^{+0.02}$ & 0.10$\pm$0.03     & 4.287$_{-0.007}^{+0.007}$ & 0.40$_{-0.01}^{+0.01}$ & 44.7$_{-2.9}^{+6.2}$      & 20.5$_{-1.1}^{+2.0}$     & 1 &        MS & L \\
	12069449 & 5750$\pm$50   & 7.39$_{-0.91}^{+0.89}$    & 0.99$_{-0.02}^{+0.02}$ & 0.05$\pm$0.02     & 4.353$_{-0.005}^{+0.007}$ & 0.31$_{-0.01}^{+0.01}$ & 34.0$_{-2.4}^{+3.0}$      & 21.2$_{-1.5}^{+1.8}$     & 0 &        MS & L \\
	12258514 & 5964$\pm$77   & 4.05$_{-0.16}^{+0.18}$    & 1.26$_{-0.01}^{+0.01}$ & 0.00$\pm$0.10     & 4.126$_{-0.003}^{+0.004}$ & 0.39$_{-0.03}^{+0.04}$ & 35.0$_{-6.6}^{+22.9}$     & 16.7$_{-3.6}^{+10.0}$    & 1 &        MS & L \\
	12317678 & 6580$\pm$77   & 2.46$_{-0.18}^{+0.22}$    & 1.34$_{-0.01}^{+0.04}$ & -0.28$\pm$0.10    & 4.048$_{-0.009}^{+0.008}$ & 1.27$_{-0.14}^{+0.13}$ & 35.3$_{-5.2}^{+10.1}$     & 5.2$_{-0.9}^{+1.8}$      & 0 &        H & L \\
\end{longtable}
\normalsize
\end{landscape}


\end{document}